\documentclass[aps,prb,twocolumn,superscriptaddress]{revtex4-2}
\usepackage{amsmath}
\usepackage{mathtools} 
\usepackage{amssymb}
\usepackage{amsthm}
\usepackage{bm} 
\usepackage{upgreek} 
\usepackage{xcolor}
\usepackage{graphicx}
\usepackage{epstopdf}
\usepackage{braket}
\usepackage{siunitx}
\usepackage{MnSymbol}
\usepackage[colorlinks,linkcolor=blue,anchorcolor=blue,citecolor=blue,urlcolor=blue]{hyperref}
\usepackage{tikz-feynman}

\newcommand{\blue}[1]{{\color{blue}#1}}

\newcommand{\nn}{{\nonumber}}

\newcommand{\eg}{\textit{e.g.{ }}}

\newcommand{\up}{\uparrow}
\newcommand{\dn}{\downarrow}

\newcommand{\av}[1]{\left\langle #1 \right\rangle}
\newcommand{\md}{\mathrm{d}}
\newcommand{\me}{\mathrm{e}}
\newcommand{\w}{\omega}

\newcommand{\mi}{\mathrm{i}}
\begin{document}
\title{Self-consistent theory of $2\times2$ pair density waves in kagome superconductors}

\author{Meng Yao} 
\affiliation{National Laboratory of Solid State Microstructures $\&$ School of Physics, Nanjing University, Nanjing 210093, China}

\author{Yan Wang} 
\affiliation{National Laboratory of Solid State Microstructures $\&$ School of Physics, Nanjing University, Nanjing 210093, China}

\author{Da Wang} \email{dawang@nju.edu.cn}
\affiliation{National Laboratory of Solid State Microstructures $\&$ School of Physics, Nanjing University, Nanjing 210093, China}
\affiliation{Collaborative Innovation Center of Advanced Microstructures, Nanjing University, Nanjing 210093, China}

\author{Jia-Xin Yin} \email{yinjx@sustech.edu.cn}
\affiliation{Department of Physics, Southern University of Science and Technology, Shenzhen, Guangdong 518005, China}

\author{Qiang-Hua Wang} \email{qhwang@nju.edu.cn}
\affiliation{National Laboratory of Solid State Microstructures $\&$ School of Physics, Nanjing University, Nanjing 210093, China}
\affiliation{Collaborative Innovation Center of Advanced Microstructures, Nanjing University, Nanjing 210093, China}

\begin{abstract}
Pair density wave (PDW) is an intriguing quantum matter proposed in the frontier of condensed matter physics.
However, the existence of PDW in microscopic models has been rare.
In this work, we obtain, by Ginzburg-Landau arguments and self-consistent mean field theory, novel $2a_0\times2a_0$ PDW on the kagome lattice arising from attractive on-bond pairing interactions and the distinct Bloch wave functions near the p-type van Hove singularity.
The PDW state carrying three independent wave-vectors, the so-called 3Q PDW, is nodeless and falls into two topological classes characterized by the Chern number $C = 0$ or $C = \pm2$.
The chiral ($C=\pm2$) PDW state presents a rare case of interaction driven topological quantum state without the requirement of spin-orbit coupling.
Finally, we analyze the stabilities and properties of these PDWs intertwining with charge orders, and discuss the relevance of our minimal model to recent experimental observations in kagome superconductors.
Our theory not only elucidates the driving force of the chiral PDW, but also predicts strongly anisotropic superconducting gap structure in the momentum space and quantized transverse thermal conductivity that can be tested in future experiments.
\end{abstract}

\maketitle

\emph{Introduction.}
Pair density wave (PDW) is an exotic superconducting (SC) order with spatially nonuniform order parameters caused by condensing the Cooper pairs with nonzero center of mass momenta. It was first conceived to exist in superconductors with a strong spin-exchange field such that two paired equal-energy electrons do not have opposite momenta anymore \cite{FFLO_Fulde, FFLO_Larkin}. Such a SC state is called FFLO state and has possibly been realized in cold atom systems \cite{cold_atom_1,cold_atom_2,cold_atom_3}.
On the other hand, in cuprates without time-reversal symmetry breaking, the PDW with nonzero pairing momentum has also been proposed phenomenologically to serve as a mother state to generate various daughter states, dubbed as intertwined orders, in the pseudogap phase \cite{mother_state1, PDW_review_2015, PDW_review_2020}.
Such a proposal subsequently triggered extensive studies of PDW in recent years \cite{PDW_review_2015, PDW_review_2020}, and some signatures of the PDW states have been reported in a series of materials such as cuprate and iron-based superconductors \cite{cuprate1,cuprate2,cuprate3,cuprate4,cuprate5,cuprate6,cuprate7,cuprate8,cuprate9,cuprate10,cuprate11, Fe1, Fe2, Fe3},
heavy fermion material UTe$_2$ \cite{UTe2_1,UTe2_2}, kagome compound $A$V$_3$Sb$_5$ \cite{Chen-2021-roton}, and other materials \cite{other_materials1,other_materials2}.
Although many theoretical successes have been achieved at the phenomenological level \cite{phenomenological1, phenomenological2, phenomenological3, topo-pdw, phenomenological4, phenomenological5, interplay_yizhou_2022, Ziqiang-2022},
the existence of spontaneous PDW is rare in microscopic models, since the usual Cooper pairing with zero momentum always enjoys perfect Fermi surface (FS) nesting in the particle-particle channel and thus leaves no room for PDW, unless the interactions are strong enough to break this weak-coupling picture of Cooper instability.
Extensive efforts have been made \cite{microscopic_1998, microscopic_2002,microscopic_2007,microscopic_2008,microscopic_2009,microscopic_2010_1,microscopic_2010_2,microscopic_2011,microscopic_2012,microscopic_2014,microscopic_2015,microscopic_2017,microscopic_2018_1,microscopic_2018_2,microscopic_2019,microscopic_2022_1,microscopic_2022_2,microscopic_2023_1_raghu_prl,microscopic_2023_2,microscopic_2023_3,microscopic_2023_4,microscopic_2023_5,microscopic_2023_6,microscopic_2023_7,microscopic_2023_8_yao,microscopic_2023_9_raghu_prb,microscopic_2023_10,microscopic_2023_11,microscopic_2023_12,microscopic_2023_13,microscopic_2024_1,microscopic_2024_2,microscopic_2024_3},
but the existence of long-range PDW order has not been conclusively demonstrated in realistic models.

In this work, we show that the PDW is favored by on-bond pairing interactions on the kagome lattice near the upper p-type van Hove (vH) filling. 
The resulting $2a_0\times2a_0$ PDW ground states and Bogoliubov-de Gennes (BdG) quasiparticles are analyzed within the Ginzburg-Landau (GL) and self-consistent mean field calculations.
Interestingly, such PDW states are found to intertwine with the charge density wave (CDW) orders and are likely to exist in the kagome superconductors $A$V$_3$Sb$_5$, which are intensively studied in the past few years \cite{Ortiz2019, CsV3Sb5-1, RbV3Sb5-1, KV3Sb5-1, review_Yin2022, review_Neupert2022, review_Jiang2023, review_Wilson2024}.

\begin{figure}
	\includegraphics[width=\linewidth]{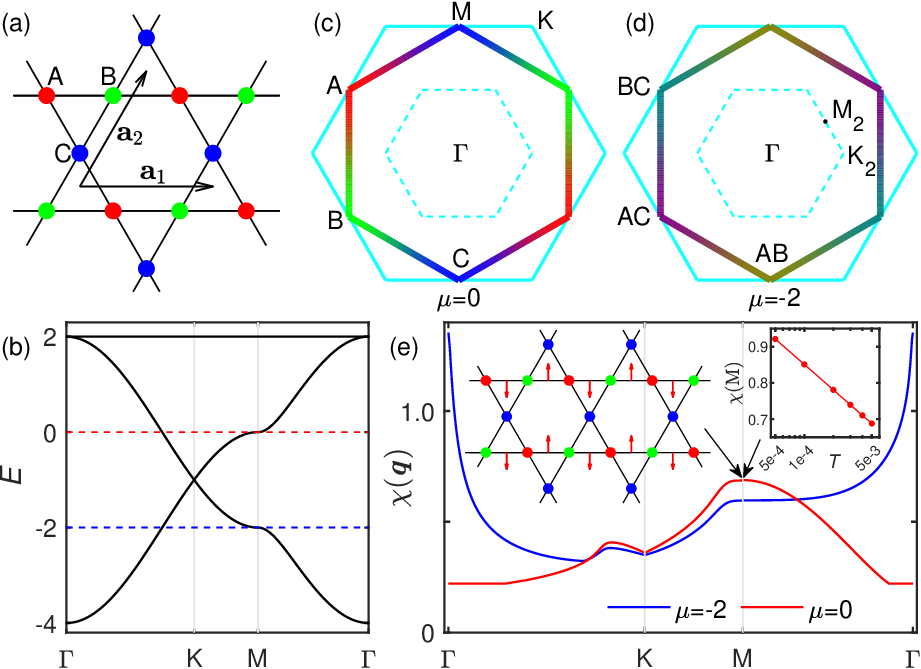}
	\caption{(a) Kagome lattice structure with the three sublattices labeled by A (red), B (green) and C (blue), respectively. ${\bf{a}}_1$ and ${\bf{a}}_2$ denote two primitive translation vectors with length $a_0$. (b) Band dispersion along $\Gamma$-$K$-$M$-$\Gamma$, with the red and blue dashed lines indicating two vH fillings. The FS at the upper vH filling is plotted in (c) in the first Brillouin zone (BZ) with color-scaled sublattice weight. The folded BZ for the $2a_0\times 2a_0$ unitcell  is given by dashed lines. (d) is similar to (c) but for the lower vH filling. (e) The largest spin-singlet pairing susceptibility $\chi(\bm{q})$ for upper (red) and lower (blue) vH fillings at temperature $T=0.005$. For one $M$-point at the upper vH filling, the left inset illustrates the eigen mode with arrow directions (lengths) representing the pairing phases (amplitudes) for relevant bonds (up to a global phase), and the right inset shows the logarithmic temperature dependence of $\chi(M)$.}
	\label{fig:primitive_cell}
\end{figure}

\emph{PDW instability near upper vH filling.}
We study the kagome lattice as shown in Fig.~\ref{fig:primitive_cell}(a), with the three sublattices labeled by A, B, and C, respectively. From the tight-binding Hamiltonian
\begin{align}
H_0=-t\sum_{\av{ij}\sigma}(c_{i\sigma}^\dag c_{j\sigma}+\text{H.c.}),
\end{align}
with only the nearest neighbor hopping $t$ (taken as energy unit), the band structure is shown in Fig.~\ref{fig:primitive_cell}(b). There are two vH fillings $(4\pm1)/12$, as indicated by the two dashed lines corresponding to chemical potential $\mu=0$ and $-2$, respectively. For these fillings, their FSs are exactly the same but their Bloch wave functions are quite different. For the upper vH filling as shown in Fig.~\ref{fig:primitive_cell}(c), the wave function at each $M$-point is contributed by purely one sublattice, hence, called p-type. Instead, for the lower vH filling as shown in Fig.~\ref{fig:primitive_cell}(d), the wave function at each $M$ is contributed by mixing two sublattices, hence called m-type. The difference between these two vH fillings leads to distinct properties as discussed in previous studies \cite{matrix-element-ef1, matrix-element-ef2, matrix-element-ef3, matrix-element-ef4, Ziqiang-2023, microscopic_2023_9_raghu_prb, QGY-MY, YQL}.
Here, we show that the particular p-type vH singularity can lead to PDW instability instead of the usual SC. If the onsite pairing interaction is suppressed by \eg Hubbard repulsion, the leading pairing interaction is expected to be on the nearest neighbor bond.
For this reason, by choosing the basis of on-bond fermion pairs $b_{iI}=c_{i+\delta,\sigma}c_{i+\delta',\sigma'}$ (indices other than unitcell position $i$ are grouped into $I$), we calculate the zero-frequency bare pairing susceptibility $\hat{\chi}(\bm{q})$ with the matrix element
$\chi_{IJ}(\bm{q})=\frac{1}{L^4}\sum_{ij}\int_0^\beta \md\tau\av{b_{iI}(\tau) b_{jJ}^\dag(0)} \me^{i\0q\cdot(\0r_j-\0r_i)}$.
For each $\bm{q}$, we diagonalize $\hat{\chi}(\bm{q})$ to obtain its eigenvalues and eigenvectors. The leading eigenvalue $\chi(\bm{q})$ in spin-singlet channel at the upper vH filling is plotted in Fig.~\ref{fig:primitive_cell}(e). One can see that the highest peak occurs at $M$, which diverges logarithmically with decreasing temperature (see right inset),
indicating the PDW instability with $2a_0\times2a_0$ period.
This instability is caused by the divergence of the dressed susceptibility $\chi/(1-g\chi)$ at finite temperature for any nonzero attraction $g$ in this pairing channel, as a direct generalization of the Cooper mechanism.
Note that for the p-type vH filling, the on-bond pairings only combine different sublattices corresponding to different $M$-points, and thus carry a finite momentum $\0M_3 = \0M_1 + \0M_2$ (mod reciprocal vectors).
This is a new feature in the kagome lattice, which is different from previous studies of PDW with on-bond pairings \cite{microscopic_2011,microscopic_2018_1}.
The eigenvector at one $M$ is illustrated in the left inset of Fig.~\ref{fig:primitive_cell}(e). The arrows indicate the pairing phases (up to a global one) on the relevant bonds. The eigen modes at the other two $M$-points can be obtained by $C_3$ rotations.
As a comparison, we also plot the susceptibility at the lower vH filling in Fig.~\ref{fig:primitive_cell}(e), where the divergence occurs at $\Gamma$, indicating the usual SC ground state with zero momentum.

\begin{figure}
\includegraphics[width=\linewidth]{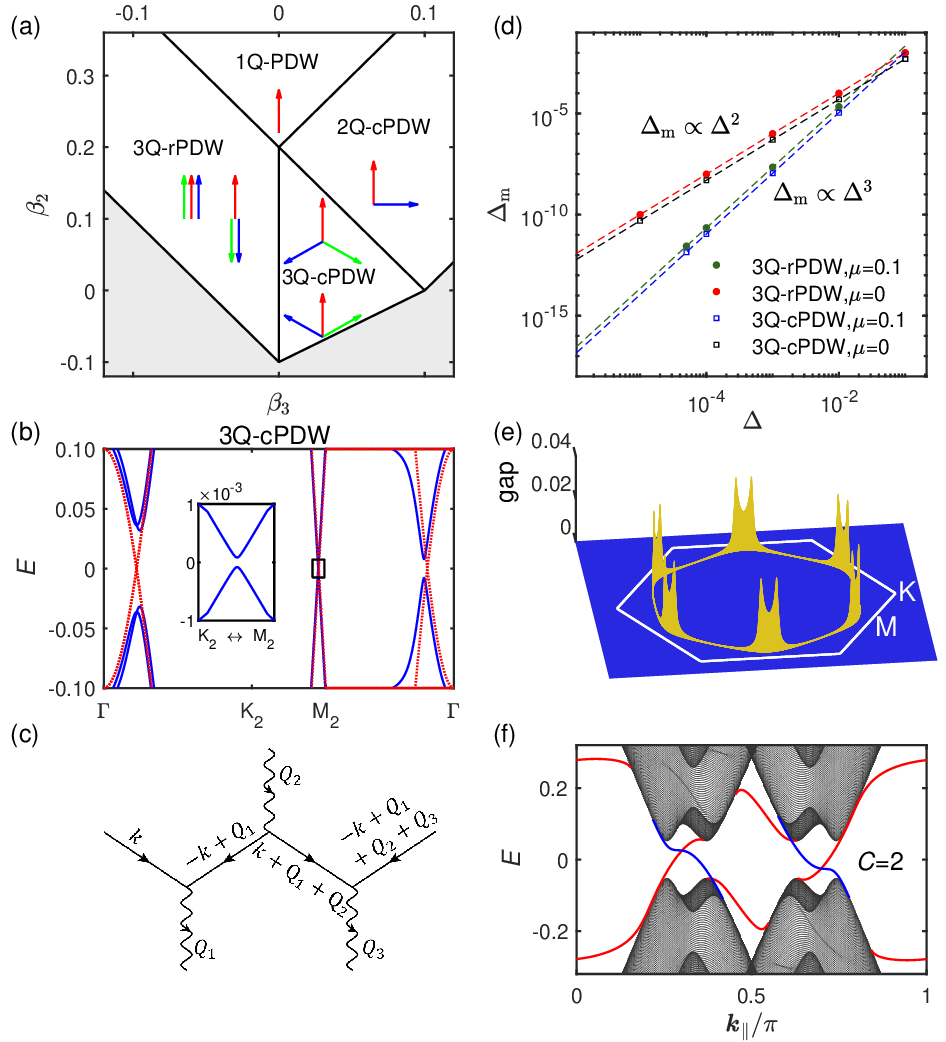}
\caption{(a) PDW phase diagram of the GL theory for $\beta_1=0.1$. The three colored arrows denote ($\psi_1$, $\psi_2$, $\psi_3$) on the complex plane, up to $C_3$-rotations and a global phase. In each 3Q phase, two degenerate patterns are given. The shaded region indicates the absence of a free energy minimum up to fourth order. For these PDW names, ``r'' means real and ``c'' means chiral.
(b) The low-energy BdG quasiparticle bands in the folded BZ for the chiral 3Q states for $\Delta=|\psi_i|=0.02$ and $\mu=0.1$, with the zoom-in near $M_2$ shown as the inset. For comparison, the normal state bands are given as red dashed lines.
(c) Example of the $3$rd-order pair scattering diagrams. The solid line denotes fermion and dashed line denotes pairing field $\psi$ ($\psi^*$) incoming (outgoing) with respect to the three-point vertices. Note that $\0Q_{1,2,3}=-\0Q_{1,2,3}$ and $\0Q_1+\0Q_2+\0Q_3=0$ up to reciprocal vectors.
(d) The gap minimum $\Delta_m$ in the real and chiral 3Q PDW states versus $\Delta$ at ($\mu=0$) or away from ($\mu=0.1$) the vH filling.
(e) The quasiparticle gap near the normal state FS for the chiral 3Q PDW state in (b).
(f) The low-energy bands for the chiral 3Q PDW state for $\Delta=0.4$ on a cylinder with open boundaries along the $\0a_1$ direction, showing two edge states on each boundary (red or blue), consistent with the total Chern number $C=2$.
}
\label{fig:gl}
\end{figure}

\emph{PDW ground states and BdG quasiparticles.}
The above analysis indicates three degenerate PDW modes at different $M$-points, which can be associated with three order parameters $\psi_{1,2,3}$, respectively. Their interplay will lead to different PDW ground states.
In terms of $\psi_{1,2,3}$, we construct a three-component GL free energy up to the fourth order by requiring the following symmetries: $C_3$-rotation, momentum conservation, time reversal (TR) and U(1) gauge symmetries.
Under these constraints, the second order term can only be $|\psi_i|^2$, and the fourth order terms can be $|\psi_i|^4$,  $|\psi_i|^2|\psi_j|^2$ and $(\psi_i^{*2}\psi_j^2+\text{c.c.})$. {Note that there is no third order term, as such a term would violate the U(1) gauge symmetry}. (See Sec.~IV in Supplementary Material \footnote{See Supplemental Material at [URL] for theoretical technique details, further numerical results, and a microscopic derivations of the GL theory.
The Supplemental Material also contains Refs. \cite{scanningJosephson,Jiang-2021-unconventional, Xu-2021-Multiband, zhao-2021-cascade, luo_unique_2023}.}
for microscopic derivation of the GL theory.)  Therefore, we write the uniform GL free energy as
\begin{align}
\label{eq:gl_pdw}
F_{\rm PDW}=-\alpha \sum_{i}|\psi_i|^2&+\beta_1 \sum_{i}|\psi_i|^4+\beta_2\sum_{i<j}|\psi_i|^2|\psi_j|^2\nonumber\\
&+\beta_3\sum_{i<j}(\psi_i^{*2}\psi_j^2+\rm{c.c.}) ,
\end{align}
where $\alpha$ and $\beta_{1,2,3}$ are real numbers.
Clearly, a large positive $\beta_2$ favors nematicity (with unequal $|\psi_i|$), while a positive $\beta_3$ causes phase frustration and favors TR breaking. By varying $\beta_2$ and $\beta_3$ under a fixed value of $\beta_1$, we obtain four possible phases, as shown in Fig.~\ref{fig:gl}\blue{(a)}. Following the usual convention, we use the number of nonzero components to label different PDW phases.
In the 1Q phase, only one component is nonzero, which is an extremely nematic PDW. In the 2Q phase, the two nonzero components are found to have the same amplitude but with a relative phase $\pi/2$, which is a chiral 2Q PDW state. The 3Q phases with $|\psi_1|=|\psi_2|=|\psi_3|$ are further divided into two classes: real (up to a global phase) and chiral (with relative phase $2\pi/3$ or $\pi/3$ between each two components). These two types of PDWs can also be denoted as $s$- and $d\pm id$ wave, respectively, in view of the phase winding along the holo hexagons and the small triangles (detouched from the holo hexagons).
Both the chiral-2Q and chiral-3Q PDW states break the time reversal symmetry.

Next, let us examine the BdG quasiparticles in these PDW states,
by solving the BdG Hamiltonian
\begin{align} \label{eq:bdg}
H=H_0+\sum_{i,\0k,ab} \Delta_i^{ab} c_{a,-\0k+\0Q_i\up}^\dag c_{b,\0k\dn}^\dag + \text{H.c.}
\end{align}
where $i$ labels the three $M$-points, $a$ and $b$ denotes sublattices, and the pairing order parameter $\Delta_i^{ab}$ is set according to the eigenvector of the pair susceptibility. For example, for the one shown in Fig.~\ref{fig:primitive_cell}(e), it only connects A and B sublattices. A common belief is that the Fermi surface cannot be fully gapped since the pairing electrons with momenta $\0k$ and $-\0k+\0Q$ cannot reside on the FS simultaneously for all momenta \cite{PDW_review_2020}. Here, however, we show that this belief is not true for the 3Q PDW states. In Fig.~\ref{fig:gl}(b), we plot the quasiparticle bands in the folded BZ for the chiral 3Q PDW state (the real 3Q state is similar and not shown).
The 3Q-PDW opens a full excitation gap, with the gap minimum $\Delta_m$ near $M_2$.
This full gap behavior is not a numerical artifact but intrinsic to the 3Q PDW states, which is fundamentally different from the 1Q and 2Q states with residual FSs (see Fig.~S1 in Supplementary Material \footnotemark[\value{footnote}]).
The full gap in the 3Q PDW states can be understood by a higher order degenerate perturbation theory, in which the single particles (described by $H_0$) are scattered by the pairing terms (in the BdG Hamiltonian Eq.~\ref{eq:bdg}), as illustrated in Fig.~\ref{fig:gl}(c): $\0k\to -\0k+\0Q_1\to \0k+\0Q_1+\0Q_2 \to -\0k+\0Q_1+\0Q_2+\0Q_3$. Since $\0Q_1+\0Q_2+\0Q_3=0$ (up to reciprocal vectors), the third order process always connects $(\0k,-\0k)$ for any $\0k$ and thus gaps out the entire FS.
In this mechanism, the minimal gap is expected to be proportional to the cube of the order parameter amplitude $\Delta=|\psi_i|$, which is in good agreement with the numerical results at $\mu=0.1$ as shown in Fig.~\ref{fig:gl}(d). Note that if the system is exactly at the upper vH filling ($\mu=0$), the second order process is enough to connect $(\0k,\0k+\0Q)$ of equal energy due to the perfect FS nesting. Therefore, the power exponent is reduced from $3$ to $2$ as seen in Fig.~\ref{fig:gl}(d).
For comparison with experiments, we plot the excitation gap near the normal state FS in the unfolded BZ as shown in Fig.~\ref{fig:gl}(e), by calculating the quasiparticle spectral function thereon (see Fig.~S2 in Supplementary Material).
The $\0k$-dependent excitation gap is maximal (minimal) in the $\Gamma-M$ ($\Gamma-K$) direction. This seems to be consistent with the gap variation on one out of several FSs (in the absence of CDW gap) in a recent ARPES experiment \cite{gapless-ARPES} in $A$V$_3$Sb$_5$ up to the resolution uncertainties.
In the fully gapped 3Q PDW states, the Chern number $C$ is well defined, which is found to be $0$ for the real PDW and $\pm2$ for the chiral ones.
In Fig.~\ref{fig:gl}(f), we show the edge states (two chiral edge modes on each boundary) for the chiral 3Q phase on a cylinder with open boundaries along the $\0a_1$ direction. (Here, a large PDW pairing is
assumed to reduce the finite size effect but without loss of the qualitative physics.)
It is remarkable that such a topological chiral PDW could emerge from an interacting model without the necessity of spin-orbit coupling.
Moreover, the chiral edge states are Weyl ferminons, and should exhibit a quantized thermal Hall conductance, $\kappa_{xy}/T = C(\pi^2/3)(k_B^2/h)$ \cite{thermalHall}, which could be probed in future experiments.

\emph{Self-consistent mean field calculations.}
The above GL analysis indicates four PDW candidate ground states depending on the phenomenological GL parameters $\beta_{1,2,3}$. In this section, we narrow down and sharpen the results by self-consistent mean field calculations with an on-bond pairing interaction
\begin{align}
\label{eq:j_term}
H_P=-J\sum_{\av{ij}\sigma\tau}(\sigma c_{i\sigma}^\dag c_{j\bar{\sigma}}^\dag) (\tau c_{j\bar{\tau}} c_{i\tau}),
\end{align}
where $\av{ij}$ denote the nearest neighbor bonds, $\bar{\sigma}=-\sigma=\pm1$ and $\bar{\tau}=-\tau=\pm1$.
The total Hamiltonian is
$H_0 +H_P$, and the quartic interaction is decoupled in the pairing channel $\sum_{\av{ij}\sigma\tau}-J\av{\sigma c_{i\sigma}^\dag c_{j\bar{\sigma}}^\dag} (\tau c_{j\bar{\tau}} c_{i\tau})-J(\sigma c_{i\sigma}^\dag c_{j\bar{\sigma}}^\dag) \av{\tau c_{j\bar{\tau}} c_{i\tau}} +J\av{\sigma c_{i\sigma}^\dag c_{j\bar{\sigma}}^\dag} \av{\tau c_{j\bar{\tau}} c_{i\tau}}$. We require self-consistency in the order parameter $\Delta_{ij}=\sum_\tau \tau \av{c_{j\bar{\tau}} c_{i\tau}}$.
Here, for simplicity, we do not explicitly distinguish the unitcell and sublattice indices anymore.
Since the PDW instability occurs at $M$, we impose the $2a_0\times2a_0$ periodicity on the lattice.
After self-consistency is achieved, we also calculate the onsite and on-bond charge densities $n_i=\av{c_{i\sigma}^\dag c_{i\sigma}}$ and $\chi_{ij}=\av{c_{j\sigma}^\dag c_{i\sigma}}$, from the latter we obtain the current $\text{Im}(\chi_{ij})$ and valence-bond strength $\text{Re}(\chi_{ij})$.

\begin{figure}
	\includegraphics[width=\linewidth]{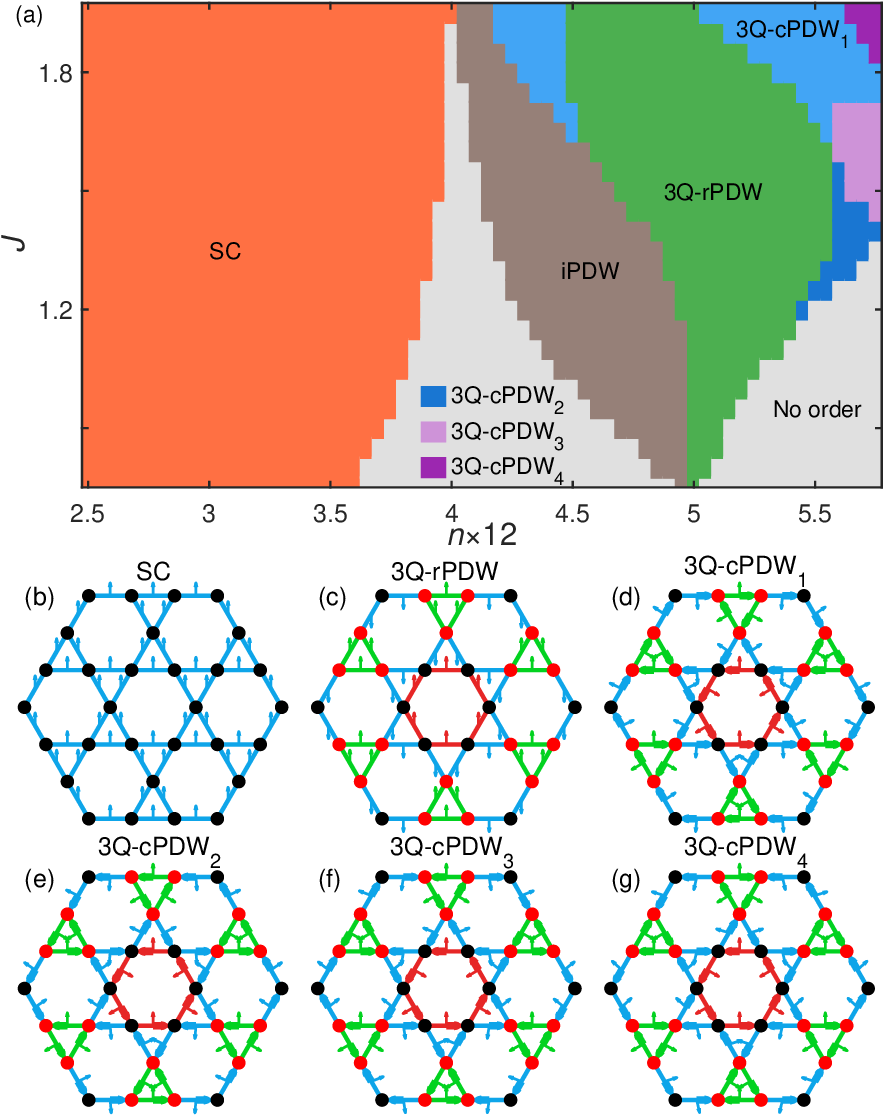}
	\caption{(a) Mean field phase diagram with respect to on-bond pairing interaction $J$ and filling level $n$ at $T=0.005$, with different phases represented by different colors. The brown region denotes incommensurate PDW (iPDW). The pairing patterns of different phases are illustrated in (b) to (g) as arrows starting from each bond center. In addition, the induced CDW patterns are also given, with black/red dots denoting $n_i$, blue/red/green bonds denoting Re$\chi_{ij}$, and arrows along each bond denoting Im$\chi_{ij}$.}
	\label{fig:phase_onlyj}
\end{figure}

Our results are summarized in the phase diagram Fig.~\ref{fig:phase_onlyj}(a). Near the lower vH filling, we find the usual SC with the pairing pattern shown in Fig.~\ref{fig:phase_onlyj}(b). While near the upper vH filling, we obtain 3Q PDW states only.
These results are consistent with the diverging pair susceptibility at $\Gamma$ ($M$) near the lower (upper) vH filling.
Larger supercells (up to $6a_0\times6a_0$) have been used to check the robustness of these $2a_0\times2a_0$ PDW.
In the real 3Q PDW phase as shown in Fig.~\ref{fig:phase_onlyj}(c), one can see that the PDW induces secondary onsite and valence-bond CDWs.
The chiral 3Q states are more complicated. It further induces different types of loop current (LC) orders, according to which we further divide it into four subphases as shown in Figs.~\ref{fig:phase_onlyj}(d-g).
For both the real and chiral 3Q PDWs, secondary CDW orders are induced, manifesting intertwining between these two types of orders.
The strong intertwining between PDW and other orders has been proposed to explain the variety of competing orders in the pseudogap phase of cuprates \cite{PDW_review_2015}. Our study here is clear evidence of the intertwining between PDW and CDW on the kagome lattice based on microscopic calculations.
On the other hand, in the brown region below the upper vH filling, we find larger unitcells may further lower the total energy. This is consistent with the fact that the momentum with the largest $\chi(\0q)$ varies gradually from $M$ to $K$ (not shown) as the filling level decreases from $\frac{5}{12}$ (upper vH) to $\frac14$ (Dirac point). Therefore, we attribute this region as incommensurate PDW.

\begin{figure*}
\includegraphics[width=0.9\linewidth]{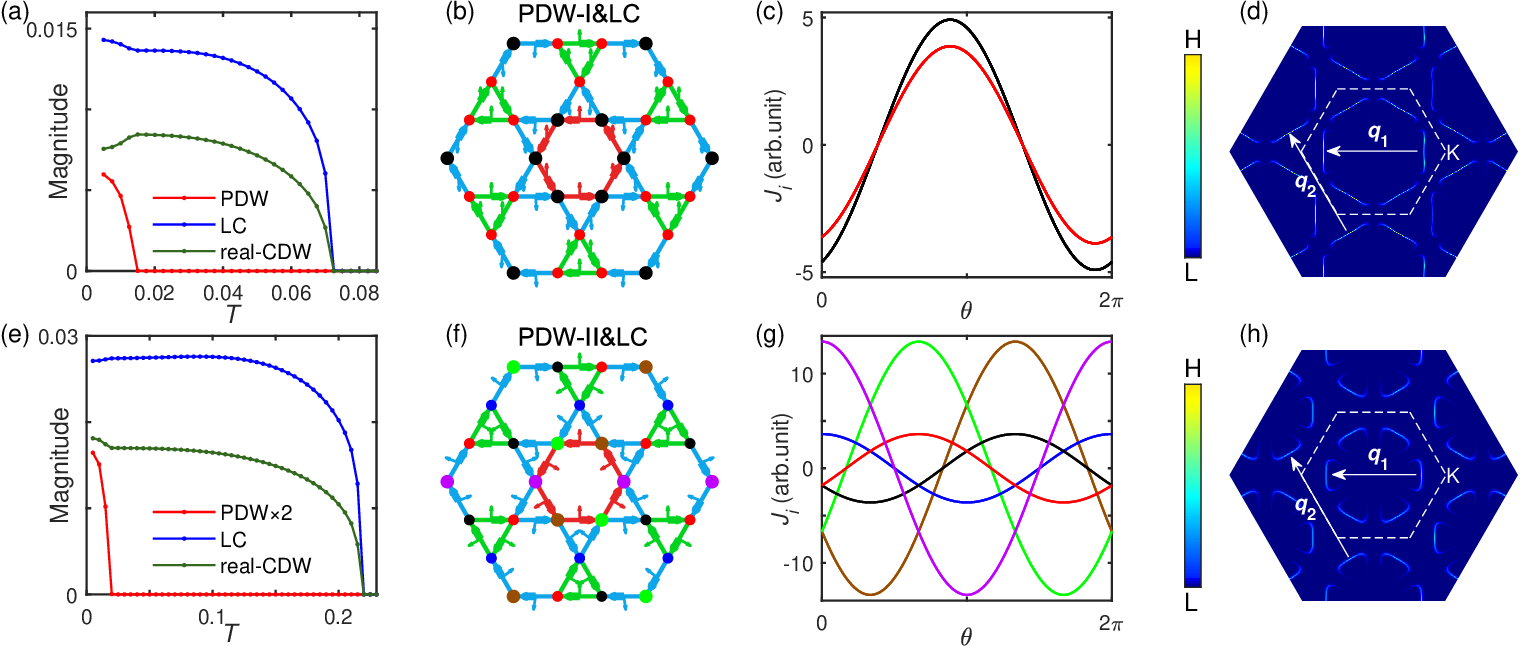}
\caption{ (a-d) mean field results for $V_1=1$, $V_2=2$, $J=1.2$, $n=0.45$, including temperature-dependence of the order parameters in (a), PDW/CDW patterns in (b), local Josephson currents $J_i(\theta)$ in (c), and unfolded spectral function at zero energy in (d). The pattern convention in (b) is similar to Fig.~\ref{fig:phase_onlyj}, except the dot size now denotes $n_i$ while the dot color indicates different line-shape of $J_{i}(\theta)$ in (c). (e-h) are similar plots to (a-d) but for $V_1=1.5$, $V_2=3$, $J=1.6$, $n=0.47$.
}
\label{fig:phase_v1v2j}
\end{figure*}

\emph{PDW on CDW: possibility in $A$V$_3$Sb$_5$.}
Finally, we examine the possibilities of these PDWs in $A$V$_3$Sb$_5$, for which the superconductivity occurs inside the CDW state \cite{Ortiz2019, CsV3Sb5-1, RbV3Sb5-1, KV3Sb5-1, review_Yin2022, review_Neupert2022, review_Jiang2023, review_Wilson2024}.
We have seen that the PDWs can induce CDWs. Conversely, the background of CDWs will inevitably affect the PDWs.
In $A$V$_3$Sb$_5$, the microscopic mechanism of (chiral) CDW is still an unsettled issue \cite{cdw-1,cdw-2,cdw-3,cdw-4,cdw-5,cdw-6}. For our purpose, in order to obtain robust LC (chiral CDW) orders, we adopt the $V_1$-$V_2$ interaction \cite{Ziqiang-2023}
\begin{align}
H_V=V_1\sum_{\av{ij}}n_in_j + V_2\sum_{\av{\av{ij}}}n_in_j ,
\end{align}
with $V_1(V_2)$ the (next) nearest neighbor Coulomb repulsion. In the following, the next nearest neighbor hopping $t'=-0.05$ is also included in $H_0$ to account for the FS revealed by angle-resolved photoemission spectroscopy (ARPES) \cite{ARPES-2022-1, ARPES-2022-2, li-prx-2023, review-1}.
The pairing interaction term $H_P$ in Eq.~\ref{eq:j_term} is also kept. The total Hamiltonian is $H_0+H_P+H_V$.
In the following mean field calculations, we decouple $H_P$ in the pairing channel, and decouple $H_V$ in the charge bond channel $\sum_{\langle i,j \rangle, \av{\av{ij}}, \sigma}-V_{ij}\langle c_{j\sigma}^{\dagger}c_{i\sigma} \rangle c_{i\sigma}^{\dagger}c_{j\sigma}-
V_{ij}c_{j\sigma}^{\dagger}c_{i\sigma}  \langle c_{i\sigma}^{\dagger}c_{j\sigma}\rangle
+V_{ij}\langle c_{j\sigma}^{\dagger}c_{i\sigma} \rangle \langle c_{i\sigma}^{\dagger}c_{j\sigma} \rangle$. Exactly speaking, both the quartic $H_P$ and $H_V$ should be decoupled in all possible channels. Here, we just take them as the effective interactions to drive PDW and CDW, respectively.

In Fig.~\ref{fig:phase_v1v2j}, we show two typical mean field solutions
For the first case ($V_1=1$, $V_2=2$, $J=1.2$, $n=0.45$), the temperature dependence of the order parameters are shown in Fig.~\ref{fig:phase_v1v2j}(a), indicating that the PDW can occur inside the CDW state with a lower $T_c$. Fig.~\ref{fig:phase_v1v2j}(b) shows the ground state with coexisting (almost) real 3Q PDW and chiral CDW. Of course, due to the chiral CDW background, the real 3Q PDW also acquires a weak chirality.
This can be understood in the effective GL theory by adding the third-order coupling term
\begin{align}
\gamma_3(\psi_1^{*}\psi_2\phi_3+\psi_2^{*}\psi_3\phi_1+\psi_3^{*}\psi_1\phi_2+{\rm{c.c.}}),
\end{align}
to $F_{\rm PDW}$, where $\psi_i$/$\phi_i$ are PDW/CDW order parameters and $\gamma_3$ is the coupling strength. Clearly, the chiral CDW (complex $\phi_i$) drives the chirality of PDW, regardless of the sign of $\gamma_3$.
Similarly, Figs.~\ref{fig:phase_v1v2j}(e,f) show the results for the other case ($V_1=1.5$, $V_2=3$, $J=1.6$, $n=0.47$) with the chiral 3Q PDW occurring inside the chiral CDW state. Due to the chiral CDW background, the relative phases between each two PDW components deviate but only slightly from $2\pi/3$.
Besides the above 3Q PDW phases, we also obtain a nematic ($\sim$1Q) PDW phase in the background of nematic CDW (see Fig.~S4 in the Supplementary Material \footnotemark[\value{footnote}]), for which a weak chirality is also observed as a result of the intertwining between the two orders. In addition, more results including a phase diagram near the upper vH filling can be found in section III of the Supplementary Material  \footnotemark[\value{footnote}].

We see that the PDW can exist at a lower temperature inside the CDW phase due to the intertwining between these two types of orders. This makes it not only of theoretical interest but also a likely candidate for the superconductivity in $A$V$_3$Sb$_5$. In particular, in a recent experiment \cite{Deng_N_2024}, some signatures of the $2a_0\times2a_0$ PDW have been reported. In view of various experimental progresses, we present some supporting theoretical discussions.
(1) We calculate the local Josephson currents $J_i(\theta)$ versus the Josephson phase difference $\theta$ (see Sec.~I in Supplementary Material  \footnotemark[\value{footnote}] for technical details), as shown in Figs.~\ref{fig:phase_v1v2j}(c,g) for the two mean field states, respectively. The critical current $J_{ic}=\text{max}[J_i(\theta)]$, as measured in the scanning Josephson spectroscopy, shows a $2a_0\times2a_0$ period, which is consistent with the experiment.
(2) The 3Q PDW states are fully gapped but the minimal gap is small, hence, behaving as gapless superconductors with ``residual FSs'' for realistic experiments with finite energy resolutions. Such a property can be used to distinguish it from the usual SC. As examples, in Figs.~\ref{fig:phase_v1v2j}(d,h), we plot the unfolded single-particle spectral function at zero energy with a realistic scattering rate $\eta=0.005$, for the two mean field states, respectively.
The ``residual FSs'' provide characteristic scattering vectors $\0q_1$ and $\0q_2$ (up to rotation symmetry), which should be visible in the quasiparticle inteference (QPI) pattern (see Fig.~S6 in Supplementary Material  \footnotemark[\value{footnote}]).
We should note, however, that in order to make closer comparison to experiments, one has to take into account the multiple orbitals (or bands) into account, which are beyond our minimal model but deserve further elaborations.
(3) The multi-Q (mQ) PDW provides a possibility to support fractional quantum flux $\Phi= \frac{1}{m}\sum_{i}W_i\Phi_0$ where $\Phi_0=h/2e$ is the Abrikosov quantum flux and $W_i$ is the winding number of $\psi_i$. This is a direct generalization of the two-component GL theory for the usual SC \cite{Sigrist1989,Babaev2002}, and provides an alternative to the charge-4e or -6e pairing to explain the experimentally-observed half or one-third quantum flux \cite{4e6e_PRX_2024}.

\emph{Summary.}
We have found robust $2a_0\times2a_0$ PDWs on the kagome lattice around the upper vH filling. The properties of these PDWs are studied in GL and mean field theories. By further considering the CDW background, these PDWs are found to be consistent with some of the intriguing superconducting phenomena in $A$V$_3$Sb$_5$.
In particular, our work provide a minimal self-consistent model for the chiral $2a_0\times2a_0$ PDW ground state in the kagome lattice with broken time-reversal symmetry and exhibiting topological nontrivial features.

This work is supported by National Key R\&D Program of China (Grant No. 2022YFA1403201), National Natural Science Foundation of China (Grant No. 12374147, No. 12274205 and No. 92365203).

\vspace{10pt}
\begin{center}
  {\Large Appendix }
\end{center}

\appendix
In this appendix, we first give technical details for the mean field theory, unfolded spectral function, local density of states (LDOS), quasiparticle interference (QPI), and local Josephson current. Then we provide further numerical results referred to but not presented in the main text, in the pure PDW states, and the PDW states on CDW backgrounds.  Finally, a microscopic derivation of the GL theory is given.

\section{Technical details}
\subsection{Mean field theory}
The mean field Hamiltonian with pairing and hopping order parameters on bonds can be written as follows, in the Nambu basis,
\begin{align}
    H=\sum_{ij}\psi_i^\dag h_{ij} \psi_j ,
\end{align}
where $\psi_i^\dag=\begin{bmatrix}
    c_{i\up}^\dag &c_{i\dn}
\end{bmatrix}$ is the Nambu spinor at site $i$, and $h_{ij}$ is a $2\times2$ single-particle Hamiltonian
\begin{align}
    h_{ij}=\begin{bmatrix}
        -t_{ij}-\chi_{ij}-\mu\delta_{ij} & -\Delta_{ij} \\
        -\Delta_{ij}^* & t_{ij}+\chi_{ij}^*+\mu\delta_{ij}
    \end{bmatrix} ,
\end{align}
where $t_{ij}=t$ on nearest neighbor (NN) bonds, $\mu$ is the chemical potential, $\chi_{ij}=V_{ij}\av{c_{j\sigma}^\dag c_{i\sigma}}$ (assumed to be independent of the spin $\sigma$) is the hopping order parameter on the bond $\av{ij}$, with Coulomb interaction $V_{ij}=V_{1}$  ($V_2$) on NN (NNN) bonds, and finally, $\Delta_{ij}=J\av{c_{j\dn} c_{i\up}-c_{j\up} c_{i\dn}}$ is the singlet pairing order parameter on NN bonds with attractive pairing interaction of strength $J$. We assume the order parameters are periodic from unitcell to unitcell, and a unitcell contains $m\times m$ primitive cells (so that there are $n=3m^2$ lattice sites within a unitcell). For the $2\times2$ PDW and CDW states, we take $m=2$. (To test whether this is indeed the ground state pattern, we can do the mean field calculations with a larger unitcell, or $m>2$). In order to take care of the periodicity explicitly, we denote a lattice site as $i=(R,r)$, where $R$ labels the unitcell and $r$ labels the sites within that unitcell, and $R_i=R+r$ is the position vector of this site. We then combine the fields within a unitcell as a spinor
\begin{align}
    \psi_R^\dag=(
        c_{\up R,1}^\dag , c_{\up R,2}^\dag ,\cdots, c_{\up R,n}^\dag , c_{\dn R,1} , c_{\dn R,2} ,\cdots ,c_{\dn R,n} ) ,
\end{align}
by which we can rewrite the Hamiltonian as
\begin{align}
    H=\sum_{R,R'}\Psi_R^\dag h(R,R') \Psi_{R'} = \sum_{k} \Psi_k^\dag h_k \Psi_k ,
\end{align}
where $h$ is now an $2n\times2n$ matrix (which actually depends on $R-R'$ only by the assumed periodicity), whose row (or column) index is composed of $(\mu,r)$, where $\mu=\up,\dn$ labels the particle- or hole-part in the Nambu space, and $h_k$ is the Fourier transformation of $h(R,R')$, defined as, by elements,
\begin{align}
    h_k^{\mu r,\mu' r'} = \sum_{R'} h^{\mu r,\mu'r'}(R,R')\me^{-ik\cdot(R+r-R'-r')},
\end{align}
where $k$ is the momentum. Note we include the phase $k\cdot(r-r')$ in the Fourier transformation for later convenience without changing the eigen spectrum (this operation is just a unitary transformation). We define the eigen states of $h_k$ as $h_k \ket{k\ell}=E_{k\ell} \ket{k\ell}$, for $\ell=1,2,\cdots,2n$, and denote the particle- and hole-components of the wave function via $\ket{k\ell}=(u_{k\ell},v_{k\ell} )^t$. We calculate the order parameters self-consistently as follows, for $R_i=R+r$ and $R_j = R'+r'$,
\begin{align}
    \chi_{ij}&=\frac{V_{ij}}{N_c}\sum_{k\ell} u_{k\ell}^*(r')u_{k\ell}(r)f(E_{k\ell})\me^{ik\cdot(R_j-R_i)}, \\
    \Delta_{ij}&=\frac{J}{N_c}\sum_{k\ell}\left[ v_{k\ell}^*(r')v_{k\ell}(r)f(E_{k\ell})\me^{ik\cdot(R_j-R_i)} \right.\nn \\
        &\left. \qquad \qquad+ (R_i\leftrightarrow R_j)\right],
\end{align}
where $N_c$ is the total number of unitcells on the lattice, $f(E)$ is the Fermi distribution function, and the summation over $k$ is limited within the reduced Brillouin zone (BZ). The local charge density at site $i=(R,r)$ is given by
\begin{align}
    n_i=\frac{1}{N_c}\sum_{k\ell}|u_{k\ell}(r)|^2 f(E_{k\ell}),
\end{align}
where we tune the chemical potential to match the average electron filling.

By the self-consistent mean field calculations, we obtain the results for pure PDW states in Fig.~3, and coexisting PDW/CDW states in Fig.~4 in the main text. Further results will be given in the following.

\subsection{Unfolded spectral function}
From the mean field eigenstates, we obtain the Green’s function (a $2n\times2n$ matrix) as a function of momentum at a complex frequency $z$,
\begin{align}
G(k,z)=\sum_{i}\frac{\ket{k\ell}\bra{k\ell}}{z-E_{k\ell}} .
\end{align}
We set $z=i\w_n$ for Matsubara Green’s function at the Matsubara frequency $\w_n$, and set $z=\w+i\eta$ for retarded Green’s function at real-frequency $\w$, and $\eta$ is a smearing factor characteristic of an underlying scattering rate. In this work, unless specified, we use $\eta=0.005$.
{A finite smearing factor is always needed to calculate the retarded Green's function. On one hand, the smearing factor may represent the elastic scattering in the real material. On the other hand, a propriate smearing factor is used to reduce the finite-size effect in the calculation. However, we would like to note that if the mini-gap is too small, it may be smeared by smearing factor or elastic scattering within the material.}

The unfolded spectral function, which is directly measured by ARPES, can be expressed simply as, for arbitrary momentum $k$ in the unfolded BZ,
\begin{align}
    A(k,\w)=-\frac{2}{\pi n}\text{Im}\left[ \sum_{r,r'} G^{r\up,r'\up}(k,\w+i\eta) \right],
\end{align}
where the factor of $2$ follows from spin summation.
For lattices (such as the kagome lattice) with many sites inside a primitive cell, the unfolded spectral function is not necessarily periodic from BZ to extended BZ any more. The spectral weight in the first BZ turns out to be weaker than that in the second BZ. Following the usual convention we plot the stronger spectral function from the second BZ back in the first BZ. Two typical results are shown in Fig.~4(d,h) in the main text.

\subsection{Local density of states and quasiparticle interference}
From the mean field eigenstates, we obtain the real-space Green’s function (a $2\times2$ matrix in the Nambu basis),
\begin{align}
    G^{\mu\nu}(i,j;z)=\frac{1}{N_c}\sum_{k\ell} \frac{ \av{r\mu|k\ell}\av{k\ell|r'\nu} }{z-E_{k\ell}} \me^{ik\cdot(R+r-R'-r')},
\end{align}
where $\mu$ and $\nu$ denote the Nambu components, $\ket{r\mu}$ denotes a Nambu-basis state at real-space position $r$, and $i=(R,r)$ and $j=(R',r')$. The local density of states can be obtained as
\begin{align}
    \rho(i,\w)=-\frac{2}{\pi}\text{Im}\left[ G^{\up\up}(i,i,\w) \right].
\end{align}
If we add scalar impurity potentials $V_I \tau_3$ (where $\tau_3$ is the Pauli matrix in the Nambu basis) on the lattice sites $I$ (henceforth upper-case letters denote impurity sites), the corrected Green’s function at an arbitrary site $i$ (which may or may not be an impurity sites) can be obtained by the T-matrix formalism, dropping the frequency argument for short,
\begin{align}
    \tilde{G}(i,i)=G(i,i)+\sum_{IJ}G(i,I)T_{IJ}G(J,i),
\end{align}
where the summation is over impurity sites, and the T-matrix is given by
\begin{align}
T_{IJ}=V_I\tau_3\delta_{IJ}+V_I\tau_3\sum_{I'}G(I,I')T_{I'J} .
\end{align}
We can use $\tilde{G}$ to obtain the LDOS in the presence of the impurities $\tilde{\rho}(i,\w)$. The QPI data is the power-spectrum of the modified LDOS $|\tilde{\rho}(q,\w)|^2$, with $\tilde{\rho}(q,\w)$ the Fourier transformation of $\tilde{\rho}(i,\w)$:
\begin{align}
\tilde{\rho}(q,\w)=\frac{1}{N}\sum_{i}\tilde{\rho}(i,\w)\me^{-iq\cdot R_i}.
\end{align}
In our calculations, we place the impurity at the center of a triangle which couples to the densities of the three nearest neighbor sites with equal strength.
The resulting LDOS and QPI will be shown and discussed in subsequent sections.

\subsection{Local Josephson current}
Suppose a conventional SC lead is connected as a probe in the tunneling limit to the lattice in the form of point-contact. We want to know the Josephson current as a function of the Josephson phase difference, which we define as the phase of the probe $\theta$, setting the overall phase of the PDW as the reference point. The onsite $2\times2$ Green’s function on the probe tip can be well approximated as
\begin{align}
    g(i\w_n,\theta)\approx -\pi N_0 \frac{i\w_0\tau_0+\Delta_0(\cos\theta\tau_1-\sin\theta\tau_2)}{\sqrt{\Delta_0^2+\w_n^2}},
\end{align}
where $N_0$ is the DOS and $\Delta_0$ is the pairing gap in the tip. The local $2\times2$ Green’s function on the lattice site $i$ is just given by $g_i (i\w_n )=G(i,i,i\w_n)$, see the previous section. With these ingredients, we can obtain the Josephson current in the tunnelling limit as,
\begin{align} \label{eq:josephson}
	J_i(\theta) =-\mi a^2 T\sum_{\mu,n}\mu g_{i}^{\bar{\mu}\mu }(\text{i}\w_n)\tilde{g}^{\mu\bar{\mu}}(\text{i}\w_n,\theta)+\text{c.c.}   ,
\end{align}
where $\mu=1(\up),-1(\dn)$, $\bar{\mu}=-\mu$, and $a$ is the tunneling matrix element. In our calculations, we set $\Delta_0=0.01$ and take the temperature $T\rightarrow 0$.
Note the Josephson current is not directly measured by STM. Since the working temperature may be well above the Josephson coupling energy, the Josephson tunneling happens in the diffusive regime. Nonetheless, this can lead to a zero-energy-bias peak in the tunneling spectrum, the height of which is proportional to the square of the theoretical Josephson critical current \cite{scanningJosephson}. It is in this sense that STM can probe the Josephson critical current.
In the main text, we presented the resulting $J_i(\theta)$ for two representative 3Q PDW states in Fig.~4(c,g). In the subsequent sections, we will present the results for another two cases for comparison and completeness.

\begin{figure*}
	\includegraphics[width=\linewidth]{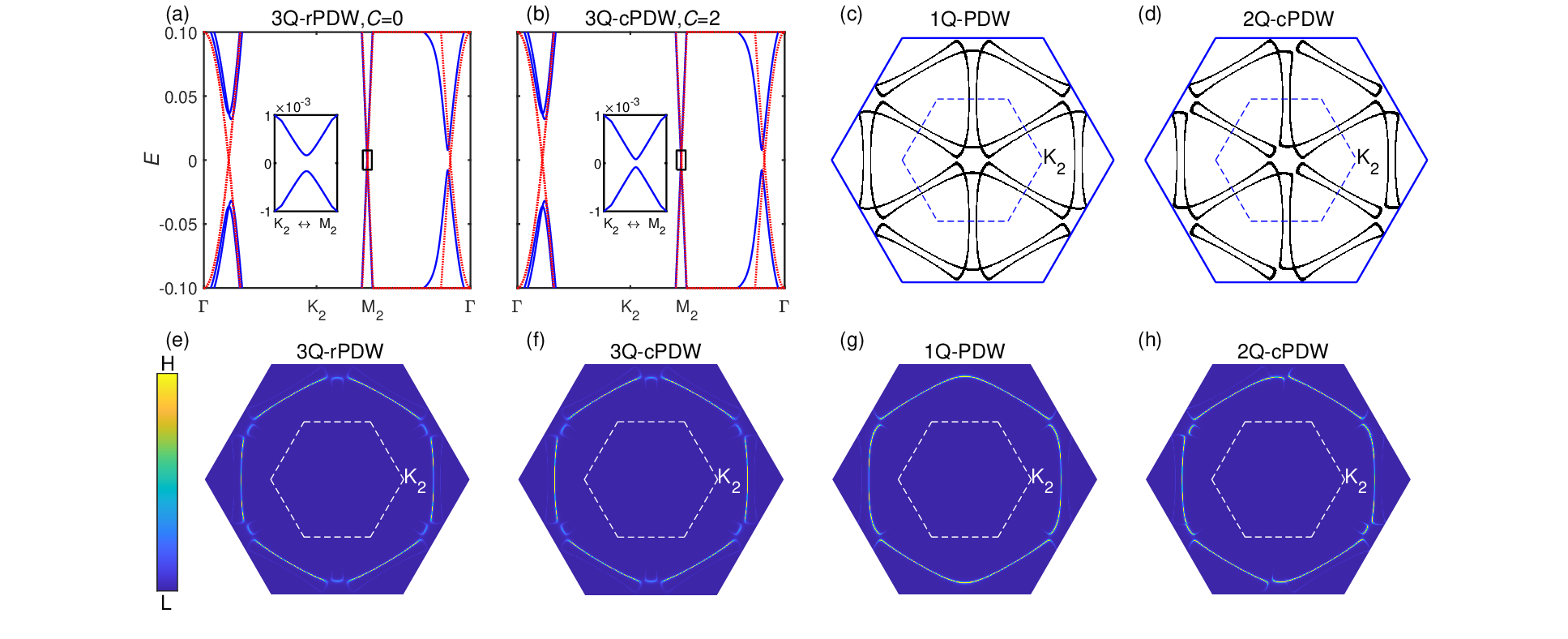}
	\caption{The low-energy BdG quasiparticle bands for the real and chiral 3Q states are plotted in (a) and (b), respectively, in the folded BZ, with the zoom-in near $M_2$ shown as the insets. For comparison, the bands without PDW are given as red dashed lines, and the total Chern numbers $C$ of the filled bands are displayed. The FSs for the 1Q and 2Q states are plotted in (c,d). In (e) to (h), the unfolded single-particle spectral functions at zero energy are plotted. In these calculations, we use the BdG parameters $\Delta=0.02$, $\mu=0.1$.
    }
	\label{fig:SM_GL_onlyj}
\end{figure*}

\begin{figure}[b]
	\includegraphics[width=0.48\textwidth]
    {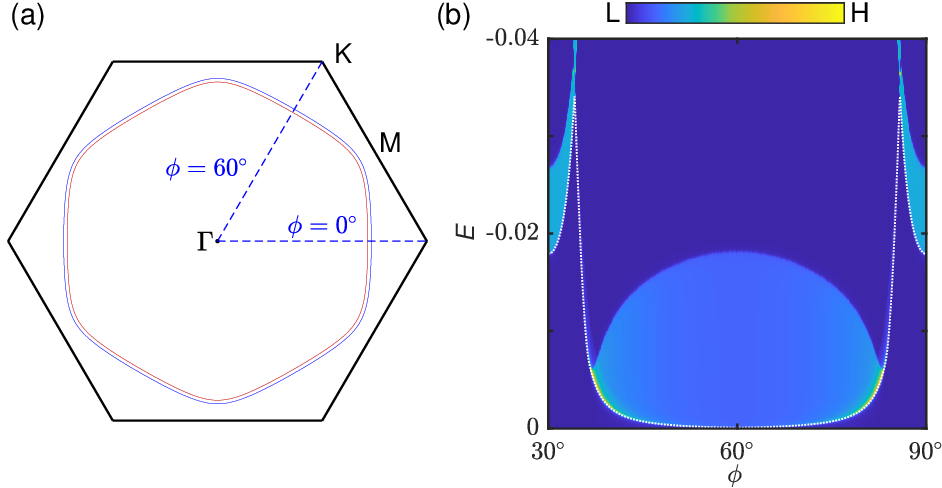}
    \caption{Around the normal state FS, within the region indicated by the area enclosed by blue (outside the FS) and red (inside the FS) lines in (a), we calculate the integrated spectral function $\bar{A}(\phi, \w)$ in the 3Q-cPDW state as shown in (b). The white dotted line represents the resulting quasiparticle gap. In the calculations, we use the parameters $\Delta=0.02$, $\mu=0.1$, $\eta=0.0001$.}
    \label{fig:SM_gap_akw}
\end{figure}

\begin{figure}[b]
	\includegraphics[width=0.48\textwidth]{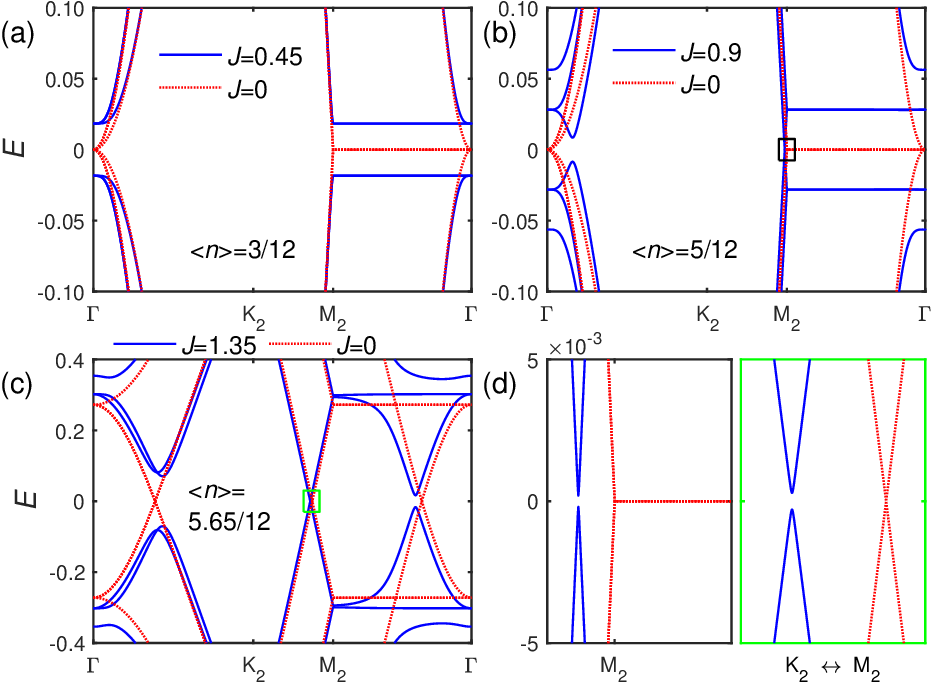}
	\caption{Low-energy BdG quasiparticle bands along the high-symmetry path in the reduced BZ obtained by mean field calculations.
	Red dashed lines corresponding to normal state at $J=0$ are plotted for comparison.
	(d) Zoom-in of low-energy band dispersions inside black and green boxes in (b) and (c), respectively.}
	\label{fig:SM_dispersion_onlyj}
\end{figure}

\begin{figure*}
	\includegraphics[width=\linewidth]{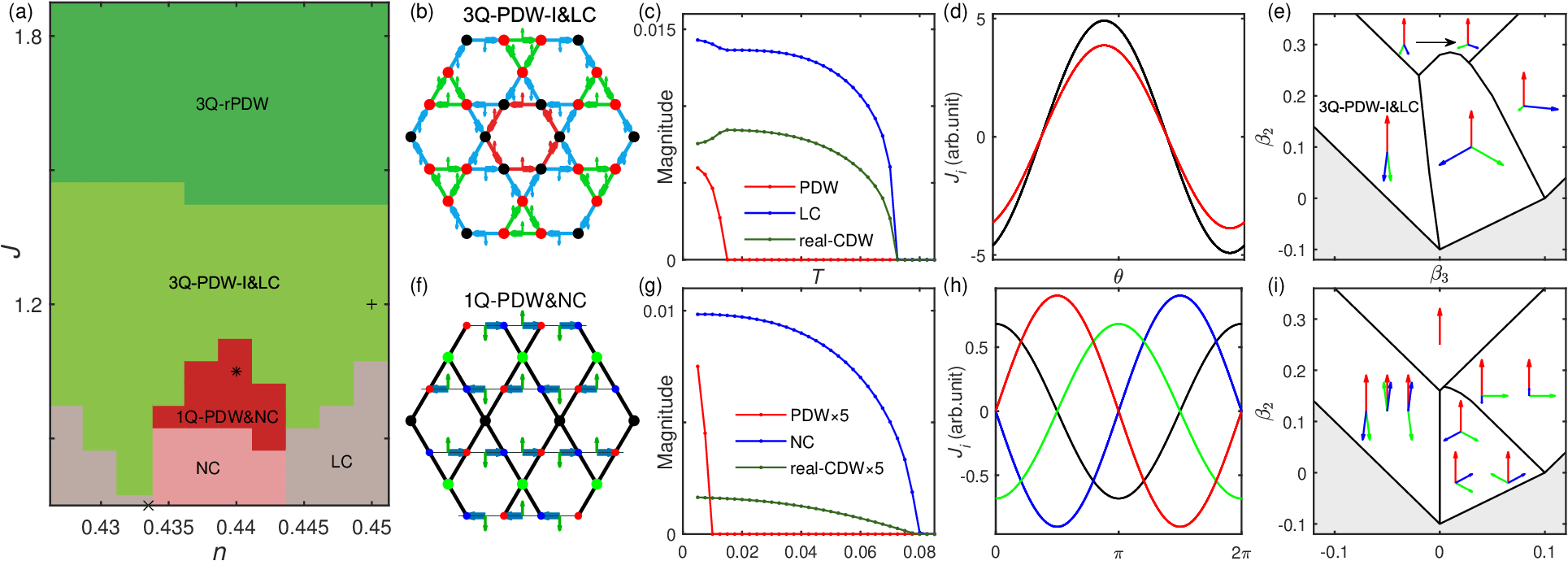}
	\caption{(a) Mean field phase diagram around the upper vH filling (marked by $\times$) for $V_1=1$, $V_2=2$, $t'=-0.05$ and $T=0.005$. For the representative case marked by $+$, the PDW and CDW patterns are plotted in (b) with the same convention in Fig.~3 in the main text, the temperature dependence of the order parameters are plotted in (c), and the local Josephson current $J_i$ versus the pairing phase $\theta$ are plotted in (d) with the line colors corresponding to different sites in (b). Under this LC background, the GL phase diagram is shown in (e) with the same convention as Fig.~2(a) in the main text, for $\alpha=1$, $\beta_1=0.1$, $\gamma_1=-2\beta_1$, and $\gamma_2=\gamma_3=\beta_1$. For the other representative case marked by $*$, the results are similarly plotted in (f to i).
    }
	\label{fig:SM_phase_v1v2j}
\end{figure*}

\section{Further results for pure PDW phases}\label{appendix:dispersion_pdw}
In the main text we obtained 3Q-PDW states in the mean field theory, which are most relevant to experiments. However, on general grounds, we also showed that the GL theory allows 1Q- and 2Q-states.  Here, for completeness and comparison, we also discuss the properties of the 1Q- and 2Q-PDW states.
We plot the quasiparticle bands in the folded BZ for the real and chiral 3Q PDW states, in Fig.~\ref{fig:SM_GL_onlyj}(a) and Fig.~\ref{fig:SM_GL_onlyj}(b), respectively.
Different from the 3Q PDWs, the 1Q- and 2Q-PDW states are not fully gapped, since not all of the three Q-vectors are present but all of them are required for pair-scattering mechanism  illustrated in Fig.~2(c) (main text). Therefore, the 1Q and 2Q PDWs are usual PDW states with genuine residual FSs. To see the residual FSs clearly, the single particle spectral function $A(\0k,\w=0)$ without unfolding is plotted in Fig.~\ref{fig:SM_GL_onlyj}(c) for the 1Q-PDW and Fig.~\ref{fig:SM_GL_onlyj}(d) for the chiral 2Q-PDW.

Corresponding to the four PDW states, the resulting unfolded $A(\0k,E=0)$ are shown in Fig.~\ref{fig:SM_GL_onlyj}(e-h) accordingly.
Although the 3Q PDW states (either real or chiral) are fully gapped, we examined in the main text that the gap minimum is small as it arises from the higher-order pair scattering processes. This might appear as gapless if the gap minimum is within the experimental resolution. We mimic the finite resolution by a scattering rate $\eta = 0.005$ in our  calculations. Here, although the ``residual FSs'' (an artifact from finite scattering rate $\eta$) are visible and show similarities to the genuine residual FS in the case of 1Q and 2Q PDWs, the spectral weight distribution around the $M$-points can still be used to distinguish the 3Q from the nematic 1Q and 2Q PDW states.

In the main text, we presented the quasiparticle gap along the normal state FS. Here, we explain how it is obtained.
Due to the finite momentum pairing, the gap does not open exactly on the normal state FS. In other words, the locus of minimal gap along the normal of the FS deviates (but only slightly) from the FS. This can already be seen in Fig.~\ref{fig:SM_GL_onlyj}(a,b). {This is, in fact, an interesting and also perhaps significant feature to be confirmed by ARPES.} Therefore, to obtain the gap minima along the normal direction of the FS, we integrate the unfolded spectral function for each direction over a small $k$-range near the FS, see the narrow shell enclosing the FS in Fig.~\ref{fig:SM_gap_akw}(a). From the integrated $\bar{A}(\phi,\w)=\int A(k,\phi,\w)\md k$ shown in Fig.~\ref{fig:SM_gap_akw}(b), we extract the gap size as the binding energy on the lower edge of the excitation spectrum for each direction. The resulting gap along the FS are plotted in Fig.~2(e) in the main text.

The BdG quasiparticle bands, obtained by self-consistent mean-field calculations, are shown in Fig.~\ref{fig:SM_dispersion_onlyj}.
For the usual SC state at the lower vH filling, see Fig.~\ref{fig:SM_dispersion_onlyj}(a), the BdG quasiparticle bands open uniform energy gaps at the Fermi energy.
While for the 3Q-rPDW state [see Fig.~\ref{fig:SM_dispersion_onlyj}(b)],
and 3Q-cPDW [see Fig.~\ref{fig:SM_dispersion_onlyj}(c)],
the gaps do not uniformly open along the high symmetry path and leaves a minigap near point $M_2$.
Fig.~\ref{fig:SM_dispersion_onlyj}(d) is the Zoom-in of band dispersions inside black and green boxes in Fig.~\ref{fig:SM_dispersion_onlyj}(b,c), demonstrating the small but finite gap minimum.

\section{Further results for PDW on CDW backgrounds}
By self-consistent mean field calculations, we obtain the phase diagram near the upper vH filling for $V_1=1$, $V_2=2$, $t'=-0.05$ and $T=0.005$ as shown in Fig.~\ref{fig:SM_phase_v1v2j}(a).
We observe that in the majority of the given parameter space the 3Q-PDW states are realized. For completeness and comparison, here we also discuss the other phases.
At small $J$, two types of CDW orders are obtained: the loop current (LC) [see Fig.~\ref{fig:SM_phase_v1v2j}(b)] and nematic current (NC) [see Fig.~\ref{fig:SM_phase_v1v2j}(f)] orders.
As $J$ increases, the 3Q PDW-I (almost real 3Q PDW) is formed on the LC state [see Fig.~\ref{fig:SM_phase_v1v2j}(b)] but with a lower $T_c$ [see Fig.~\ref{fig:SM_phase_v1v2j}(c)].
On the other hand, in the NC state, increasing $J$ firstly induces the 1Q PDW order [see Fig.~\ref{fig:SM_phase_v1v2j}(f), the current and PDW orders on other bonds are very small] also with a lower $T_c$ [see Fig.~\ref{fig:SM_phase_v1v2j}(g)], and then the 3Q PDW-I occurs and coexists with the LC order.
For sufficiently large $J$, the LC order vanishes and only the 3Q real PDW (with secondary real CDW) is left.
For the above two representative cases, the local Josephson currents $J_i(\theta)$ are plotted in Fig.~\ref{fig:SM_phase_v1v2j}(d) and (h), respectively. Clearly, the critical current shows $a_0\times a_0$ periodicity in the nematic 1Q (same for 2Q, not shown) PDW state, but $2a_0\times2a_0$ periodicity in the real 3Q (same for chiral 3Q, already shown in the main text) PDW state.

\begin{figure}
	\includegraphics[width=0.48\textwidth]{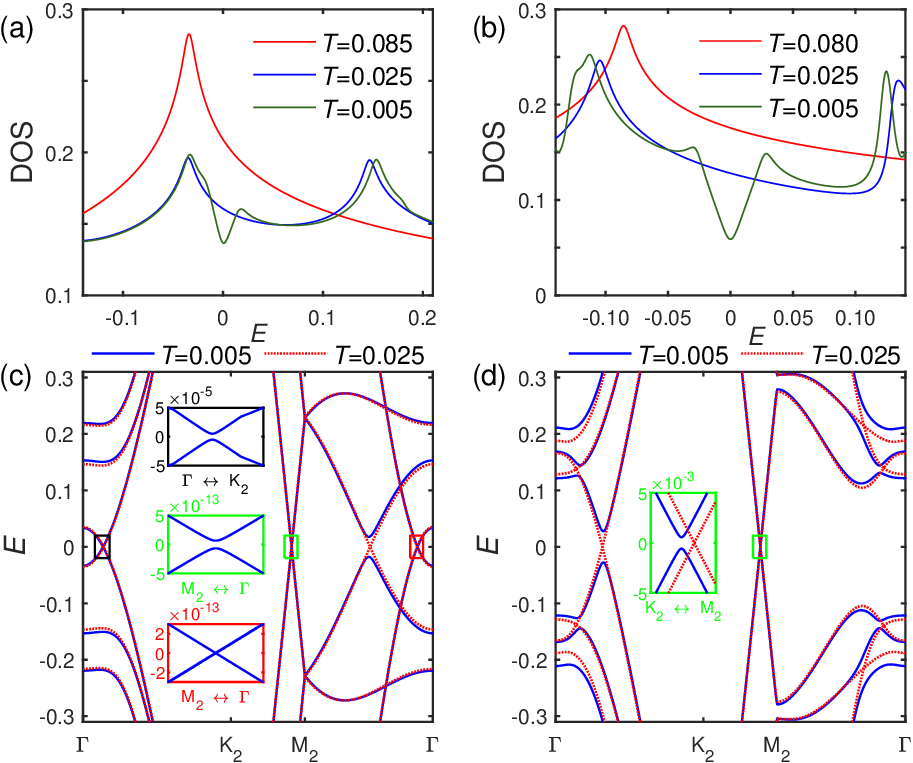}
	\caption{For $J=1.05$, $ n =0.44$ marked by $\ast$ in Fig.~\ref{fig:SM_phase_v1v2j}(a),
	(a) shows the DOS at different temperatures.
	(c) displays the low-energy BdG quasiparticle bands along one of three non-equivalent $\Gamma-\rm{K}_2-\rm{M}_2-\Gamma$, with insets zooming in on band dispersions.
	(b) and (d) is similar to (a) and (c), respectively, but for $J=1.2$, $ n =0.45$ marked by $+$ in Fig.~\ref{fig:SM_phase_v1v2j}(a).}
	\label{fig:SM_v1v2j_fig1}
\end{figure}

\begin{figure}
	\includegraphics[width=0.48\textwidth]{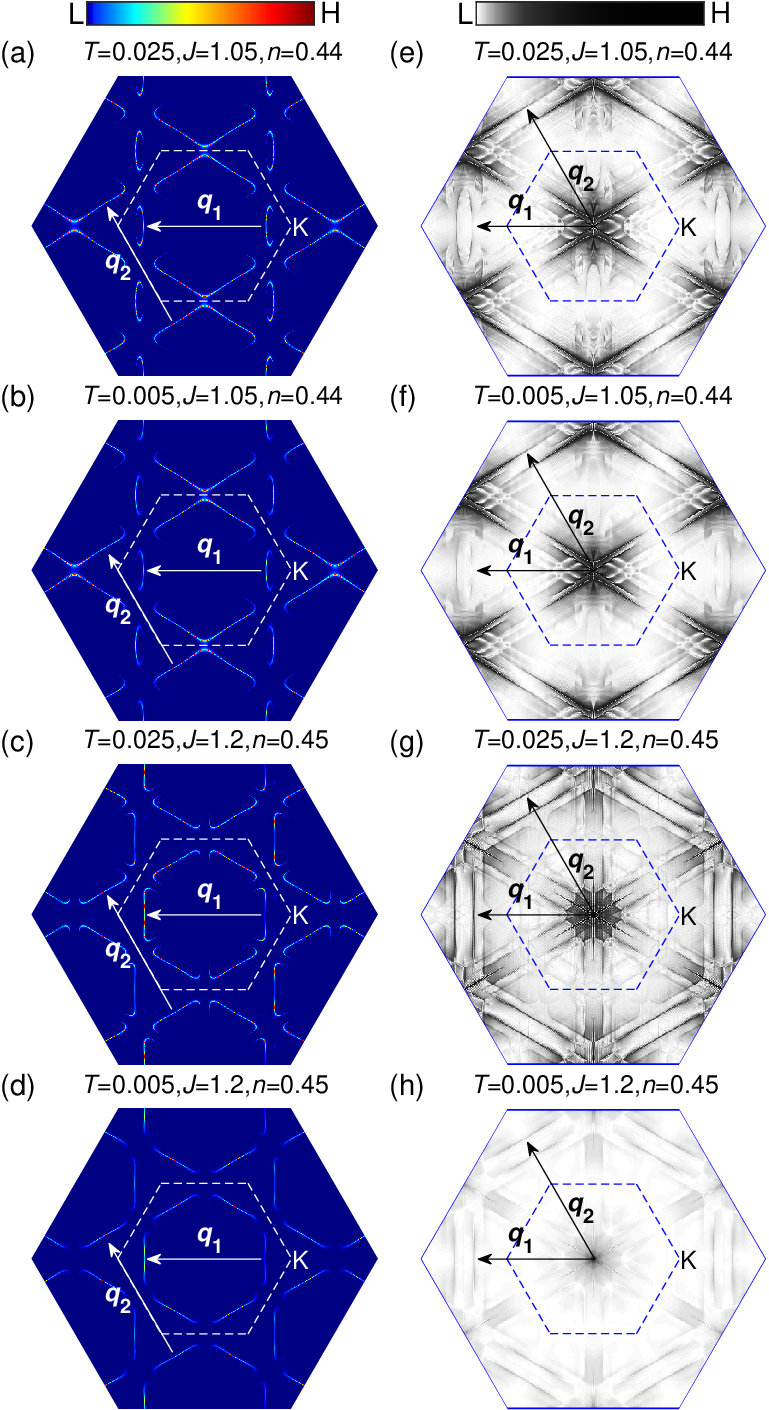}
	\caption{The unfolded spectral functions $A(\bm{k},E=0)$ at two temperatures are plotted in (a,b) for the state marked by $\ast$ in Fig.~\ref{fig:SM_phase_v1v2j}(a), and (c,d) for the state marked by $+$, respectively.
	(e-h) QPI maps at Fermi energy corresponding to (a-d).}
	\label{fig:SM_v1v2j_fig2}
\end{figure}

To gain qualitative insights into the interplay between PDW and CDW, we resort to GL theory in a given CDW background. We add symmetry-allowed couplings between the PDW  $\psi_i$ and CDW order parameters $\phi_i$ to $F_{\rm PDW}$ as
\begin{align}
  F=F_{\rm PDW}&+\gamma_1\sum_{i}|\psi_i|^{2}|\phi_i|^2+\gamma_2\sum_{i<j}|\psi_i|^{2}|\phi_j|^2\nonumber\\
  &+\gamma_3(\psi_1^{*}\psi_2\phi_3+\psi_2^{*}\psi_3\phi_1+\psi_3^{*}\psi_1\phi_2+{\rm{c.c.}})\label{eq:free_energy_cdw}
\end{align}
where $\gamma_{1,2,3}$ are real numbers. For a given CDW $\phi_{1,2,3}$ and the couplings $\gamma_{1,2,3}$, we can find all possible PDW ground states.
As an illustration, we choose $\gamma_1=-2\beta_1$ and $\gamma_2=\gamma_3=\beta_1$.
(Here, to compare with the mean field results,
we set $\gamma_1<0$ and $\gamma_2>0$ to realize the 1Q PDW on the 1Q nematic CDW state, and set $\gamma_3>0$ to realize the real 3Q PDW on the 3Q LC state.)
For the chiral CDW, we set $\phi_1=\phi_2=\phi_3=\me^{i\alpha}$ ($\alpha\approx0.7\pi$) as obtained from mean field calculations.
The phase diagram is shown in Fig.~\ref{fig:SM_phase_v1v2j}(e).
We can identify the above two mean field ground states as inheriting from the real and chiral 3Q PDWs, respectively, by comparing with Fig.~2(a) in the main text.
Due to the chiral CDW background, the chiral 3Q PDW phase (PDW-II) is enlarged, and
the real 3Q PDW state (PDW-I) acquires a weak chirality as a result of the $\gamma_3$-term, consistent with our mean field result in Fig.~\ref{fig:SM_phase_v1v2j}(b).
In addition, there are another two nematic phases inheriting from the 1Q and 2Q PDWs. The former is realized on a nematic CDW background, while the latter may be realized in other microscopic models.

We further study the 1Q-PDW$\&$NC and 3Q-PDW-I$\&$LC phases in Fig.~\ref{fig:SM_phase_v1v2j}(a).
Fig.~\ref{fig:SM_v1v2j_fig1}(a,b) show the DOS $\rho(E)$ at different temperatures for two sets of values of $J$ and $n$, which correspond to the two representative states marked by $\ast$ and $+$ in Fig.~\ref{fig:SM_phase_v1v2j}(a).
The DOS exhibits a CDW gap at $T=0.025$ compared to the normal state.
Furthermore, at $T=0.005$, a V-shaped superconducting gap opens in the CDW state,
which corresponds to the gapless BdG quasiparticle bands for 1Q-PDW$\&$NC
and to the quasiparticle bands with a very small gap for 3Q-PDW-I$\&$LC as shown in Fig.~\ref{fig:SM_v1v2j_fig1}(c,d).
It is worth noting that the V-shaped superconducting gap has been observed in several experiments \cite{Chen-2021-roton, Jiang-2021-unconventional, Xu-2021-Multiband, zhao-2021-cascade, luo_unique_2023} in kagome superconductors.

In the main text, we presented the ``residual FSs'' for the 3Q PDWs. Here, we provide further results in the pure-CDW state and the QPI spectrum.
Fig.~\ref{fig:SM_v1v2j_fig2} shows the unfolded spectral functions $A(\bm{k}, E)$ and the QPI at the Fermi energy for the two representative states marked by $\ast$ and $+$ in Fig.~\ref{fig:SM_phase_v1v2j}(a).
In the CDW states, two bar-like signals are observed at $\bm{q}_1$ and $\bm{q}_2$ in Fig.~\ref{fig:SM_v1v2j_fig2}(e,g).
As the temperature decreases, PDWs emerge, but the bar-like signals persist at $\bm{q}_1$ and $\bm{q}_2$, as depicted in Fig.~\ref{fig:SM_v1v2j_fig2}(f,h).
The persistence of such features in the SC state serves as an indicator of the ``residual FS'' in the PDW state.


\section{LDOS of PDW orders}

In particular, we focus on LDOS in different cases.

\subsection{The LDOS of PDW and CDW-PDW state}
\begin{figure}
	\includegraphics[width=0.48\textwidth]{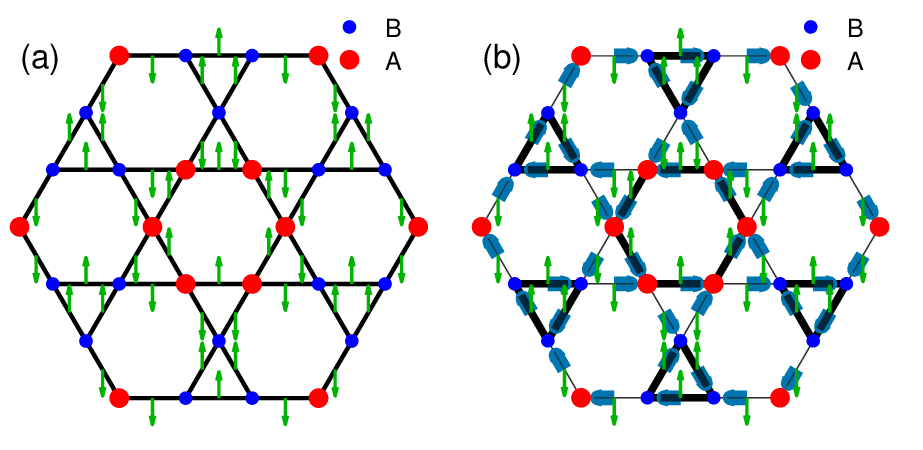}
	\caption{Schematics of the $2a_{0}\times 2a_{0}$ state. The size of the circle on the site denotes particle number $\langle n_i\rangle$, the thickness of bonds denote the real CDW order parameters $t-\text{Re}(\chi_{ij})$, arrows starting from the middle of bonds indicate the PDW order parameters $\Delta_{ij}$ and  arrows on the bonds indicate the imaginary CDW order parameters $\text{Imag}(\chi_{ij})$. }
	\label{fig:phase}
\end{figure}

\begin{figure}
	\includegraphics[width=0.4\textwidth]{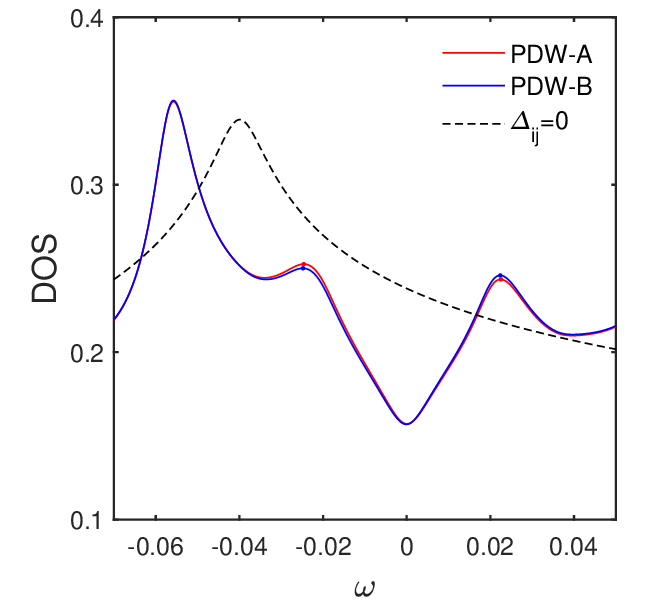}
	\caption{The LDOS for $\chi_{ij}=0$, $|\Delta_{ij}|=0.02$ and $\mu=0.04$. Dots denote local maxima.}
	\label{fig:only_pdw}
\end{figure}

First, we only consider PDW state. We set $\chi_{ij}=0$ and $\Delta_{ij}=|\Delta_{ij}|\exp(\text{i}\phi_{ij})$, which are shown in Fig.~\ref{fig:phase}(a). The length and direction of the arrows denote magnitude $|\Delta_{ij}|=0.02$ and phase $\phi_{ij}$, respectively. As shown in Fig.~\ref{fig:only_pdw}, the corresponding LDOS on sites A and B show small differences around superconducting coherence peaks.

Experimentally, CDW and PDW are co-existing at low temperatures and the superconducting gap is smaller than CDW gap, so we consider $|\text{Re}(\chi_{ij})|=0.1$, $|\text{Imag}(\chi_{ij})|=0.1$ and $|\Delta_{ij}|=0.01$. As shown in Fig.~\ref{fig:phase}(b), the thick (thin) bonds denote $\text{Re}(\chi_{ij})=-0.1 (0.1)$ and the direction of arrows on the bonds denotes the flowing direction of imaginary CDW (loop current) order.

\begin{figure}
	\includegraphics[width=0.4\textwidth]{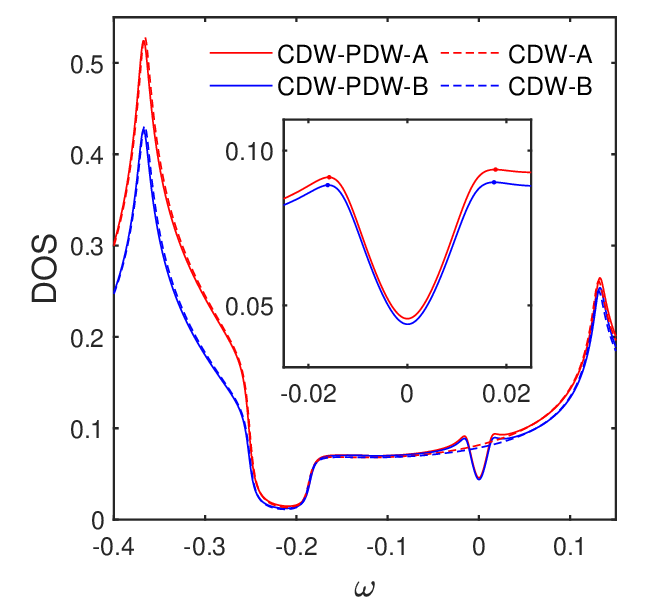}
	\caption{The LDOS for $|\text{Re}(\chi_{ij})|=0.1$, $|\text{Imag}(\chi_{ij})|=0.1$, $|\Delta_{ij}|=0.01$ ($|\text{Re}(\chi_{ij})|=0.1$, $|\text{Imag}(\chi_{ij})|=0.1$, $|\Delta_{ij}|=0$) and $\mu=0.04$ are represented by solid (dashed) lines. The inset shows the zoom-in of LDOS around the superconducting coherence peaks and dots denote local maxima.}
	\label{fig:cdw_pdw}
\end{figure}

Fig.~\ref{fig:cdw_pdw} shows the LDOS on sites A and B, the superconducting coherence peaks are around $\w=\pm0.015$ and the difference between $A$ and $B$ is small.

\subsection{The LDOS of Nematic-PDW and CDW-Nematic-PDW state}

\begin{figure}
	\includegraphics[width=0.4\textwidth]{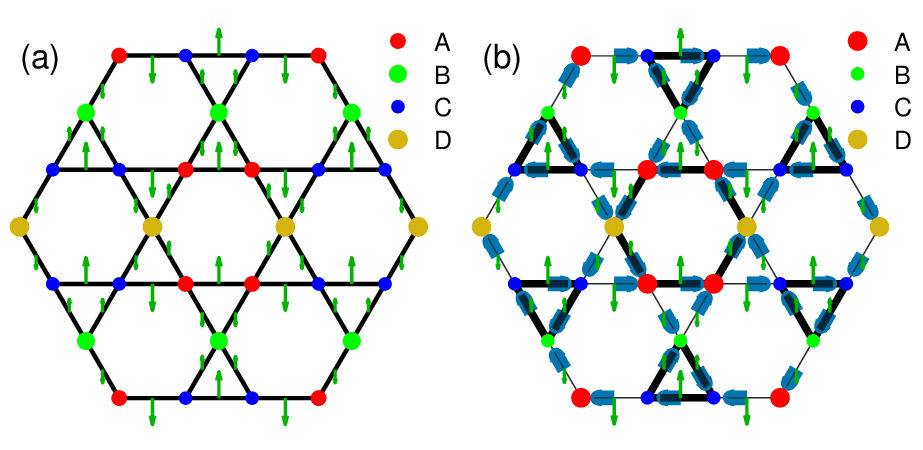}
	\caption{Schematics of the $2a_{0}\times 2a_{0}$ state. The size of the circle on the site denotes particle number $\langle n_i\rangle$, the thickness of bonds denote the real CDW order parameters $t-\text{Re}(\chi_{ij})$, arrows starting from the middle of bonds indicate the PDW order parameters $\Delta_{ij}$ and arrows on the bonds indicate the imaginary CDW order parameters $\text{Imag}(\chi_{ij})$. }
	\label{fig:nphase}
\end{figure}

\begin{figure}
	\includegraphics[width=0.4\textwidth]{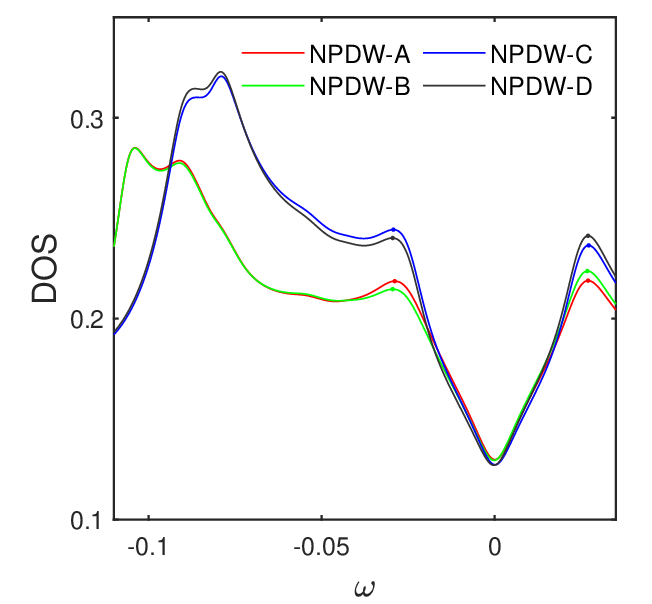}
	\caption{The LDOS for $\chi_{ij}=0$, $|\Delta_{ij}|_{\text {max}}=0.04$, $|\Delta_{ij}|_{\text {min}}=0.024$ and $\mu=0.07$. Dots denote local maxima.}
	\label{fig:npdw}
\end{figure}

Experimentally, PDW may be nematic. We consider the Nematic-PDW (NPDW) as Fig.~\ref{fig:nphase}(a), $\chi_{ij}=0$, $|\Delta_{ij}|_{\text {max}}=0.04$ and $|\Delta_{ij}|_{\text {min}}=0.024$. The corresponding LDOS on sites A,B,C and D are shown in Fig.~\ref{fig:npdw}, and the difference between the locations of local maxima for different sites is larger than that in Fig.~\ref{fig:only_pdw}.

\begin{figure}
	\includegraphics[width=0.4\textwidth]{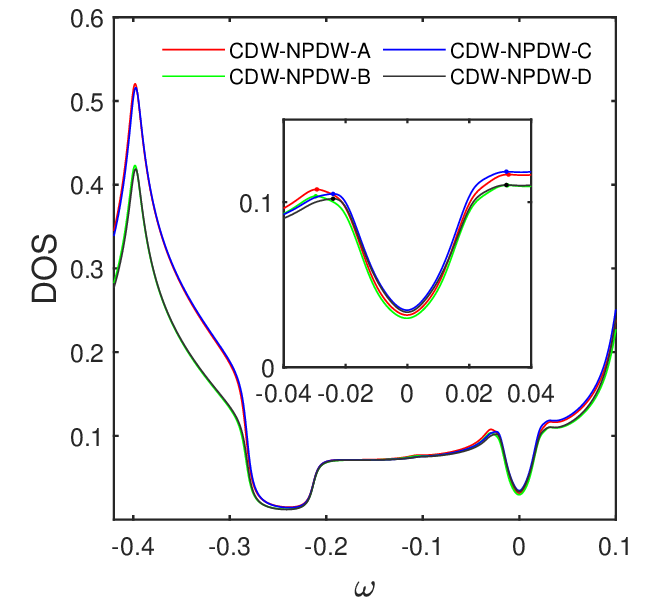}
	\caption{The LDOS for $|\text{Re}(\chi_{ij})|=0.1$, $|\text{Imag}(\chi_{ij})|=0.1$, $|\Delta_{ij}|_{\text{max}}=0.02$, $|\Delta_{ij}|_{\text{min}}=0.012$ and $\mu=0.07$. The inset shows the zoom-in of LDOS around the superconducting coherence peaks and dots denote local maxima.}
	\label{fig:cdw_npdw}
\end{figure}

For CDW and Nematic-PDW coexisting, as shown in Fig.~\ref{fig:nphase}(b), we set $|\text{Re}(\chi_{ij})|=0.1$, $|\text{Imag}(\chi_{ij})|=0.1$, $|\Delta_{ij}|_{\text{max}}=0.02$, $|\Delta_{ij}|_{\text{min}}=0.012$. Fig.~\ref{fig:cdw_npdw} shows the LDOS on different sites, there are obvious changes in the locations of local maxima as the sites shift.

\section{Microscopic derivation of the Ginzburg-Landau theory}
We first consider $H_P$, as given in Eq. (4) of the main text, the path integral representation of the partition function is
\begin{align}
    Z&=\int D \overline{\varphi}D\varphi \mathrm{e}^{-S};\quad S=S_0+S_I\\
    S_0&=\int_0^{\frac{1}{T}} \mathrm{d}\tau \sum_{i\sigma}\overline{\varphi}_{i\sigma}(\partial_\tau-\mu)\varphi_{i\sigma}-\sum_{ ij \sigma}(t_{ij}\overline{\varphi}_{i\sigma}\varphi_{j\sigma}+\text{H.c.})\\
    S_I&=\int_0^{\frac{1}{T}}\mathrm{d}\tau (-J)\sum_{\langle ij \rangle}(\overline{\varphi}_{i\uparrow} \overline{\varphi}_{j\downarrow}-\overline{\varphi}_{i\downarrow}\overline{\varphi}_{j\uparrow})(\varphi_{j\downarrow}\varphi_{i\uparrow}-\varphi_{j\uparrow}\varphi_{i\downarrow})
\end{align}
\begin{figure}
	\includegraphics[width=\linewidth]{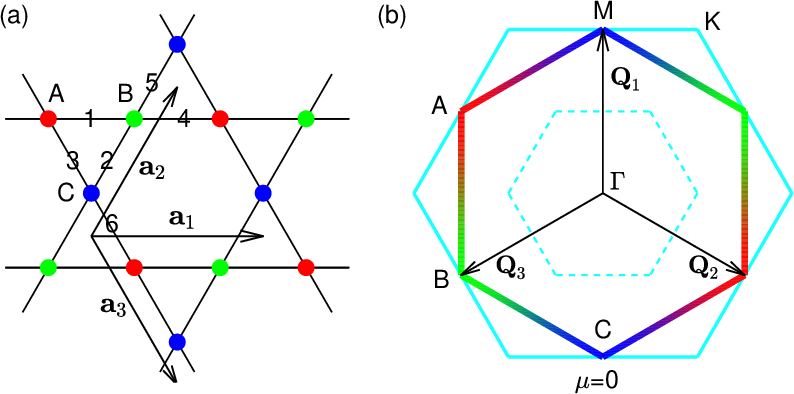}
	\caption{(a) Kagome lattice structure with the three sublattices labeled by A (red), B (green) and C (blue), respectively. $\mathbf{a}_{1,2,3}$ denote three translation vectors with length $a_0$.The numbers 1 to 6 represent the six bonds within a unit cell. (b) The
    three vectors $\bm{Q}_{b=1,2,3}$ are depicted in the first BZ, the color-scaled line represents the FS at the upper vH filling.}
	\label{fig:SM_GLT_lattice}
\end{figure}
By Hubbard-Stratonovich decomposition to introduce the pairing field, we obtain
\begin{align}
    &Z= \int D \overline{\varphi}D\varphi  \int D\overline{\psi}D\psi \mathrm{e}^{-S}\\
    &S=\int_0^{\frac{1}{T}} \mathrm{d}\tau \sum_{i\sigma}\overline{\varphi}_{i\sigma}(\partial_\tau-\mu)\varphi_{i\sigma}-\sum_{ij \sigma}(t_{ij}\overline{\varphi}_{i\sigma}\varphi_{j\sigma}+\text{H.c.})\nonumber\\
    &\quad+\sum_{R,\tilde{b}}(-[\psi^{\tilde{b}}_{R}
    \sum_{\alpha_{\tilde{b}}}( \overline{\varphi}^{\alpha_{\tilde{b}}}_{R+\delta({\alpha_{\tilde{b}}}),\uparrow} \overline{\varphi}^{\overline{\alpha_{\tilde{b}}}}_{R+\delta({\overline{\alpha_{\tilde{b}}}}),\downarrow})+\text{H.c.}]
    +\frac{1}{J}|\psi^{\tilde{b}}_{R}|^2)
\end{align}
where $R$ labels the unitcell, $\tilde{b}=1,2,\cdots,6$ specifies the bond within an unitcell, as shown in Fig.~\ref{fig:SM_GLT_lattice}(a), and $(\alpha_{\tilde{b}},\overline{\alpha_{\tilde{b}}})$ correspond to the sublattice indices at the two ends of bond $\tilde{b}$. For instance, when $\tilde{b}=1$, $\alpha_{\tilde{b}}(\overline{\alpha_{\tilde{b}}})=\text{A}(\text{B}),\text{B}(\text{A})$. $\psi^{\tilde{b}}$ represents the bosonic auxiliary pairing field located on bond
$\tilde{b}$ within the unit cell $R$. $\delta(\alpha_{\tilde{b}})$ represents the offset between the unit cell containing the endpoint $\alpha(\tilde{b})$ of bond $\tilde{b}$ and the unit cell $R$. For example, when $\tilde{b}=2$, $\delta(\text{A}_2)=\mathbf{a}_1$ and $\delta(\text{B}_2)=0$.
\par
We now introduce the Fourier transformation
\begin{align}
    \varphi_{\bm{R},\tau}^{\alpha}&=\sqrt{\frac{T}{N}}\sum_{\bm{k},\omega_n}\mathrm{e}^{-\text{i}\omega_n\tau+\text{i}\bm{k}\cdot \bm{R}}\varphi_{\bm{k},\omega_n}^{\alpha}\\
    \psi_{\bm{R},\tau}^{\tilde{b}}&=\sum_{\bm{q},\nu_n}\mathrm{e}^{-\text{i}\nu_n\tau+\text{i}\bm{q}\cdot \bm{R}}\psi_{\bm{q},\nu_n}^{\tilde{b}}
\end{align}
and obtain:
\begin{align}
    &S=\sum_{k,\sigma}\sum_{\alpha\beta} \overline{\varphi}^\alpha_{k,\sigma}(-\text{i}\omega_n\delta_{\alpha\beta}+h_{\bm{k}}^{\alpha\beta})\varphi_{k,\sigma}^\beta+\sum_{q,b}\frac{N}{JT}|\psi_q^{\tilde{b}}|^2+\nonumber\\
    &\quad\sum_{k,q,\tilde{b}}-[\psi_q^{\tilde{b}} \sum_{\alpha_{\tilde{b}}}\overline{\varphi}_{k,\uparrow}^{\alpha_{\tilde{b}}} \overline{\varphi}_{-k+q,\downarrow}^{\overline{\alpha_{\tilde{b}}}}
    \text{e}^{-\text{i} \left(\bm{k}\cdot \bm{\delta}({\alpha_{\tilde{b}}})+[-\bm{k}+\bm{q}]\cdot \bm{\delta}({\overline{\alpha_{\tilde{b}}}}) \right)}+\text{H.c}]
\end{align}
where $k=(\bm{k},\omega_n)$, $q=(\bm{q},\nu_n)$, and $\alpha,\beta=\mathrm{A,B,C}$. We will neglect the temporal fluctuation of $\psi$. Following the discussion in the main text, we simplify our notation with $\psi_{\tilde{b}}=\psi^{\tilde{b}}_{q=\pm Q_{\tilde{b}}}$, where $Q_{\tilde{b}}=(\bm{Q}_{\tilde{b}},\nu_n=0)$, with $Q_4=Q_1,Q_5=Q_2,Q_6=Q_3$, and $\psi_4=-\psi_1, \psi_5=-\psi_2, \psi_6=-\psi_3$. The vectors $\bm{Q}_{b=1,2,3}$ are illustrated in Fig.~\ref{fig:SM_GLT_lattice}(b). Next, we turn to the Nambu spinor representation $\overline{\varphi}=(\overline{\varphi}_\uparrow,\varphi_\downarrow)$, where $\varphi_\sigma$ list all $k$ and $\alpha$. The effective action is then given by
\begin{align}
    S=\frac{2N}{JT}(|\psi_1|^2+|\psi_2|^2+|\psi_3|^2)-\overline{\varphi}(\mathcal{G}^{-1}_0+\Delta)\varphi
\end{align}
where $\mathcal{G}_0$ is the Gov'kov function in Nambu space. It is related to the normal state single particle function $G_0$ as follows
\begin{align}
    [\mathcal{G}_{0,\uparrow\uparrow}]^{\alpha\beta}_{k}&=[(\text{i}\omega_n-\tilde{h}_{\bm{k}})^{-1}]^{\alpha\beta}\equiv G_{0,k}^{\alpha\beta}\\
    [\mathcal{G}_{0,\downarrow\downarrow}]^{\alpha\beta}_{-k}&=[(\text{i}\omega_n+\tilde{h}^{T}_{\bm{-k}})^{-1}]^{\alpha\beta}\nonumber\\
    &=-[(\text{i}(-\omega_n)-\tilde{h}_{\bm{-k}})^{-1}]^{\beta\alpha}\equiv -G_{0,-k}^{\beta\alpha}\nonumber
\end{align}
$\Delta$ is the pairing matrix, and it satisfies the following relations
\begin{align}
  \Delta_{k\uparrow,-k\pm Q_b\downarrow}^{\alpha_{b},\overline{\alpha_{b}}}=\psi_b \cdot &f_{\Delta}(k,\alpha_{b};-k\pm Q_{b},\overline{\alpha_{b}})\\
  \Delta_{-k\pm Q_b\downarrow,k\uparrow}^{\overline{\alpha_{b}},\alpha_{b}}=\psi_b^* \cdot &f^{*}_{\Delta}(k,\alpha_{b};-k\pm Q_{b},\overline{\alpha_{b}})\\
  f_{\Delta}(k,\alpha_{b};-k\pm Q_b,\overline{\alpha_{b}})&\equiv \text{e}^{-\text{i} \left(\bm{k}\cdot \bm{\delta}({\alpha_{b}})+[-\bm{k}+\bm{Q}_b]\cdot \bm{\delta}({\overline{\alpha_{b}}}) \right)}\nonumber\\
  &-\text{e}^{-\text{i} \left(\bm{k}\cdot \bm{\delta}({\alpha_{b+3}})+[-\bm{k}+\bm{Q}_b]\cdot \bm{\delta}({\overline{\alpha_{b+3}}}) \right)}
  \\
  f_{\Delta}(k,\alpha_{b};-k\pm Q_b,\overline{\alpha_{b}})&=f_{\Delta}(-k\pm Q_b,\overline{\alpha_{b}};k,\alpha_{b})\\
  f_{\Delta}(k,\alpha_{b};-k + Q_b,\overline{\alpha_{b}})&=f_{\Delta}(k,\alpha_{b};-k- Q_b,\overline{\alpha_{b}})
\end{align}
where $b=1,2,3$.
\par
The path integral can be completed exactly, giving us the GL free energy
\begin{equation}
    F=-\frac{T\text{ln}Z}{N}=\frac{2}{J}(|\psi_1|^2+|\psi_2|^2+|\psi_3|^2)-\frac{T}{N}\text{Trln}(\mathcal{G}_0^{-1}+\Delta)
\end{equation}
The order parameters $\psi_{b=1,2,3}$ are needed to be optimized to minimize $F$. Our task thus reduced to computing the expansion of $\text{Trln}(\mathcal{G}_0^{-1}+\Delta)$
\begin{align}
    \text{Trln}(\mathcal{G}_0^{-1}+\Delta)&=\text{Trln}(\mathcal{G}_0^{-1}[1+\mathcal{G}_0\Delta])\nonumber\\
    &=\text{Trln}(\mathcal{G}_0^{-1})+\text{Trln}(1+\mathcal{G}_0\Delta)\nonumber\\
    &=\text{const.}-\sum_{n=\text{even}}^{\infty}\frac{1}{n}\text{Tr}(\mathcal{G}_0\Delta)^{n}
\end{align}
Note that $[\mathcal{G}_0\Delta]$ is off-diagonal in Nambu space, only even orders are nonzero after the trace operation.
\par
The second-order term is
\begin{align}
    -\frac{1}{2}&\text{Tr}(\mathcal{G}_0\Delta)^{2}=-\text{Tr}\left([\mathcal{G}_{0,\uparrow\uparrow}]\Delta_{\uparrow\downarrow}[\mathcal{G}_{0,\downarrow\downarrow}]\Delta_{\downarrow\uparrow}\right)\nonumber\\
    &=-\sum_{b=1,2,3}\sum_{\alpha_b\beta_b}\sum_{k}\nonumber\\
    &\quad[\mathcal{G}_{0,\uparrow\uparrow}]_k^{\overline{\beta_b},\alpha_b}
    \Delta_{k\uparrow,-k+Q_b\downarrow}^{\alpha_b,\overline{\alpha_b}}
    [\mathcal{G}_{0,\downarrow\downarrow}]_{-k+Q_b}^{\overline{\alpha_b},\beta_b}
    \Delta_{-k+Q_b\downarrow,k\uparrow}^{\beta_b,\overline{\beta_b}}\\
    &=\sum_{b}|\psi_b|^2\sum_{\alpha_b\beta_b}\sum_{k}
    [G_{0,k}^{\overline{\beta_b},\alpha_b}
    G_{0,-k+Q_b}^{\beta_b,\overline{\alpha_b}}\nonumber\\
    &\qquad\quad\cdot f_{\Delta}(k,\alpha_b;-k+Q_{b},\overline{\alpha_{b}})f_{\Delta}^*(k,\overline{\beta_{b}};-k+Q_b,\beta_{b})]
\end{align}
The corresponding Feynman diagram is shown in Fig.~\ref{fig:feynman_fig1}(a).

\begin{widetext}
The fourth-order term is
\begin{align}
    -\frac{1}{4}&\text{Tr}(\mathcal{G}_0\Delta)^{4}=-\frac{1}{2}\text{Tr}\left([\mathcal{G}_{0,\uparrow\uparrow}]\Delta_{\uparrow\downarrow}[\mathcal{G}_{0,\downarrow\downarrow}]\Delta_{\downarrow\uparrow}[\mathcal{G}_{0,\uparrow\uparrow}]\Delta_{\uparrow\downarrow}[\mathcal{G}_{0,\downarrow\downarrow}]\Delta_{\downarrow\uparrow}\right)\nonumber\\
    &=-\frac{1}{2}\sum_{b1,b2,b3,b4}\sum_{\alpha_{b1},\beta_{b2},\gamma_{b3},\eta_{b4}}\sum_{k}
    [\mathcal{G}_{0,\uparrow\uparrow}]_k^{\overline{\eta _{b4}},\alpha_{b1}}
    \Delta_{k\uparrow,-k+Q_{b1}\downarrow}^{\alpha_{b1},\overline{\alpha_{b1}}}
    [\mathcal{G}_{0,\downarrow\downarrow}]_{-k+Q_{b1}}^{\overline{\alpha_{b1}},\beta_{b2}}
    \Delta_{-k+Q_{b1}\downarrow,k-Q_{b1}+Q_{b2}\uparrow}^{\beta_{b2},\overline{\beta_{b2}}}\nonumber\\
    &\qquad\cdot[\mathcal{G}_{0,\uparrow\uparrow}]_{k-Q_{b1}+Q_{b2}}^{\overline{\beta_{b2}},\gamma_{b3}}
    \Delta_{k-Q_{b1}+Q_{b2}\uparrow,-k+Q_{b1}-Q_{b2}+Q_{b3}\downarrow}^{\gamma_{b3},\overline{\gamma_{b3}}}
    [\mathcal{G}_{0,\downarrow\downarrow}]_{-k+Q_{b1}-Q_{b2}+Q_{b3}}^{\overline{\gamma_{b3}},\eta_{b4}}
    \Delta_{-k+Q_{b1}-Q_{b2}+Q_{b3}\downarrow,k\uparrow}^{\eta_{b4},\overline{\eta_{b4}}}\\
    &=-\frac{1}{2}\sum_{b1,b2,b3,b4}\psi_{b1}\psi_{b2}^*\psi_{b3}\psi_{b4}^*
    \sum_{\alpha_{b1},\beta_{b2},\gamma_{b3},\eta_{b4}}\sum_{k}
    G_{0,k}^{\overline{\eta _{b4}},\alpha_{b1}}G_{0,-k+Q_{b1}}^{\beta_{b2},\overline{\alpha_{b1}}}
    G_{0,k-Q_{b1}+Q_{b2}}^{\overline{\beta_{b2}},\gamma_{b3}}G_{0,-k+Q_{b1}-Q_{b2}+Q_{b3}}^{\eta_{b4},\overline{\gamma_{b3}}}\nonumber\\
    &\qquad\qquad\qquad\qquad\cdot f_{\Delta}(k,\alpha_{b1};-k+Q_{b1},\overline{\alpha_{b1}})\cdot
    f^*_{\Delta}(k-Q_{b1}+Q_{b2},\overline{\beta_{b2}};-k+Q_{b1},\beta_{b2})\nonumber\\
    &\qquad\qquad\qquad\qquad\cdot f_{\Delta}(k-Q_{b1}+Q_{b2},\gamma_{b3};-k+Q_{b1}-Q_{b2}+Q_{b3},\overline{\gamma_{b3}})\cdot
    f^*_{\Delta}(k,\overline{\eta_{b4}};-k+Q_{b1}-Q_{b2}+Q_{b3},\eta_{b4})
\end{align}
With the constraint $Q_{b1}-Q_{b2}+Q_{b3}-Q_{b4}=0$, we can divide the fourth-order term into several parts. When $b1=b2=b3=b4=b$:
\begin{align}
    &-\frac{1}{2}\sum_{b}|\psi_b|^4
    \sum_{\alpha_{b},\beta_{b},\gamma_{b},\eta_{b}}\sum_{k}
    G_{0,k}^{\overline{\eta _{b}},\alpha_{b}}G_{0,-k+Q_{b}}^{\beta_{b},\overline{\alpha_{b}}}G_{0,k}^{\overline{\beta_{b}},\gamma_{b}}G_{0,-k+Q_{b}}^{\eta_{b},\overline{\gamma_{b}}}\nonumber\\
    &\qquad\qquad\qquad\cdot f_{\Delta}(k,\alpha_{b};-k+Q_{b},\overline{\alpha_{b}})\cdot
    f^*_{\Delta}(k,\overline{\beta_{b}};-k+Q_{b},\beta_{b})
    \cdot f_{\Delta}(k,\gamma_{b};-k+Q_{b},\overline{\gamma_{b}})\cdot
    f^*_{\Delta}(k,\overline{\eta_{b}};-k+Q_{b},\eta_{b})
    \label{eq:gl_4_1}
\end{align}
When $b1=b2;b3=b4$ and $b1\ne b3$:
\begin{align}
    &-\frac{1}{2}\sum_{b1\ne b3}|\psi_{b1}|^2|\psi_{b3}|^2
    \sum_{\alpha_{b1},\beta_{b1},\gamma_{b3},\eta_{b3}}\sum_{k}
    G_{0,k}^{\overline{\eta _{b3}},\alpha_{b1}}G_{0,-k+Q_{b1}}^{\beta_{b1},\overline{\alpha_{b1}}}
    G_{0,k}^{\overline{\beta_{b1}},\gamma_{b3}}G_{0,-k+Q_{b3}}^{\eta_{b3},\overline{\gamma_{b3}}}\nonumber\\
    &\qquad\qquad\cdot f_{\Delta}(k,\alpha_{b1};-k+Q_{b1},\overline{\alpha_{b1}})\cdot
    f^*_{\Delta}(k,\overline{\beta_{b1}};-k+Q_{b1},\beta_{b1})
    \cdot f_{\Delta}(k,\gamma_{b3};-k+Q_{b3},\overline{\gamma_{b3}})\cdot
    f^*_{\Delta}(k,\overline{\eta_{b3}};-k+Q_{b3},\eta_{b3})
    \label{eq:gl_4_2_1}
\end{align}
When $b1=b4;b3=b2$ and $b1\ne b3$:
\begin{align}
    &-\frac{1}{2}\sum_{b1 \ne b3}|\psi_{b1}|^2|\psi_{b3}|^2
    \sum_{\alpha_{b1},\beta_{b3},\gamma_{b3},\eta_{b1}}\sum_{k}
    G_{0,k}^{\overline{\eta _{b1}},\alpha_{b1}}G_{0,-k+Q_{b1}}^{\beta_{b3},\overline{\alpha_{b1}}}
    G_{0,k-Q_{b1}+Q_{b3}}^{\overline{\beta_{b3}},\gamma_{b3}}G_{0,-k+Q_{b1}}^{\eta_{b1},\overline{\gamma_{b3}}}\nonumber\\
    &\qquad\qquad\qquad\qquad\cdot f_{\Delta}(k,\alpha_{b1};-k+Q_{b1},\overline{\alpha_{b1}})\cdot
    f^*_{\Delta}(k-Q_{b1}+Q_{b3},\overline{\beta_{b3}};-k+Q_{b1},\beta_{b3})\nonumber\\
    &\qquad\qquad\qquad\qquad\cdot f_{\Delta}(k-Q_{b1}+Q_{b3},\gamma_{b3};-k+Q_{b1},\overline{\gamma_{b3}})\cdot
    f^*_{\Delta}(k,\overline{\eta_{b1}};-k+Q_{b1},\eta_{b1})
    \label{eq:gl_4_2_2}
\end{align}
It is easy to prove that Eq.~\ref{eq:gl_4_2_1} and Eq.~\ref{eq:gl_4_2_2} are the same, and they can be combined as follows:
\begin{align}
    &-\sum_{b\ne b'}|\psi_{b}|^2|\psi_{b'}|^2\sum_{\alpha_{b},\beta_{b},\gamma_{b'},\eta_{b'}}\sum_{k}
    G_{0,k}^{\overline{\eta _{b'}},\alpha_{b}}G_{0,-k+Q_{b}}^{\beta_{b},\overline{\alpha_{b}}}
    G_{0,k}^{\overline{\beta_{b}},\gamma_{b'}}G_{0,-k+Q_{b'}}^{\eta_{b'},\overline{\gamma_{b'}}}\nonumber\\
    &\qquad\qquad\cdot f_{\Delta}(k,\alpha_{b};-k+Q_{b},\overline{\alpha_{b}})\cdot
    f^*_{\Delta}(k,\overline{\beta_{b}};-k+Q_{b},\beta_{b})
    \cdot f_{\Delta}(k,\gamma_{b'};-k+Q_{b'},\overline{\gamma_{b'}})\cdot
    f^*_{\Delta}(k,\overline{\eta_{b'}};-k+Q_{b'},\eta_{b'})
    \label{eq:gl_4_2}
\end{align}
When $b1=b3=b;b4=b2=b'$ and $b\ne b'$:
\begin{align}
    &-\frac{1}{2}\sum_{b\ne b'}\psi_{b}^2\psi_{b'}^{*2}\sum_{\alpha_{b},\beta_{b'},\gamma_{b},\eta_{b'}}\sum_{k}
    G_{0,k}^{\overline{\eta _{b'}},\alpha_{b}}G_{0,-k+Q_{b}}^{\beta_{b'},\overline{\alpha_{b}}}
    G_{0,k-Q_{b}+Q_{b'}}^{\overline{\beta_{b'}},\gamma_{b}}G_{0,-k+Q_{b'}}^{\eta_{b'},\overline{\gamma_{b}}}\nonumber\\
    &\qquad\qquad\qquad\qquad\cdot f_{\Delta}(k,\alpha_{b};-k+Q_{b},\overline{\alpha_{b}})\cdot
    f^*_{\Delta}(k-Q_{b}+Q_{b'},\overline{\beta_{b'}};-k+Q_{b},\beta_{b'})\nonumber\\
    &\qquad\qquad\qquad\qquad\cdot f_{\Delta}(k-Q_{b}+Q_{b'},\gamma_{b};-k+Q_{b'},\overline{\gamma_{b}})\cdot
    f^*_{\Delta}(k,\overline{\eta_{b'}};-k+Q_{b'},\eta_{b'})
    \label{eq:gl_4_3}
\end{align}
The fourth-order terms given in Eq.~\ref{eq:gl_4_1}, Eq.~\ref{eq:gl_4_2}, and Eq.~\ref{eq:gl_4_3}, correspond to the Feynman diagrams shown in Fig.~\ref{fig:feynman_fig1}(b),(c) and (d), respectively.
\begin{figure*}
  \begin{tikzpicture}
    \begin{feynman}
      \vertex (a) at (-1.5,0) {$ \psi_{b}$};
      \vertex (b) at (0,0);
      \vertex (c) at (2,0);
      \vertex (d) at (3.5,0) {$\psi^*_{b}$};
      \diagram* {
        (a) -- [charged scalar,  edge label=$Q_{b}$] (b),
        (b) -- [fermion, bend left=45, looseness=1.5,  edge label=$-k+Q_{b}$] (c),
        (b) -- [fermion, bend right=45, looseness=1.5, edge label'=$k$] (c),
        (c) -- [charged scalar,  edge label=$Q_{b}$] (d),
      };
      \node at (0.05, 0.34) {$\overline{\alpha_b}$};
      \node at (0.05, -0.35) {$\alpha_b$};
      \node at (1.95, 0.38) {$\beta_b$};
      \node at (1.95, -0.42) {$\overline{\beta_b}$};
      \node at (-1,3) {(a)};
    \end{feynman}

    \begin{feynman}
      \vertex (a) at (-1.5+8,0) {$ \psi_{b}$};
      \vertex (b) at (0+8,0);
      \vertex (c) at (1.5+8,1.5);
      \vertex (d) at (1.5+8,-1.5);
      \vertex (e) at (3+8,0);
      \vertex (f) at (4.5+8,0) {$\psi_{b}$};
      \vertex (g) at (1.5+8,3) {$\psi^*_{b}$};
      \vertex (h) at (1.5+8,-3) {$\psi^*_{b}$};

      \diagram* {
        (a) -- [charged scalar, edge label=$Q_{b}$] (b),
        (f) -- [charged scalar, edge label=$Q_{b}$] (e),
        (c) -- [charged scalar, edge label=$Q_{b}$] (g),
        (d) -- [charged scalar, edge label=$Q_{b}$] (h),

        (b) -- [fermion, bend left=40, looseness=1,  edge label=$-k+Q_{b}$] (c),
        (b) -- [fermion, bend right=40, looseness=1,  edge label'=$k$] (d),
        (c) -- [anti fermion, bend left=40, looseness=1,  edge label=$k$] (e),
        (d) -- [anti fermion, bend right=40, looseness=1,  edge label'=$-k+Q_{b}$] (e),

      };
      \node at (0.3+8, 0.3) {$\overline{\alpha_b}$};
      \node at (0.3+8, -0.35) {${\alpha_b}$};
      \node at (1.2+8, 1.15) {${\beta_b}$};
      \node at (1.8+8, 1.15) {$\overline{\beta_b}$};
      \node at (2.7+8, 0.3) {${\gamma_b}$};
      \node at (2.7+8, -0.35) {$\overline{\gamma_b}$};
      \node at (1.8+8, -1.2) {${\eta_b}$};
      \node at (1.2+8, -1.2) {$\overline{\eta_b}$};
      \node at (-1+8,3) {(b)};
    \end{feynman}

    \begin{feynman}
      \vertex (a) at (-1.5,-7) {$ \psi_{b}$};
      \vertex (b) at (0,-7);

      \vertex (c) at (1.5,1.5-7);
      \vertex (d) at (1.5,-1.5-7);
      \vertex (e) at (3,0-7);
      \vertex (f) at (4.5,0-7) {$\psi_{b'}$};
      \vertex (g) at (1.5,3-7) {$\psi^*_{b}$};
      \vertex (h) at (1.5,-3-7) {$\psi^*_{b'}$};

      \diagram* {
        (a) -- [charged scalar, edge label=$Q_{b}$] (b),
        (c) -- [charged scalar, edge label=$Q_{b}$] (g),
        (f) -- [charged scalar, edge label=$Q_{b'}$] (e),
        (d) -- [charged scalar, edge label=$Q_{b'}$] (h),

        (b) -- [fermion, bend left=40, looseness=1,  edge label=$-k+Q_{b}$] (c),
        (b) -- [fermion, bend right=40, looseness=1,  edge label'=$k$] (d),
        (c) -- [anti fermion, bend left=40, looseness=1,  edge label=$k$] (e),
        (d) -- [anti fermion, bend right=40, looseness=1,  edge label'=$-k+Q_{b'}$] (e),

      };
      \node at (0.3, 0.3-7) {$\overline{\alpha_b}$};
      \node at (0.3, -0.35-7) {${\alpha_b}$};
      \node at (1.2, 1.15-7) {${\beta_{b}}$};
      \node at (1.8, 1.15-7) {$\overline{\beta_{b}}$};
      \node at (2.7, 0.3-7) {${\gamma_{b'}}$};
      \node at (2.7, -0.35-7) {$\overline{\gamma_{b'}}$};
      \node at (1.8, -1.2-7) {${\eta_{b'}}$};
      \node at (1.2, -1.2-7) {$\overline{\eta_{b'}}$};
      \node at (-1,3-7) {(c)};
    \end{feynman}

    \begin{feynman}
      \vertex (a) at (-1.5+8,-7) {$ \psi_{b}$};
      \vertex (b) at (0+8,-7);

      \vertex (c) at (1.5+8,1.5-7);
      \vertex (d) at (1.5+8,-1.5-7);
      \vertex (e) at (3+8,0-7);
      \vertex (f) at (4.5+8,0-7) {$\psi_{b}$};
      \vertex (g) at (1.5+8,3-7) {$\psi^*_{b'}$};
      \vertex (h) at (1.5+8,-3-7) {$\psi^*_{b'}$};

      \diagram* {
        (a) -- [charged scalar, edge label=$Q_{b}$] (b),
        (f) -- [charged scalar, edge label=$Q_{b}$] (e),
        (c) -- [charged scalar, edge label=$Q_{b'}$] (g),
        (d) -- [charged scalar, edge label=$Q_{b'}$] (h),

        (b) -- [fermion, bend left=40, looseness=1,  edge label=$-k+Q_{b}$] (c),
        (b) -- [fermion, bend right=40, looseness=1,  edge label'=$k$] (d),
        (c) -- [anti fermion, bend left=40, looseness=1,  edge label=$k-Q_{b}+Q_{b'}$] (e),
        (d) -- [anti fermion, bend right=40, looseness=1,  edge label'=$-k+Q_{b'}$] (e),
      };
      \node at (0.3+8, 0.3-7) {$\overline{\alpha_b}$};
      \node at (0.3+8, -0.35-7) {${\alpha_b}$};
      \node at (1.2+8, 1.15-7) {${\beta_{b'}}$};
      \node at (1.8+8, 1.15-7) {$\overline{\beta_{b'}}$};
      \node at (2.7+8, 0.3-7) {${\gamma_b}$};
      \node at (2.7+8, -0.35-7) {$\overline{\gamma_b}$};
      \node at (1.8+8, -1.2-7) {${\eta_{b'}}$};
      \node at (1.2+8, -1.2-7) {$\overline{\eta_{b'}}$};
      \node at (-1+8,3-7) {(d)};
    \end{feynman}
  \end{tikzpicture}
  \caption{The Feynman diagrams corresponding to the second-order (a) and fourth-order (b-d) terms in the GL theory, associated with Eq.~(2) of the main text, consider only the PDW order. In the diagrams, the dashed lines represent bosonic fields, with their direction indicating either $\psi^*_b$ or $\psi_b$, and their contribution is given by $f^*_{\Delta}(k,\alpha_{b};k',\overline{\alpha_{b}})$ or $f_{\Delta}(k,\alpha_{b};k',\overline{\alpha_{b}})$. The solid lines represent the fermionic Green's function $G_0^{\beta\alpha}(k)$, where the sublattice indices of the Green's function are indicated at the endpoints of the lines. The sign of each diagram is determined by $(-1)^{m/2 + 1}$, where $m$ is the total number of PDW bosonic fields. The coefficient for each diagram corresponds $\frac{c}{m}$, where $c$ is the number of distinct diagrams that can be obtained by performing an overall exchange based on the rotational relation and an overall direction reversal of the bosonic lines labeled $b$. Diagrams in which $b$ and $b'$ are exchanged are considered equivalent.
  }
  \label{fig:feynman_fig1}
  \end{figure*}
\par
Next, let us add CDW, for which we define the auxiliary fields $\phi_b$ along the specific bond directions $\bm{a}_b$ with $b=1,2,3$. That is $\phi_1^{\text{AB}}=[\phi_1^{\text{BA}}]^{*} \equiv \phi_1$, $\phi_2^{\text{CB}}=[\phi_2^{\text{BC}}]^{*} \equiv \phi_2$, $\phi_3^{\text{AC}}=[\phi_3^{\text{CA}}]^{*} \equiv \phi_3$. In Nambu spinor representation, the CDW matrix $\chi$ satisfies the following relations:
\begin{align}
    &\chi_{k\uparrow,k\pm Q_b\uparrow}^{\alpha_b,\overline{\alpha_b}}=\phi_b^{\alpha_b\overline{\alpha_b}} \cdot f_{\chi}(k,\alpha_{b};k\pm Q_{b},\overline{\alpha_{b}})\\
    &\chi_{k\downarrow,k\pm Q_b\downarrow}^{\alpha_b,\overline{\alpha_b}}=[\phi_b^{*}]^{\alpha_b\overline{\alpha_b}} \cdot \left[-f^*_{\chi}(k,\alpha_{b};k\pm Q_{b},\overline{\alpha_{b}})\right]\\
    &f_{\chi}(k\pm Q_{b},\overline{\alpha_{b}};k,\alpha_{b})=f^*_{\chi}(k,\alpha_{b};k\pm Q_{b},\overline{\alpha_{b}})\\
    &f_{\chi}(k,\alpha_{b};k+ Q_{b},\overline{\alpha_{b}})=f_{\chi}(k,\alpha_{b};k- Q_{b},\overline{\alpha_{b}})
\end{align}
Here, for simplicity, we neglect the CDW order on the next-nearest-neighbor bond, and the factor $f_{\chi}(k\alpha_{b}, k'\overline{\alpha_{b}})$ can be determined through the specific internal structure of the CDW order under $Q_{b}$.
\par
Following the above procedure, we get the GL free energy
\begin{equation}
    F=-\frac{T\text{ln}Z}{N}=\frac4V\sum_b |\phi_b|^2+\frac{2}{J}\sum_b|\psi_b|^2-\frac{T}{N}\text{Trln}(\mathcal{G}_0^{-1}+\chi+\Delta)
\end{equation}
The last term $\text{Trln}(\mathcal{G}_0^{-1}+\chi+\Delta)$ can be expanded as
\begin{align}
    \text{Trln}(\mathcal{G}_0^{-1}+\chi+\Delta)&=\text{Trln}(\mathcal{G}_0^{-1}[1+\mathcal{G}_0(\chi+\Delta)])\nonumber\\
    &=\text{Trln}(\mathcal{G}_0^{-1})+\text{Trln}(1+\mathcal{G}_0(\chi+\Delta))\nonumber\\
    &=\text{const.}+\sum_{n=1}^{\infty}\frac{(-1)^{n+1}}{n}\text{Tr}(\mathcal{G}_0\chi+\mathcal{G}_0\Delta)^{n}
\end{align}
Again, only even orders of $\Delta$ are allowed to be nonzero since only $\Delta$ is off-diagonal in Nambu space.
Therefore, the lowest-order coupling between PDW and CDW is the third-order,
\begin{align}
    &\frac{1}{3}\cdot 3 \cdot 2\text{Tr}[\mathcal{G}_{0,\uparrow\uparrow}]\Delta_{\uparrow\downarrow}[\mathcal{G}_{0,\downarrow\downarrow}]\chi_{\downarrow\downarrow}[\mathcal{G}_{0,\downarrow\downarrow}]\Delta_{\downarrow\uparrow}\\
    &=2\sum_{b1\ne b2 \ne b3}\sum_{\alpha_{b1}\beta_{b2},\gamma_{b3}}\sum_{k}
    [\mathcal{G}_{0,\uparrow\uparrow}]_k^{\overline{\gamma_{b3}},\alpha_{b1}}
    \Delta_{k\uparrow,-k+Q_{b1}\downarrow}^{\alpha_{b1},\overline{\alpha_{b1}}}
    [\mathcal{G}_{0,\downarrow\downarrow}]_{-k+Q_{b1}}^{\overline{\alpha_{b1}},\beta_{b2}}
    \chi_{-k+Q_{b1}\downarrow,-k+Q_{b1}-Q_{b2}\downarrow}^{\beta_{b2},\overline{\beta_{b2}}}
    [\mathcal{G}_{0,\downarrow\downarrow}]_{-k+Q_{b1}-Q_{b2}}^{\overline{\beta_{b2}},\gamma_{b3}}
    \Delta_{-k+Q_{b1}-Q_{b2}\downarrow,k\uparrow}^{\gamma_{b3},\overline{\gamma_{b3}}}\nonumber\\
    &=-2\sum_{b\ne b' \ne b''}\sum_{\alpha_{b},\beta_{b'},\gamma_{b''}}\psi_b\psi_{b''}^{*}[\phi^*_{b'}]^{\beta_{b'}\overline{\beta_{b'}}}\sum_{k}
    G_{0,k}^{\overline{\gamma_{b''}},\alpha_{b}}G_{0,-k+Q_{b}}^{\beta_{b'},\overline{\alpha_{b}}}G_{0,-k+Q_{b}-Q_{b'}}^{\gamma_{b''},\overline{\beta_{b'}}}\nonumber\\
    &\qquad\qquad\qquad\cdot f_\Delta(k,\alpha_b;-k+Q_b,\overline{\alpha_b})\cdot f_{\chi}(-k+Q_b-Q_{b'},\overline{\beta_{b'}};-k+Q_{b},\beta_{b'})\cdot f^*_\Delta(k,\overline{\gamma_{b''}};-k+Q_{b}-Q_{b'},\gamma_{b''})
\end{align}
The corresponding Feynman diagram is shown in Fig.~\ref{fig:feynman_fig2}(a). The third-order terms of $\phi$ contribute to the GL free energy for pure CDW, which have been studied before so we will not discuss them any more.
\par
The fourth-order term is
\begin{align}
    &-\frac{1}{4}\left[2\cdot2\cdot\text{Tr}[\mathcal{G}_{0,\uparrow\uparrow}]\Delta_{\uparrow\downarrow}[\mathcal{G}_{0,\downarrow\downarrow}]\chi_{\downarrow\downarrow}[\mathcal{G}_{0,\downarrow\downarrow}]\Delta_{\downarrow\uparrow}[\mathcal{G}_{0,\uparrow\uparrow}]\chi_{\uparrow\uparrow}+4\cdot2 \text{Tr}[\mathcal{G}_{0,\uparrow\uparrow}]\Delta_{\uparrow\downarrow}[\mathcal{G}_{0,\downarrow\downarrow}]\Delta_{\downarrow\uparrow}[\mathcal{G}_{0,\uparrow\uparrow}]\chi_{\uparrow\uparrow}[\mathcal{G}_{0,\uparrow\uparrow}]\chi_{\uparrow\uparrow}\right]\\
    &=-\sum_{b1,b2,b3,b4}\sum_{\alpha_{b1}\beta_{b2},\gamma_{b3},\eta_{b4}}\sum_{k}
    [\mathcal{G}_{0,\uparrow\uparrow}]^{\overline{\eta_{b4}},\alpha_{b1}}_k\Delta^{\alpha_{b1},\overline{\alpha_{b1}}}_{k\uparrow,-k+Q_{b1}\downarrow}[\mathcal{G}_{0,\downarrow\downarrow}]^{\overline{\alpha_{b1}},\beta_{b2}}_{-k+Q_{b1}}\chi^{\beta_{b2},\overline{\beta_{b2}}}_{-k+Q_{b1}\downarrow,-k+Q_{b1}-Q_{b2}\downarrow}\nonumber\\
    &\qquad\qquad\qquad\cdot[\mathcal{G}_{0,\downarrow\downarrow}]^{\overline{\beta_{b2}},\gamma_{b3}}_{-k+Q_{b1}-Q_{b2}}\Delta^{\gamma_{b3},\overline{\gamma_{b3}}}_{-k+Q_{b1}-Q_{b2}\downarrow,k-Q_{b1}+Q_{b2}+Q_{b3}\uparrow}[\mathcal{G}_{0,\uparrow\uparrow}]^{\overline{\gamma_{b3}},\eta_{b4}}_{k-Q_{b1}+Q_{b2}+Q_{b3}}\chi^{\eta_{b4},\overline{\eta_{b4}}}_{k-Q_{b1}+Q_{b2}+Q_{b3}\uparrow,k\uparrow}\nonumber\\
    &\qquad-2\sum_{b1,b2,b3,b4}\sum_{\alpha_{b1}\beta_{b2},\gamma_{b3},\eta_{b4}}\sum_{k}
    [\mathcal{G}_{0,\uparrow\uparrow}]^{\overline{\eta_{b4}},\alpha_{b1}}_k\Delta^{\alpha_{b1},\overline{\alpha_{b1}}}_{k\uparrow,-k+Q_{b1}\downarrow}
    [\mathcal{G}_{0,\downarrow\downarrow}]^{\overline{\alpha_{b1}},\beta_{b2}}_{-k+Q_{b1}}\Delta^{\beta_{b2},\overline{\beta_{b2}}}_{-k+Q_{b1}\downarrow,k-Q_{b1}+Q_{b2}\uparrow}\nonumber\\
    &\qquad\qquad\qquad\cdot[\mathcal{G}_{0,\uparrow\uparrow}]^{\overline{\beta_{b2}},\gamma_{b3}}_{k-Q_{b1}+Q_{b2}}\chi^{\gamma_{b3},\overline{\gamma_{b3}}}_{k-Q_{b1}+Q_{b2}\uparrow,k-Q_{b1}+Q_{b2}-Q_{b3}\uparrow}
    [\mathcal{G}_{0,\uparrow\uparrow}]^{\overline{\gamma_{b3}},\eta_{b4}}_{k-Q_{b1}+Q_{b2}-Q_{b3}}\chi^{\eta_{b4},\overline{\eta_{b4}}}_{k-Q_{b1}+Q_{b2}-Q_{b3}\uparrow,k\uparrow}\nonumber\\
\end{align}
In the following, the fourth-order terms in Eq.~\ref{eq:free_energy_cdw} are given as follows.
\par
When $b1=b2=b3=b4=b$, the resulting term of $|\psi_b|^2|\phi_b|^2$ is given by
\begin{align}
    &\sum_{b}\sum_{\alpha_{b}\beta_{b},\gamma_{b}}|\psi_b|^2|\phi_b|^2 \sum_{k}
    G_{0,k}^{\overline{\beta_{b}},\alpha_{b}}
    G^{\beta_{b},\overline{\alpha_{b}}}_{0,-k+Q_{b}}
    G^{\gamma_{b},\overline{\beta_{b}}}_{0,-k}
    G^{\overline{\gamma_{b}},\beta_{b}}_{0,k+Q_{b}}\nonumber\\
    &\qquad\qquad \cdot f_\Delta(k,\alpha_{b};-k+Q_{b},\overline{\alpha_{b}})
    f_{\chi}(-k,\overline{\beta_{b}};-k+Q_{b},\beta_{b})
    f^*_\Delta(k+Q_{b},\overline{\gamma_{b}};-k,\gamma_{b})
    f_\chi(k+Q_{b},\beta_{b};k,\overline{\beta_{b}}) \label{eq:gl_cdw_4_1}
    \\
    &+2\sum_{b}\sum_{\alpha_{b}\beta_{b},\gamma_{b}}|\psi_b|^2|\phi_b|^2\sum_{k}
    G^{\gamma_{b},\alpha_{b}}_k
    G^{\beta_{b},\overline{\alpha_{b}}}_{0,-k+Q_{b}}
    G^{\overline{\beta_{b}},\gamma_{b}}_{0,k}
    G^{\overline{\gamma_{b}},\overline{\gamma_{b}}}_{0,k-Q_{b}}\nonumber\\
    &\qquad\qquad \cdot f_\Delta(k,\alpha_{b};-k+Q_{b},\overline{\alpha_{b}})
    f^*_\Delta(k,\overline{\beta_{b}};-k+Q_{b},\beta_{b})
    f_\chi(k,\gamma_{b},k-Q_{b},\overline{\gamma_{b}})
    f_\chi(k-Q_{b},\overline{\gamma_{b}};k,\gamma_{b})\label{eq:gl_cdw_4_2}
\end{align}
For the term of $|\psi_b|^2|\phi_{b'}|^2$ with $b\ne b'$, the results are
\begin{align}
    &\sum_{b \ne b'}\sum_{\alpha_{b}\beta_{b'},\gamma_{b}}|\psi_b|^2|\phi_{b'}|^2\sum_{k}
    G^{\overline{\beta_{b'}},\alpha_{b}}_{0,k}
    G^{\beta_{b'},\overline{\alpha_{b}}}_{0,-k+Q_{b}}
    G^{\gamma_{b},\overline{\beta_{b'}}}_{0,-k+Q_{b}-Q_{b'}}
    G^{\overline{\gamma_{b}},\beta_{b'}}_{0,k+Q_{b'}}\nonumber\\
    &\quad \cdot f_\Delta(k,\alpha_{b};-k+Q_{b},\overline{\alpha_{b}})
    f_{\chi}(-k+Q_{b}-Q_{b'},\overline{\beta_{b'}};-k+Q_{b},\beta_{b'})
    f^*_\Delta(k+Q_{b'},\overline{\gamma_{b}};-k+Q_{b}-Q_{b'},\gamma_{b})
    f_\chi(k+Q_{b'},\beta_{b'};k,\overline{\beta_{b'}})\label{eq:gl_cdw_4_3}\\
    &+2\sum_{b \ne b'}\sum_{\alpha_{b}\beta_{b},\gamma_{b'}} |\psi_b|^2|\phi_{b'}|^2 \sum_{k}
    G^{\gamma_{b'},\alpha_{b}}_{0,k}\
    G^{\beta_{b},\overline{\alpha_{b}}}_{0,-k+Q_{b}}
    G^{\overline{\beta_{b}},\gamma_{b'}}_{0,k}
    G^{\overline{\gamma_{b'}},\overline{\gamma_{b'}}}_{0,k-Q_{b'}}\nonumber\\
    &\qquad\qquad \cdot f_\Delta(k,\alpha_{b};-k+Q_{b},\overline{\alpha_{b}})
    f^*_\Delta(k,\overline{\beta_{b}};-k+Q_{b},\beta_{b})
    f_\chi(k,\gamma_{b'},k-Q_{b'},\overline{\gamma_{b'}})
    f_\chi(k-Q_{b'},\overline{\gamma_{b'}};k,\gamma_{b'})\label{eq:gl_cdw_4_4}
\end{align}
The Eq.~\ref{eq:gl_cdw_4_1}, \ref{eq:gl_cdw_4_2}, \ref{eq:gl_cdw_4_3} and \ref{eq:gl_cdw_4_4} corresponding
to the Feynman diagrams shown in Fig.~\ref{fig:feynman_fig2}(b), (c), (d) and (e) respectively.
In addition, there are also other terms like $\psi_1^*\psi_2 \phi_1\phi_2$, but this effect on PDW has been captured by the third-order coupling term, so we neglect them for simplicity.
\begin{figure*}
\begin{tikzpicture}
  \begin{feynman}
    \vertex (a) at (-1.5+4,+5) {$ \psi_{b}$};
    \vertex (b) at (0+4,0+5);
    \vertex (c) at (1.5+4,1.5+5);
    \vertex (d) at (1.5+4,-1.5+5);
    \vertex (e) at (3+4,0+5);
    \vertex (f) at (4.5+4,0+5) {$\psi^*_{b''}$};
    \vertex (g) at (1.5+4,3+5) {$[\phi^*_{b'}]^{\beta_{b'}\overline{\beta_{b'}}}$};

    \diagram* {
      (a) -- [charged scalar, edge label=$Q_{b}$] (b),
      (e) -- [charged scalar, edge label=$Q_{b''}$] (f),
      (c) -- [charged boson, edge label=$Q_{b'}$] (g),
      (b) -- [fermion, bend left=40, looseness=1,  edge label=$-k+Q_{b}$] (c),
      (c) -- [fermion, bend left=40, looseness=1,  edge label=$-k+Q_{b}-Q_{b'}$] (e),
      (b) -- [fermion, bend right=50, looseness=1.4,  edge label'=$k$] (e),

    };
    \node at (0.3+4, 0.3+5) {$\overline{\alpha_b}$};
    \node at (0.6+4, -0.3+5) {${\alpha_b}$};
    \node at (1.2+4, 1.15+5) {${\beta_{b'}}$};
    \node at (1.8+4, 1.15+5) {$\overline{\beta_{b'}}$};
    \node at (2.7+4, 0.3+5) {${\gamma_b''}$};
    \node at (2.4+4, -0.3+5) {$\overline{\gamma_b''}$};
    \node at (-1+4,3+5) {(a)};
  \end{feynman}

  \begin{feynman}
    \vertex (a) at (-1.5,0) {$ \psi_{b}$};
    \vertex (b) at (0,0);
    \vertex (c) at (1.5,1.5);
    \vertex (d) at (1.5,-1.5);
    \vertex (e) at (3,0);
    \vertex (f) at (4.5,0) {$\psi_{b}^*$};
    \vertex (g) at (1.5,3) {$[\phi^*_{b}]^{\beta_{b}\overline{\beta_{b}}}$};
    \vertex (h) at (1.5,-3) {$\phi_{b}^{\beta_{b}\overline{\beta_{b}}}$};

    \diagram* {
      (a) -- [charged scalar, edge label=$Q_{b}$] (b),
      (e) -- [charged scalar, edge label=$Q_{b}$] (f),
      (c) -- [charged boson, edge label=$Q_{b}$] (g),
      (d) -- [anti charged boson, edge label=$Q_{b}$] (h),

      (b) -- [fermion, bend left=40, looseness=1,  edge label=$-k+Q_{b}$] (c),
      (b) -- [fermion, bend right=40, looseness=1,  edge label'=$k$] (d),
      (c) -- [fermion, bend left=40, looseness=1,  edge label=$-k$] (e),
      (d) -- [fermion, bend right=40, looseness=1,  edge label'=$k+Q_{b}$] (e),
    };
    \node at (0.3, 0.3) {$\overline{\alpha_b}$};
    \node at (0.3, -0.35) {${\alpha_b}$};
    \node at (1.2, 1.15) {${\beta_b}$};
    \node at (1.8, 1.15) {$\overline{\beta_b}$};
    \node at (2.7, 0.3) {${\gamma_b}$};
    \node at (2.7, -0.35) {$\overline{\gamma_b}$};
    \node at (1.8, -1.2) {${\beta_b}$};
    \node at (1.2, -1.2) {$\overline{\beta_b}$};
    \node at (-1,3) {(b)};
  \end{feynman}

  \begin{feynman}
    \vertex (a) at (-1.5+8,0) {$ \psi_{b}$};
    \vertex (b) at (0+8,0);
    \vertex (c) at (1.5+8,1.5);
    \vertex (d) at (1.5+8,-1.5);
    \vertex (e) at (3+8,0);
    \vertex (f) at (4.5+8,0) {$\phi_{b}^{\gamma_{b}\overline{\gamma_{b}}}$};
    \vertex (g) at (1.5+8,3) {$\psi_b^*$};
    \vertex (h) at (1.5+8,-3) {$\phi_{b}^{\overline{\gamma_{b}}\gamma_{b}}$};

    \diagram* {
      (a) -- [charged scalar, edge label=$Q_{b}$] (b),
      (f) -- [charged boson, edge label=$Q_{b}$] (e),
      (c) -- [charged scalar, edge label=$Q_{b}$] (g),
      (d) -- [anti charged boson, edge label=$Q_{b}$] (h),

      (b) -- [fermion, bend left=40, looseness=1,  edge label=$-k+Q_{b}$] (c),
      (b) -- [fermion, bend right=40, looseness=1,  edge label'=$k$] (d),
      (c) -- [anti fermion, bend left=40, looseness=1,  edge label=$k$] (e),
      (d) -- [fermion, bend right=40, looseness=1,  edge label'=$k-Q_{b}$] (e),

    };
    \node at (0.3+8, 0.3) {$\overline{\alpha_b}$};
    \node at (0.3+8, -0.35) {${\alpha_b}$};
    \node at (1.2+8, 1.15) {${\beta_b}$};
    \node at (1.8+8, 1.15) {$\overline{\beta_b}$};
    \node at (2.7+8, 0.3) {${\gamma_b}$};
    \node at (2.7+8, -0.35) {$\overline{\gamma_b}$};
    \node at (1.8+8, -1.2) {${\overline{\gamma_b}}$};
    \node at (1.2+8, -1.2) {$\gamma_b$};
    \node at (-1+8,3) {(c)};
  \end{feynman}

  \begin{feynman}
    \vertex (a) at (-1.5,0-7) {$ \psi_{b}$};
    \vertex (b) at (0,0-7);
    \vertex (c) at (1.5,1.5-7);
    \vertex (d) at (1.5,-1.5-7);
    \vertex (e) at (3,0-7);
    \vertex (f) at (4.5,0-7) {$\psi_{b}^*$};
    \vertex (g) at (1.5,3-7) {$[\phi^*_{b'}]^{\beta_{b'}\overline{\beta_{b'}}}$};
    \vertex (h) at (1.5,-3-7) {$\phi_{b'}^{\beta_{b'}\overline{\beta_{b'}}}$};

    \diagram* {
      (a) -- [charged scalar, edge label=$Q_{b}$] (b),
      (e) -- [charged scalar, edge label=$Q_{b}$] (f),
      (c) -- [charged boson, edge label=$Q_{b'}$] (g),
      (d) -- [anti charged boson, edge label=$Q_{b'}$] (h),

      (b) -- [fermion, bend left=40, looseness=1,  edge label=$-k+Q_{b}$] (c),
      (b) -- [fermion, bend right=40, looseness=1,  edge label'=$k$] (d),
      (c) -- [fermion, bend left=40, looseness=1,  edge label=$-k+Q_{b}-Q_{b'}$] (e),
      (d) -- [fermion, bend right=40, looseness=1,  edge label'=$k+Q_{b'}$] (e),
    };
    \node at (0.3, 0.3-7) {$\overline{\alpha_b}$};
    \node at (0.3, -0.35-7) {${\alpha_b}$};
    \node at (1.2, 1.15-7) {${\beta_{b'}}$};
    \node at (1.8, 1.15-7) {$\overline{\beta_{b'}}$};
    \node at (2.7, 0.3-7) {${\gamma_b}$};
    \node at (2.7, -0.35-7) {$\overline{\gamma_b}$};
    \node at (1.8, -1.2-7) {${\beta_{b'}}$};
    \node at (1.2, -1.2-7) {$\overline{\beta_{b'}}$};
    \node at (-1,3-7) {(d)};
  \end{feynman}

  \begin{feynman}
    \vertex (a) at (-1.5+8,0-7) {$ \psi_{b}$};
    \vertex (b) at (0+8,0-7);
    \vertex (c) at (1.5+8,1.5-7);
    \vertex (d) at (1.5+8,-1.5-7);
    \vertex (e) at (3+8,0-7);
    \vertex (f) at (4.5+8,0-7) {$\phi_{b'}^{\gamma_{b'}\overline{\gamma_{b'}}}$};
    \vertex (g) at (1.5+8,3-7) {$\psi_b^*$};
    \vertex (h) at (1.5+8,-3-7) {$\phi_{b'}^{\overline{\gamma_{b'}}\gamma_{b'}}$};

    \diagram* {
      (a) -- [charged scalar, edge label=$Q_{b}$] (b),
      (f) -- [charged boson, edge label=$Q_{b'}$] (e),
      (c) -- [charged scalar, edge label=$Q_{b}$] (g),
      (d) -- [anti charged boson, edge label=$Q_{b'}$] (h),

      (b) -- [fermion, bend left=40, looseness=1,  edge label=$-k+Q_{b}$] (c),
      (b) -- [fermion, bend right=40, looseness=1,  edge label'=$k$] (d),
      (c) -- [anti fermion, bend left=40, looseness=1,  edge label=$k$] (e),
      (d) -- [fermion, bend right=40, looseness=1,  edge label'=$k-Q_{b'}$] (e),

    };
    \node at (0.3+8, 0.3-7) {$\overline{\alpha_b}$};
    \node at (0.3+8, -0.35-7) {${\alpha_b}$};
    \node at (1.2+8, 1.15-7) {${\beta_b}$};
    \node at (1.8+8, 1.15-7) {$\overline{\beta_b}$};
    \node at (2.7+8, 0.3-7) {${\gamma_{b'}}$};
    \node at (2.7+8, -0.35-7) {$\overline{\gamma_{b'}}$};
    \node at (1.8+8, -1.2-7) {${\overline{\gamma_{b'}}}$};
    \node at (1.2+8, -1.2-7) {$\gamma_{b'}$};
    \node at (-1+8,3-7) {(e)};
  \end{feynman}
\end{tikzpicture}
\caption{The Feynman diagrams corresponding to the third-order (a) and fourth-order (b-e) terms in the GL theory, associated with Eq.~\ref{eq:free_energy_cdw}, consider both CDW and PDW orders. In the diagrams, the solid lines represent the fermionic Green's function $G_0^{\beta\alpha}(k)$, where the sublattice indices of the Green's function are indicated at the endpoints of the lines. The dashed lines represent PDW bosonic fields, with their direction indicating either $\psi^*_b$ or $\psi_b$, contributing $f_{\Delta}(k,\alpha_{b};k',\overline{\alpha_{b}})$ or $f_{\Delta}(k,\alpha_{b};k',\overline{\alpha_{b}})$. The wavy lines represent CDW bosonic fields, with their direction indicating either $\phi^*_b$ or $\phi_b$, contributing $f_{\chi}(k,\alpha_{b};k',\overline{\alpha_{b}})$, where $k$ denotes the outgoing momentum. The sign of each diagram is determined by $(-1)^{m/2 +n+ 1}$, where $m$ is the total number of PDW bosonic fields and $n$ is the total number of PDW and CDW bosonic fields. The coefficient for each diagram corresponds to $\frac{c}{n}$, where $c$ is the number of distinct diagrams that can be obtained by performing an overall exchange based on the rotational relation and an overall direction reversal of the PDW and CDW bosonic lines labeled $b$. Diagrams in which $b$ and $b'$ are exchanged are considered equivalent.
}
\label{fig:feynman_fig2}
\end{figure*}
\end{widetext}

\bibliography{reference}

\begin{thebibliography}{105}%
\makeatletter
\providecommand \@ifxundefined [1]{%
 \@ifx{#1\undefined}
}%
\providecommand \@ifnum [1]{%
 \ifnum #1\expandafter \@firstoftwo
 \else \expandafter \@secondoftwo
 \fi
}%
\providecommand \@ifx [1]{%
 \ifx #1\expandafter \@firstoftwo
 \else \expandafter \@secondoftwo
 \fi
}%
\providecommand \natexlab [1]{#1}%
\providecommand \enquote  [1]{``#1''}%
\providecommand \bibnamefont  [1]{#1}%
\providecommand \bibfnamefont [1]{#1}%
\providecommand \citenamefont [1]{#1}%
\providecommand \href@noop [0]{\@secondoftwo}%
\providecommand \href [0]{\begingroup \@sanitize@url \@href}%
\providecommand \@href[1]{\@@startlink{#1}\@@href}%
\providecommand \@@href[1]{\endgroup#1\@@endlink}%
\providecommand \@sanitize@url [0]{\catcode `\\12\catcode `\$12\catcode
  `\&12\catcode `\#12\catcode `\^12\catcode `\_12\catcode `\%12\relax}%
\providecommand \@@startlink[1]{}%
\providecommand \@@endlink[0]{}%
\providecommand \url  [0]{\begingroup\@sanitize@url \@url }%
\providecommand \@url [1]{\endgroup\@href {#1}{\urlprefix }}%
\providecommand \urlprefix  [0]{URL }%
\providecommand \Eprint [0]{\href }%
\providecommand \doibase [0]{https://doi.org/}%
\providecommand \selectlanguage [0]{\@gobble}%
\providecommand \bibinfo  [0]{\@secondoftwo}%
\providecommand \bibfield  [0]{\@secondoftwo}%
\providecommand \translation [1]{[#1]}%
\providecommand \BibitemOpen [0]{}%
\providecommand \bibitemStop [0]{}%
\providecommand \bibitemNoStop [0]{.\EOS\space}%
\providecommand \EOS [0]{\spacefactor3000\relax}%
\providecommand \BibitemShut  [1]{\csname bibitem#1\endcsname}%
\let\auto@bib@innerbib\@empty
\bibitem [{\citenamefont {Fulde}\ and\ \citenamefont
  {Ferrell}(1964)}]{FFLO_Fulde}%
  \BibitemOpen
  \bibfield  {author} {\bibinfo {author} {\bibfnamefont {P.}~\bibnamefont
  {Fulde}}\ and\ \bibinfo {author} {\bibfnamefont {R.~A.}\ \bibnamefont
  {Ferrell}},\ }\bibfield  {title} {\bibinfo {title} {Superconductivity in a
  strong spin-exchange field},\ }\href
  {https://doi.org/10.1103/PhysRev.135.A550} {\bibfield  {journal} {\bibinfo
  {journal} {Phys. Rev.}\ }\textbf {\bibinfo {volume} {135}},\ \bibinfo {pages}
  {A550} (\bibinfo {year} {1964})}\BibitemShut {NoStop}%
\bibitem [{\citenamefont {Larkin}\ and\ \citenamefont
  {Ovchinnikov}(1964)}]{FFLO_Larkin}%
  \BibitemOpen
  \bibfield  {author} {\bibinfo {author} {\bibfnamefont {A.~I.}\ \bibnamefont
  {Larkin}}\ and\ \bibinfo {author} {\bibfnamefont {Y.~N.}\ \bibnamefont
  {Ovchinnikov}},\ }\bibfield  {title} {\bibinfo {title} {{Nonuniform state of
  superconductors}},\ }\href@noop {} {\bibfield  {journal} {\bibinfo  {journal}
  {Zh. Eksp. Teor. Fiz.}\ }\textbf {\bibinfo {volume} {47}},\ \bibinfo {pages}
  {1136} (\bibinfo {year} {1964})}\BibitemShut {NoStop}%
\bibitem [{\citenamefont {Zwierlein}\ \emph {et~al.}(2006)\citenamefont
  {Zwierlein}, \citenamefont {Schirotzek}, \citenamefont {Schunck},\ and\
  \citenamefont {Ketterle}}]{cold_atom_1}%
  \BibitemOpen
  \bibfield  {author} {\bibinfo {author} {\bibfnamefont {M.~W.}\ \bibnamefont
  {Zwierlein}}, \bibinfo {author} {\bibfnamefont {A.}~\bibnamefont
  {Schirotzek}}, \bibinfo {author} {\bibfnamefont {C.~H.}\ \bibnamefont
  {Schunck}},\ and\ \bibinfo {author} {\bibfnamefont {W.}~\bibnamefont
  {Ketterle}},\ }\bibfield  {title} {\bibinfo {title} {Fermionic superfluidity
  with imbalanced spin populations},\ }\href
  {https://doi.org/10.1126/science.1122318} {\bibfield  {journal} {\bibinfo
  {journal} {Science}\ }\textbf {\bibinfo {volume} {311}},\ \bibinfo {pages}
  {492} (\bibinfo {year} {2006})}\BibitemShut {NoStop}%
\bibitem [{\citenamefont {Partridge}\ \emph {et~al.}(2006)\citenamefont
  {Partridge}, \citenamefont {Li}, \citenamefont {Kamar}, \citenamefont
  {an~Liao},\ and\ \citenamefont {Hulet}}]{cold_atom_2}%
  \BibitemOpen
  \bibfield  {author} {\bibinfo {author} {\bibfnamefont {G.~B.}\ \bibnamefont
  {Partridge}}, \bibinfo {author} {\bibfnamefont {W.}~\bibnamefont {Li}},
  \bibinfo {author} {\bibfnamefont {R.~I.}\ \bibnamefont {Kamar}}, \bibinfo
  {author} {\bibfnamefont {Y.}~\bibnamefont {an~Liao}},\ and\ \bibinfo {author}
  {\bibfnamefont {R.~G.}\ \bibnamefont {Hulet}},\ }\bibfield  {title} {\bibinfo
  {title} {Pairing and phase separation in a polarized fermi gas},\ }\href
  {https://doi.org/10.1126/science.1122876} {\bibfield  {journal} {\bibinfo
  {journal} {Science}\ }\textbf {\bibinfo {volume} {311}},\ \bibinfo {pages}
  {503} (\bibinfo {year} {2006})}\BibitemShut {NoStop}%
\bibitem [{\citenamefont {Radzihovsky}\ and\ \citenamefont
  {Sheehy}(2010)}]{cold_atom_3}%
  \BibitemOpen
  \bibfield  {author} {\bibinfo {author} {\bibfnamefont {L.}~\bibnamefont
  {Radzihovsky}}\ and\ \bibinfo {author} {\bibfnamefont {D.~E.}\ \bibnamefont
  {Sheehy}},\ }\bibfield  {title} {\bibinfo {title} {Imbalanced
  feshbach-resonant fermi gases},\ }\href
  {https://doi.org/10.1088/0034-4885/73/7/076501} {\bibfield  {journal}
  {\bibinfo  {journal} {Rep. Prog. Phys.}\ }\textbf {\bibinfo {volume} {73}},\
  \bibinfo {pages} {076501} (\bibinfo {year} {2010})}\BibitemShut {NoStop}%
\bibitem [{\citenamefont {Lee}(2014{\natexlab{a}})}]{mother_state1}%
  \BibitemOpen
  \bibfield  {author} {\bibinfo {author} {\bibfnamefont {P.~A.}\ \bibnamefont
  {Lee}},\ }\bibfield  {title} {\bibinfo {title} {Amperean pairing and the
  pseudogap phase of cuprate superconductors},\ }\href
  {https://doi.org/10.1103/PhysRevX.4.031017} {\bibfield  {journal} {\bibinfo
  {journal} {Phys. Rev. X}\ }\textbf {\bibinfo {volume} {4}},\ \bibinfo {pages}
  {031017} (\bibinfo {year} {2014}{\natexlab{a}})}\BibitemShut {NoStop}%
\bibitem [{\citenamefont {Fradkin}\ \emph {et~al.}(2015)\citenamefont
  {Fradkin}, \citenamefont {Kivelson},\ and\ \citenamefont
  {Tranquada}}]{PDW_review_2015}%
  \BibitemOpen
  \bibfield  {author} {\bibinfo {author} {\bibfnamefont {E.}~\bibnamefont
  {Fradkin}}, \bibinfo {author} {\bibfnamefont {S.~A.}\ \bibnamefont
  {Kivelson}},\ and\ \bibinfo {author} {\bibfnamefont {J.~M.}\ \bibnamefont
  {Tranquada}},\ }\bibfield  {title} {\bibinfo {title} {Colloquium: Theory of
  intertwined orders in high temperature superconductors},\ }\href
  {https://doi.org/10.1103/RevModPhys.87.457} {\bibfield  {journal} {\bibinfo
  {journal} {Rev. Mod. Phys.}\ }\textbf {\bibinfo {volume} {87}},\ \bibinfo
  {pages} {457} (\bibinfo {year} {2015})}\BibitemShut {NoStop}%
\bibitem [{\citenamefont {Agterberg}\ \emph {et~al.}(2020)\citenamefont
  {Agterberg}, \citenamefont {Davis}, \citenamefont {Edkins}, \citenamefont
  {Fradkin}, \citenamefont {Van~Harlingen}, \citenamefont {Kivelson},
  \citenamefont {Lee}, \citenamefont {Radzihovsky}, \citenamefont {Tranquada},\
  and\ \citenamefont {Wang}}]{PDW_review_2020}%
  \BibitemOpen
  \bibfield  {author} {\bibinfo {author} {\bibfnamefont {D.~F.}\ \bibnamefont
  {Agterberg}}, \bibinfo {author} {\bibfnamefont {J.~S.}\ \bibnamefont
  {Davis}}, \bibinfo {author} {\bibfnamefont {S.~D.}\ \bibnamefont {Edkins}},
  \bibinfo {author} {\bibfnamefont {E.}~\bibnamefont {Fradkin}}, \bibinfo
  {author} {\bibfnamefont {D.~J.}\ \bibnamefont {Van~Harlingen}}, \bibinfo
  {author} {\bibfnamefont {S.~A.}\ \bibnamefont {Kivelson}}, \bibinfo {author}
  {\bibfnamefont {P.~A.}\ \bibnamefont {Lee}}, \bibinfo {author} {\bibfnamefont
  {L.}~\bibnamefont {Radzihovsky}}, \bibinfo {author} {\bibfnamefont {J.~M.}\
  \bibnamefont {Tranquada}},\ and\ \bibinfo {author} {\bibfnamefont
  {Y.}~\bibnamefont {Wang}},\ }\bibfield  {title} {\bibinfo {title} {The
  physics of pair-density waves: Cuprate superconductors and beyond},\ }\href
  {https://doi.org/https://doi.org/10.1146/annurev-conmatphys-031119-050711}
  {\bibfield  {journal} {\bibinfo  {journal} {Annu. Rev. Condens. Matter
  Phys.}\ }\textbf {\bibinfo {volume} {11}},\ \bibinfo {pages} {231} (\bibinfo
  {year} {2020})}\BibitemShut {NoStop}%
\bibitem [{\citenamefont {Li}\ \emph {et~al.}(2007)\citenamefont {Li},
  \citenamefont {H\"ucker}, \citenamefont {Gu}, \citenamefont {Tsvelik},\ and\
  \citenamefont {Tranquada}}]{cuprate1}%
  \BibitemOpen
  \bibfield  {author} {\bibinfo {author} {\bibfnamefont {Q.}~\bibnamefont
  {Li}}, \bibinfo {author} {\bibfnamefont {M.}~\bibnamefont {H\"ucker}},
  \bibinfo {author} {\bibfnamefont {G.~D.}\ \bibnamefont {Gu}}, \bibinfo
  {author} {\bibfnamefont {A.~M.}\ \bibnamefont {Tsvelik}},\ and\ \bibinfo
  {author} {\bibfnamefont {J.~M.}\ \bibnamefont {Tranquada}},\ }\bibfield
  {title} {\bibinfo {title} {Two-dimensional superconducting fluctuations in
  stripe-ordered
  {${\mathrm{La}}_{1.875}{\mathrm{Ba}}_{0.125}{\mathrm{CuO}}_{4}$}},\ }\href
  {https://doi.org/10.1103/PhysRevLett.99.067001} {\bibfield  {journal}
  {\bibinfo  {journal} {Phys. Rev. Lett.}\ }\textbf {\bibinfo {volume} {99}},\
  \bibinfo {pages} {067001} (\bibinfo {year} {2007})}\BibitemShut {NoStop}%
\bibitem [{\citenamefont {Berg}\ \emph
  {et~al.}(2007{\natexlab{a}})\citenamefont {Berg}, \citenamefont {Fradkin},
  \citenamefont {Kim}, \citenamefont {Kivelson}, \citenamefont {Oganesyan},
  \citenamefont {Tranquada},\ and\ \citenamefont {Zhang}}]{cuprate2}%
  \BibitemOpen
  \bibfield  {author} {\bibinfo {author} {\bibfnamefont {E.}~\bibnamefont
  {Berg}}, \bibinfo {author} {\bibfnamefont {E.}~\bibnamefont {Fradkin}},
  \bibinfo {author} {\bibfnamefont {E.-A.}\ \bibnamefont {Kim}}, \bibinfo
  {author} {\bibfnamefont {S.~A.}\ \bibnamefont {Kivelson}}, \bibinfo {author}
  {\bibfnamefont {V.}~\bibnamefont {Oganesyan}}, \bibinfo {author}
  {\bibfnamefont {J.~M.}\ \bibnamefont {Tranquada}},\ and\ \bibinfo {author}
  {\bibfnamefont {S.~C.}\ \bibnamefont {Zhang}},\ }\bibfield  {title} {\bibinfo
  {title} {Dynamical layer decoupling in a stripe-ordered high-${T}_{c}$
  superconductor},\ }\href {https://doi.org/10.1103/PhysRevLett.99.127003}
  {\bibfield  {journal} {\bibinfo  {journal} {Phys. Rev. Lett.}\ }\textbf
  {\bibinfo {volume} {99}},\ \bibinfo {pages} {127003} (\bibinfo {year}
  {2007}{\natexlab{a}})}\BibitemShut {NoStop}%
\bibitem [{\citenamefont {Agterberg}\ and\ \citenamefont
  {Tsunetsugu}(2008)}]{cuprate3}%
  \BibitemOpen
  \bibfield  {author} {\bibinfo {author} {\bibfnamefont {D.~F.}\ \bibnamefont
  {Agterberg}}\ and\ \bibinfo {author} {\bibfnamefont {H.}~\bibnamefont
  {Tsunetsugu}},\ }\bibfield  {title} {\bibinfo {title} {Dislocations and
  vortices in pair-density-wave superconductors},\ }\href
  {https://doi.org/10.1038/nphys999} {\bibfield  {journal} {\bibinfo  {journal}
  {Nat. Phys.}\ }\textbf {\bibinfo {volume} {4}},\ \bibinfo {pages} {639}
  (\bibinfo {year} {2008})}\BibitemShut {NoStop}%
\bibitem [{\citenamefont {Berg}\ \emph
  {et~al.}(2009{\natexlab{a}})\citenamefont {Berg}, \citenamefont {Fradkin},\
  and\ \citenamefont {Kivelson}}]{cuprate4}%
  \BibitemOpen
  \bibfield  {author} {\bibinfo {author} {\bibfnamefont {E.}~\bibnamefont
  {Berg}}, \bibinfo {author} {\bibfnamefont {E.}~\bibnamefont {Fradkin}},\ and\
  \bibinfo {author} {\bibfnamefont {S.~A.}\ \bibnamefont {Kivelson}},\
  }\bibfield  {title} {\bibinfo {title} {Charge-4e superconductivity from
  pair-density-wave order in certain high-temperature superconductors},\ }\href
  {https://doi.org/10.1038/nphys1389} {\bibfield  {journal} {\bibinfo
  {journal} {Nat. Phys.}\ }\textbf {\bibinfo {volume} {5}},\ \bibinfo {pages}
  {830} (\bibinfo {year} {2009}{\natexlab{a}})}\BibitemShut {NoStop}%
\bibitem [{\citenamefont {Hamidian}\ \emph {et~al.}(2016)\citenamefont
  {Hamidian}, \citenamefont {Edkins}, \citenamefont {Joo}, \citenamefont
  {Kostin}, \citenamefont {Eisaki}, \citenamefont {Uchida}, \citenamefont
  {Lawler}, \citenamefont {Kim}, \citenamefont {Mackenzie}, \citenamefont
  {Fujita}, \citenamefont {Lee},\ and\ \citenamefont {Davis}}]{cuprate5}%
  \BibitemOpen
  \bibfield  {author} {\bibinfo {author} {\bibfnamefont {M.~H.}\ \bibnamefont
  {Hamidian}}, \bibinfo {author} {\bibfnamefont {S.~D.}\ \bibnamefont
  {Edkins}}, \bibinfo {author} {\bibfnamefont {S.~H.}\ \bibnamefont {Joo}},
  \bibinfo {author} {\bibfnamefont {A.}~\bibnamefont {Kostin}}, \bibinfo
  {author} {\bibfnamefont {H.}~\bibnamefont {Eisaki}}, \bibinfo {author}
  {\bibfnamefont {S.}~\bibnamefont {Uchida}}, \bibinfo {author} {\bibfnamefont
  {M.~J.}\ \bibnamefont {Lawler}}, \bibinfo {author} {\bibfnamefont {E.-A.}\
  \bibnamefont {Kim}}, \bibinfo {author} {\bibfnamefont {A.~P.}\ \bibnamefont
  {Mackenzie}}, \bibinfo {author} {\bibfnamefont {K.}~\bibnamefont {Fujita}},
  \bibinfo {author} {\bibfnamefont {J.}~\bibnamefont {Lee}},\ and\ \bibinfo
  {author} {\bibfnamefont {J.~C.~S.}\ \bibnamefont {Davis}},\ }\bibfield
  {title} {\bibinfo {title} {Detection of a {Cooper}-pair density wave in
  {Bi$_2$Sr$_2$CaCu$_2$O$_{8+x}$}},\ }\href
  {https://doi.org/10.1038/nature17411} {\bibfield  {journal} {\bibinfo
  {journal} {Nature}\ }\textbf {\bibinfo {volume} {532}},\ \bibinfo {pages}
  {343} (\bibinfo {year} {2016})}\BibitemShut {NoStop}%
\bibitem [{\citenamefont {Ruan}\ \emph {et~al.}(2018)\citenamefont {Ruan},
  \citenamefont {Li}, \citenamefont {Hu}, \citenamefont {Hao}, \citenamefont
  {Li}, \citenamefont {Cai}, \citenamefont {Zhou}, \citenamefont {Lee},\ and\
  \citenamefont {Wang}}]{cuprate6}%
  \BibitemOpen
  \bibfield  {author} {\bibinfo {author} {\bibfnamefont {W.}~\bibnamefont
  {Ruan}}, \bibinfo {author} {\bibfnamefont {X.}~\bibnamefont {Li}}, \bibinfo
  {author} {\bibfnamefont {C.}~\bibnamefont {Hu}}, \bibinfo {author}
  {\bibfnamefont {Z.}~\bibnamefont {Hao}}, \bibinfo {author} {\bibfnamefont
  {H.}~\bibnamefont {Li}}, \bibinfo {author} {\bibfnamefont {P.}~\bibnamefont
  {Cai}}, \bibinfo {author} {\bibfnamefont {X.}~\bibnamefont {Zhou}}, \bibinfo
  {author} {\bibfnamefont {D.-H.}\ \bibnamefont {Lee}},\ and\ \bibinfo {author}
  {\bibfnamefont {Y.}~\bibnamefont {Wang}},\ }\bibfield  {title} {\bibinfo
  {title} {Visualization of the periodic modulation of {Cooper} pairing in a
  cuprate superconductor},\ }\href {https://doi.org/10.1038/s41567-018-0276-8}
  {\bibfield  {journal} {\bibinfo  {journal} {Nat. Phys.}\ }\textbf {\bibinfo
  {volume} {14}},\ \bibinfo {pages} {1178} (\bibinfo {year}
  {2018})}\BibitemShut {NoStop}%
\bibitem [{\citenamefont {Edkins}\ \emph {et~al.}(2019)\citenamefont {Edkins},
  \citenamefont {Kostin}, \citenamefont {Fujita}, \citenamefont {Mackenzie},
  \citenamefont {Eisaki}, \citenamefont {Uchida}, \citenamefont {Sachdev},
  \citenamefont {Lawler}, \citenamefont {Kim}, \citenamefont {Davis},\ and\
  \citenamefont {Hamidian}}]{cuprate7}%
  \BibitemOpen
  \bibfield  {author} {\bibinfo {author} {\bibfnamefont {S.~D.}\ \bibnamefont
  {Edkins}}, \bibinfo {author} {\bibfnamefont {A.}~\bibnamefont {Kostin}},
  \bibinfo {author} {\bibfnamefont {K.}~\bibnamefont {Fujita}}, \bibinfo
  {author} {\bibfnamefont {A.~P.}\ \bibnamefont {Mackenzie}}, \bibinfo {author}
  {\bibfnamefont {H.}~\bibnamefont {Eisaki}}, \bibinfo {author} {\bibfnamefont
  {S.}~\bibnamefont {Uchida}}, \bibinfo {author} {\bibfnamefont
  {S.}~\bibnamefont {Sachdev}}, \bibinfo {author} {\bibfnamefont {M.~J.}\
  \bibnamefont {Lawler}}, \bibinfo {author} {\bibfnamefont {E.-A.}\
  \bibnamefont {Kim}}, \bibinfo {author} {\bibfnamefont {J.~C.~S.}\
  \bibnamefont {Davis}},\ and\ \bibinfo {author} {\bibfnamefont {M.~H.}\
  \bibnamefont {Hamidian}},\ }\bibfield  {title} {\bibinfo {title} {Magnetic
  field–induced pair density wave state in the cuprate vortex halo},\ }\href
  {https://doi.org/10.1126/science.aat1773} {\bibfield  {journal} {\bibinfo
  {journal} {Science}\ }\textbf {\bibinfo {volume} {364}},\ \bibinfo {pages}
  {976} (\bibinfo {year} {2019})}\BibitemShut {NoStop}%
\bibitem [{\citenamefont {Shi}\ \emph {et~al.}(2020)\citenamefont {Shi},
  \citenamefont {Baity}, \citenamefont {Terzic}, \citenamefont {Sasagawa},\
  and\ \citenamefont {Popović}}]{cuprate8}%
  \BibitemOpen
  \bibfield  {author} {\bibinfo {author} {\bibfnamefont {Z.}~\bibnamefont
  {Shi}}, \bibinfo {author} {\bibfnamefont {P.~G.}\ \bibnamefont {Baity}},
  \bibinfo {author} {\bibfnamefont {J.}~\bibnamefont {Terzic}}, \bibinfo
  {author} {\bibfnamefont {T.}~\bibnamefont {Sasagawa}},\ and\ \bibinfo
  {author} {\bibfnamefont {D.}~\bibnamefont {Popović}},\ }\bibfield  {title}
  {\bibinfo {title} {Pair density wave at high magnetic fields in cuprates with
  charge and spin orders},\ }\href {https://doi.org/10.1038/s41467-020-17138-z}
  {\bibfield  {journal} {\bibinfo  {journal} {Nat. Commun.}\ }\textbf {\bibinfo
  {volume} {11}},\ \bibinfo {pages} {3323} (\bibinfo {year}
  {2020})}\BibitemShut {NoStop}%
\bibitem [{\citenamefont {Du}\ \emph {et~al.}(2020)\citenamefont {Du},
  \citenamefont {Li}, \citenamefont {Joo}, \citenamefont {Donoway},
  \citenamefont {Lee}, \citenamefont {Davis}, \citenamefont {Gu}, \citenamefont
  {Johnson},\ and\ \citenamefont {Fujita}}]{cuprate9}%
  \BibitemOpen
  \bibfield  {author} {\bibinfo {author} {\bibfnamefont {Z.}~\bibnamefont
  {Du}}, \bibinfo {author} {\bibfnamefont {H.}~\bibnamefont {Li}}, \bibinfo
  {author} {\bibfnamefont {S.~H.}\ \bibnamefont {Joo}}, \bibinfo {author}
  {\bibfnamefont {E.~P.}\ \bibnamefont {Donoway}}, \bibinfo {author}
  {\bibfnamefont {J.}~\bibnamefont {Lee}}, \bibinfo {author} {\bibfnamefont
  {J.~C.~S.}\ \bibnamefont {Davis}}, \bibinfo {author} {\bibfnamefont
  {G.}~\bibnamefont {Gu}}, \bibinfo {author} {\bibfnamefont {P.~D.}\
  \bibnamefont {Johnson}},\ and\ \bibinfo {author} {\bibfnamefont
  {K.}~\bibnamefont {Fujita}},\ }\bibfield  {title} {\bibinfo {title} {Imaging
  the energy gap modulations of the cuprate pair-density-wave state},\ }\href
  {https://doi.org/10.1038/s41586-020-2143-x} {\bibfield  {journal} {\bibinfo
  {journal} {Nature}\ }\textbf {\bibinfo {volume} {580}},\ \bibinfo {pages}
  {65} (\bibinfo {year} {2020})}\BibitemShut {NoStop}%
\bibitem [{\citenamefont {Li}\ \emph {et~al.}(2021)\citenamefont {Li},
  \citenamefont {Zou}, \citenamefont {Ding}, \citenamefont {Yan}, \citenamefont
  {Ye}, \citenamefont {Li}, \citenamefont {Hao}, \citenamefont {Zhao},
  \citenamefont {Zhou},\ and\ \citenamefont {Wang}}]{cuprate10}%
  \BibitemOpen
  \bibfield  {author} {\bibinfo {author} {\bibfnamefont {X.}~\bibnamefont
  {Li}}, \bibinfo {author} {\bibfnamefont {C.}~\bibnamefont {Zou}}, \bibinfo
  {author} {\bibfnamefont {Y.}~\bibnamefont {Ding}}, \bibinfo {author}
  {\bibfnamefont {H.}~\bibnamefont {Yan}}, \bibinfo {author} {\bibfnamefont
  {S.}~\bibnamefont {Ye}}, \bibinfo {author} {\bibfnamefont {H.}~\bibnamefont
  {Li}}, \bibinfo {author} {\bibfnamefont {Z.}~\bibnamefont {Hao}}, \bibinfo
  {author} {\bibfnamefont {L.}~\bibnamefont {Zhao}}, \bibinfo {author}
  {\bibfnamefont {X.}~\bibnamefont {Zhou}},\ and\ \bibinfo {author}
  {\bibfnamefont {Y.}~\bibnamefont {Wang}},\ }\bibfield  {title} {\bibinfo
  {title} {Evolution of charge and pair density modulations in overdoped
  ${\mathrm{bi}}_{2}{\mathrm{sr}}_{2}{\mathrm{cuo}}_{6+\ensuremath{\delta}}$},\
  }\href {https://doi.org/10.1103/PhysRevX.11.011007} {\bibfield  {journal}
  {\bibinfo  {journal} {Phys. Rev. X}\ }\textbf {\bibinfo {volume} {11}},\
  \bibinfo {pages} {011007} (\bibinfo {year} {2021})}\BibitemShut {NoStop}%
\bibitem [{\citenamefont {Wang}\ \emph {et~al.}(2021)\citenamefont {Wang},
  \citenamefont {Choubey}, \citenamefont {Chong}, \citenamefont {Chen},
  \citenamefont {Ren}, \citenamefont {Eisaki}, \citenamefont {Uchida},
  \citenamefont {Hirschfeld},\ and\ \citenamefont {Davis}}]{cuprate11}%
  \BibitemOpen
  \bibfield  {author} {\bibinfo {author} {\bibfnamefont {S.}~\bibnamefont
  {Wang}}, \bibinfo {author} {\bibfnamefont {P.}~\bibnamefont {Choubey}},
  \bibinfo {author} {\bibfnamefont {Y.~X.}\ \bibnamefont {Chong}}, \bibinfo
  {author} {\bibfnamefont {W.}~\bibnamefont {Chen}}, \bibinfo {author}
  {\bibfnamefont {W.}~\bibnamefont {Ren}}, \bibinfo {author} {\bibfnamefont
  {H.}~\bibnamefont {Eisaki}}, \bibinfo {author} {\bibfnamefont
  {S.}~\bibnamefont {Uchida}}, \bibinfo {author} {\bibfnamefont {P.~J.}\
  \bibnamefont {Hirschfeld}},\ and\ \bibinfo {author} {\bibfnamefont
  {J.~C.~S.}\ \bibnamefont {Davis}},\ }\bibfield  {title} {\bibinfo {title}
  {Scattering interference signature of a pair density wave state in the
  cuprate pseudogap phase},\ }\href
  {https://doi.org/10.1038/s41467-021-26028-x} {\bibfield  {journal} {\bibinfo
  {journal} {Nat. Commun.}\ }\textbf {\bibinfo {volume} {12}},\ \bibinfo
  {pages} {6087} (\bibinfo {year} {2021})}\BibitemShut {NoStop}%
\bibitem [{\citenamefont {Cho}\ \emph {et~al.}(2017)\citenamefont {Cho},
  \citenamefont {Yang}, \citenamefont {Yuan}, \citenamefont {Shen},
  \citenamefont {Wolf},\ and\ \citenamefont {Lortz}}]{Fe1}%
  \BibitemOpen
  \bibfield  {author} {\bibinfo {author} {\bibfnamefont {C.-W.}\ \bibnamefont
  {Cho}}, \bibinfo {author} {\bibfnamefont {J.~H.}\ \bibnamefont {Yang}},
  \bibinfo {author} {\bibfnamefont {N.~F.~Q.}\ \bibnamefont {Yuan}}, \bibinfo
  {author} {\bibfnamefont {J.}~\bibnamefont {Shen}}, \bibinfo {author}
  {\bibfnamefont {T.}~\bibnamefont {Wolf}},\ and\ \bibinfo {author}
  {\bibfnamefont {R.}~\bibnamefont {Lortz}},\ }\bibfield  {title} {\bibinfo
  {title} {Thermodynamic evidence for the fulde-ferrell-larkin-ovchinnikov
  state in the {${\mathrm{KFe}}_{2}{\mathrm{As}}_{2}$} superconductor},\ }\href
  {https://doi.org/10.1103/PhysRevLett.119.217002} {\bibfield  {journal}
  {\bibinfo  {journal} {Phys. Rev. Lett.}\ }\textbf {\bibinfo {volume} {119}},\
  \bibinfo {pages} {217002} (\bibinfo {year} {2017})}\BibitemShut {NoStop}%
\bibitem [{\citenamefont {Liu}\ \emph {et~al.}(2023)\citenamefont {Liu},
  \citenamefont {Wei}, \citenamefont {He}, \citenamefont {Zhang}, \citenamefont
  {Wang},\ and\ \citenamefont {Wang}}]{Fe2}%
  \BibitemOpen
  \bibfield  {author} {\bibinfo {author} {\bibfnamefont {Y.}~\bibnamefont
  {Liu}}, \bibinfo {author} {\bibfnamefont {T.}~\bibnamefont {Wei}}, \bibinfo
  {author} {\bibfnamefont {G.}~\bibnamefont {He}}, \bibinfo {author}
  {\bibfnamefont {Y.}~\bibnamefont {Zhang}}, \bibinfo {author} {\bibfnamefont
  {Z.}~\bibnamefont {Wang}},\ and\ \bibinfo {author} {\bibfnamefont
  {J.}~\bibnamefont {Wang}},\ }\bibfield  {title} {\bibinfo {title} {Pair
  density wave state in a monolayer high-{Tc} iron-based superconductor},\
  }\href {https://doi.org/10.1038/s41586-023-06072-x} {\bibfield  {journal}
  {\bibinfo  {journal} {Nature}\ }\textbf {\bibinfo {volume} {618}},\ \bibinfo
  {pages} {934} (\bibinfo {year} {2023})}\BibitemShut {NoStop}%
\bibitem [{\citenamefont {Zhao}\ \emph {et~al.}(2023)\citenamefont {Zhao},
  \citenamefont {Blackwell}, \citenamefont {Thinel}, \citenamefont {Handa},
  \citenamefont {Ishida}, \citenamefont {Zhu}, \citenamefont {Iyo},
  \citenamefont {Eisaki}, \citenamefont {Pasupathy},\ and\ \citenamefont
  {Fujita}}]{Fe3}%
  \BibitemOpen
  \bibfield  {author} {\bibinfo {author} {\bibfnamefont {H.}~\bibnamefont
  {Zhao}}, \bibinfo {author} {\bibfnamefont {R.}~\bibnamefont {Blackwell}},
  \bibinfo {author} {\bibfnamefont {M.}~\bibnamefont {Thinel}}, \bibinfo
  {author} {\bibfnamefont {T.}~\bibnamefont {Handa}}, \bibinfo {author}
  {\bibfnamefont {S.}~\bibnamefont {Ishida}}, \bibinfo {author} {\bibfnamefont
  {X.}~\bibnamefont {Zhu}}, \bibinfo {author} {\bibfnamefont {A.}~\bibnamefont
  {Iyo}}, \bibinfo {author} {\bibfnamefont {H.}~\bibnamefont {Eisaki}},
  \bibinfo {author} {\bibfnamefont {A.~N.}\ \bibnamefont {Pasupathy}},\ and\
  \bibinfo {author} {\bibfnamefont {K.}~\bibnamefont {Fujita}},\ }\bibfield
  {title} {\bibinfo {title} {Smectic pair-density-wave order in
  {EuRbFe$_4$As$_4$}},\ }\href {https://doi.org/10.1038/s41586-023-06103-7}
  {\bibfield  {journal} {\bibinfo  {journal} {Nature}\ }\textbf {\bibinfo
  {volume} {618}},\ \bibinfo {pages} {940} (\bibinfo {year}
  {2023})}\BibitemShut {NoStop}%
\bibitem [{\citenamefont {Gu}\ \emph {et~al.}(2023)\citenamefont {Gu},
  \citenamefont {Carroll}, \citenamefont {Wang}, \citenamefont {Ran},
  \citenamefont {Broyles}, \citenamefont {Siddiquee}, \citenamefont {Butch},
  \citenamefont {Saha}, \citenamefont {Paglione}, \citenamefont {Davis},\ and\
  \citenamefont {Liu}}]{UTe2_1}%
  \BibitemOpen
  \bibfield  {author} {\bibinfo {author} {\bibfnamefont {Q.}~\bibnamefont
  {Gu}}, \bibinfo {author} {\bibfnamefont {J.~P.}\ \bibnamefont {Carroll}},
  \bibinfo {author} {\bibfnamefont {S.}~\bibnamefont {Wang}}, \bibinfo {author}
  {\bibfnamefont {S.}~\bibnamefont {Ran}}, \bibinfo {author} {\bibfnamefont
  {C.}~\bibnamefont {Broyles}}, \bibinfo {author} {\bibfnamefont
  {H.}~\bibnamefont {Siddiquee}}, \bibinfo {author} {\bibfnamefont {N.~P.}\
  \bibnamefont {Butch}}, \bibinfo {author} {\bibfnamefont {S.~R.}\ \bibnamefont
  {Saha}}, \bibinfo {author} {\bibfnamefont {J.}~\bibnamefont {Paglione}},
  \bibinfo {author} {\bibfnamefont {J.~C.~S.}\ \bibnamefont {Davis}},\ and\
  \bibinfo {author} {\bibfnamefont {X.}~\bibnamefont {Liu}},\ }\bibfield
  {title} {\bibinfo {title} {Detection of a pair density wave state in
  {UTe$_2$}},\ }\href {https://doi.org/10.1038/s41586-023-05919-7} {\bibfield
  {journal} {\bibinfo  {journal} {Nature}\ }\textbf {\bibinfo {volume} {618}},\
  \bibinfo {pages} {921} (\bibinfo {year} {2023})}\BibitemShut {NoStop}%
\bibitem [{\citenamefont {Aishwarya}\ \emph {et~al.}(2023)\citenamefont
  {Aishwarya}, \citenamefont {May-Mann}, \citenamefont {Raghavan},
  \citenamefont {Nie}, \citenamefont {Romanelli}, \citenamefont {Ran},
  \citenamefont {Saha}, \citenamefont {Paglione}, \citenamefont {Butch},
  \citenamefont {Fradkin},\ and\ \citenamefont {Madhavan}}]{UTe2_2}%
  \BibitemOpen
  \bibfield  {author} {\bibinfo {author} {\bibfnamefont {A.}~\bibnamefont
  {Aishwarya}}, \bibinfo {author} {\bibfnamefont {J.}~\bibnamefont {May-Mann}},
  \bibinfo {author} {\bibfnamefont {A.}~\bibnamefont {Raghavan}}, \bibinfo
  {author} {\bibfnamefont {L.}~\bibnamefont {Nie}}, \bibinfo {author}
  {\bibfnamefont {M.}~\bibnamefont {Romanelli}}, \bibinfo {author}
  {\bibfnamefont {S.}~\bibnamefont {Ran}}, \bibinfo {author} {\bibfnamefont
  {S.~R.}\ \bibnamefont {Saha}}, \bibinfo {author} {\bibfnamefont
  {J.}~\bibnamefont {Paglione}}, \bibinfo {author} {\bibfnamefont {N.~P.}\
  \bibnamefont {Butch}}, \bibinfo {author} {\bibfnamefont {E.}~\bibnamefont
  {Fradkin}},\ and\ \bibinfo {author} {\bibfnamefont {V.}~\bibnamefont
  {Madhavan}},\ }\bibfield  {title} {\bibinfo {title} {Magnetic-field-sensitive
  charge density waves in the superconductor {UTe$_2$}},\ }\href
  {https://doi.org/10.1038/s41586-023-06005-8} {\bibfield  {journal} {\bibinfo
  {journal} {Nature}\ }\textbf {\bibinfo {volume} {618}},\ \bibinfo {pages}
  {928} (\bibinfo {year} {2023})}\BibitemShut {NoStop}%
\bibitem [{\citenamefont {Chen}\ \emph {et~al.}(2021)\citenamefont {Chen},
  \citenamefont {Yang}, \citenamefont {Hu}, \citenamefont {Zhao}, \citenamefont
  {Yuan}, \citenamefont {Xing}, \citenamefont {Qian}, \citenamefont {Huang},
  \citenamefont {Li}, \citenamefont {Ye}, \citenamefont {Ma}, \citenamefont
  {Ni}, \citenamefont {Zhang}, \citenamefont {Yin}, \citenamefont {Gong},
  \citenamefont {Tu}, \citenamefont {Lei}, \citenamefont {Tan}, \citenamefont
  {Zhou}, \citenamefont {Shen}, \citenamefont {Dong}, \citenamefont {Yan},
  \citenamefont {Wang},\ and\ \citenamefont {Gao}}]{Chen-2021-roton}%
  \BibitemOpen
  \bibfield  {author} {\bibinfo {author} {\bibfnamefont {H.}~\bibnamefont
  {Chen}}, \bibinfo {author} {\bibfnamefont {H.}~\bibnamefont {Yang}}, \bibinfo
  {author} {\bibfnamefont {B.}~\bibnamefont {Hu}}, \bibinfo {author}
  {\bibfnamefont {Z.}~\bibnamefont {Zhao}}, \bibinfo {author} {\bibfnamefont
  {J.}~\bibnamefont {Yuan}}, \bibinfo {author} {\bibfnamefont {Y.}~\bibnamefont
  {Xing}}, \bibinfo {author} {\bibfnamefont {G.}~\bibnamefont {Qian}}, \bibinfo
  {author} {\bibfnamefont {Z.}~\bibnamefont {Huang}}, \bibinfo {author}
  {\bibfnamefont {G.}~\bibnamefont {Li}}, \bibinfo {author} {\bibfnamefont
  {Y.}~\bibnamefont {Ye}}, \bibinfo {author} {\bibfnamefont {S.}~\bibnamefont
  {Ma}}, \bibinfo {author} {\bibfnamefont {S.}~\bibnamefont {Ni}}, \bibinfo
  {author} {\bibfnamefont {H.}~\bibnamefont {Zhang}}, \bibinfo {author}
  {\bibfnamefont {Q.}~\bibnamefont {Yin}}, \bibinfo {author} {\bibfnamefont
  {C.}~\bibnamefont {Gong}}, \bibinfo {author} {\bibfnamefont {Z.}~\bibnamefont
  {Tu}}, \bibinfo {author} {\bibfnamefont {H.}~\bibnamefont {Lei}}, \bibinfo
  {author} {\bibfnamefont {H.}~\bibnamefont {Tan}}, \bibinfo {author}
  {\bibfnamefont {S.}~\bibnamefont {Zhou}}, \bibinfo {author} {\bibfnamefont
  {C.}~\bibnamefont {Shen}}, \bibinfo {author} {\bibfnamefont {X.}~\bibnamefont
  {Dong}}, \bibinfo {author} {\bibfnamefont {B.}~\bibnamefont {Yan}}, \bibinfo
  {author} {\bibfnamefont {Z.}~\bibnamefont {Wang}},\ and\ \bibinfo {author}
  {\bibfnamefont {H.-J.}\ \bibnamefont {Gao}},\ }\bibfield  {title} {\bibinfo
  {title} {Roton pair density wave in a strong-coupling kagome
  superconductor},\ }\href {https://doi.org/10.1038/s41586-021-03983-5}
  {\bibfield  {journal} {\bibinfo  {journal} {Nature}\ }\textbf {\bibinfo
  {volume} {599}},\ \bibinfo {pages} {222} (\bibinfo {year}
  {2021})}\BibitemShut {NoStop}%
\bibitem [{\citenamefont {Agosta}(2018)}]{other_materials1}%
  \BibitemOpen
  \bibfield  {author} {\bibinfo {author} {\bibfnamefont {C.~C.}\ \bibnamefont
  {Agosta}},\ }\bibfield  {title} {\bibinfo {title} {Inhomogeneous
  superconductivity in organic and related superconductors},\ }\href
  {https://doi.org/10.3390/cryst8070285} {\bibfield  {journal} {\bibinfo
  {journal} {Crystals}\ }\textbf {\bibinfo {volume} {8}},\ \bibinfo {pages}
  {285} (\bibinfo {year} {2018})}\BibitemShut {NoStop}%
\bibitem [{\citenamefont {Liu}\ \emph {et~al.}(2021)\citenamefont {Liu},
  \citenamefont {Chong}, \citenamefont {Sharma},\ and\ \citenamefont
  {Davis}}]{other_materials2}%
  \BibitemOpen
  \bibfield  {author} {\bibinfo {author} {\bibfnamefont {X.}~\bibnamefont
  {Liu}}, \bibinfo {author} {\bibfnamefont {Y.~X.}\ \bibnamefont {Chong}},
  \bibinfo {author} {\bibfnamefont {R.}~\bibnamefont {Sharma}},\ and\ \bibinfo
  {author} {\bibfnamefont {J.~C.~S.}\ \bibnamefont {Davis}},\ }\bibfield
  {title} {\bibinfo {title} {Discovery of a cooper-pair density wave state in a
  transition-metal dichalcogenide},\ }\href
  {https://doi.org/10.1126/science.abd4607} {\bibfield  {journal} {\bibinfo
  {journal} {Science}\ }\textbf {\bibinfo {volume} {372}},\ \bibinfo {pages}
  {1447} (\bibinfo {year} {2021})}\BibitemShut {NoStop}%
\bibitem [{\citenamefont {Berg}\ \emph
  {et~al.}(2007{\natexlab{b}})\citenamefont {Berg}, \citenamefont {Fradkin},
  \citenamefont {Kim}, \citenamefont {Kivelson}, \citenamefont {Oganesyan},
  \citenamefont {Tranquada},\ and\ \citenamefont {Zhang}}]{phenomenological1}%
  \BibitemOpen
  \bibfield  {author} {\bibinfo {author} {\bibfnamefont {E.}~\bibnamefont
  {Berg}}, \bibinfo {author} {\bibfnamefont {E.}~\bibnamefont {Fradkin}},
  \bibinfo {author} {\bibfnamefont {E.-A.}\ \bibnamefont {Kim}}, \bibinfo
  {author} {\bibfnamefont {S.~A.}\ \bibnamefont {Kivelson}}, \bibinfo {author}
  {\bibfnamefont {V.}~\bibnamefont {Oganesyan}}, \bibinfo {author}
  {\bibfnamefont {J.~M.}\ \bibnamefont {Tranquada}},\ and\ \bibinfo {author}
  {\bibfnamefont {S.~C.}\ \bibnamefont {Zhang}},\ }\bibfield  {title} {\bibinfo
  {title} {Dynamical {Layer} {Decoupling} in a {Stripe}-{Ordered} {High}- {T} c
  {Superconductor}},\ }\href {https://doi.org/10.1103/PhysRevLett.99.127003}
  {\bibfield  {journal} {\bibinfo  {journal} {Phys. Rev. Lett.}\ }\textbf
  {\bibinfo {volume} {99}},\ \bibinfo {pages} {127003} (\bibinfo {year}
  {2007}{\natexlab{b}})}\BibitemShut {NoStop}%
\bibitem [{\citenamefont {Berg}\ \emph
  {et~al.}(2009{\natexlab{b}})\citenamefont {Berg}, \citenamefont {Fradkin},
  \citenamefont {Kivelson},\ and\ \citenamefont
  {Tranquada}}]{phenomenological2}%
  \BibitemOpen
  \bibfield  {author} {\bibinfo {author} {\bibfnamefont {E.}~\bibnamefont
  {Berg}}, \bibinfo {author} {\bibfnamefont {E.}~\bibnamefont {Fradkin}},
  \bibinfo {author} {\bibfnamefont {S.~A.}\ \bibnamefont {Kivelson}},\ and\
  \bibinfo {author} {\bibfnamefont {J.~M.}\ \bibnamefont {Tranquada}},\
  }\bibfield  {title} {\bibinfo {title} {Striped superconductors: how spin,
  charge and superconducting orders intertwine in the cuprates},\ }\href
  {https://doi.org/10.1088/1367-2630/11/11/115004} {\bibfield  {journal}
  {\bibinfo  {journal} {New J. Phys.}\ }\textbf {\bibinfo {volume} {11}},\
  \bibinfo {pages} {115004} (\bibinfo {year} {2009}{\natexlab{b}})}\BibitemShut
  {NoStop}%
\bibitem [{\citenamefont {Lee}(2014{\natexlab{b}})}]{phenomenological3}%
  \BibitemOpen
  \bibfield  {author} {\bibinfo {author} {\bibfnamefont {P.~A.}\ \bibnamefont
  {Lee}},\ }\bibfield  {title} {\bibinfo {title} {Amperean {Pairing} and the
  {Pseudogap} {Phase} of {Cuprate} {Superconductors}},\ }\href
  {https://doi.org/10.1103/PhysRevX.4.031017} {\bibfield  {journal} {\bibinfo
  {journal} {Phys. Rev. X}\ }\textbf {\bibinfo {volume} {4}},\ \bibinfo {pages}
  {031017} (\bibinfo {year} {2014}{\natexlab{b}})}\BibitemShut {NoStop}%
\bibitem [{\citenamefont {Cho}\ \emph {et~al.}(2014)\citenamefont {Cho},
  \citenamefont {Soto-Garrido},\ and\ \citenamefont {Fradkin}}]{topo-pdw}%
  \BibitemOpen
  \bibfield  {author} {\bibinfo {author} {\bibfnamefont {G.~Y.}\ \bibnamefont
  {Cho}}, \bibinfo {author} {\bibfnamefont {R.}~\bibnamefont {Soto-Garrido}},\
  and\ \bibinfo {author} {\bibfnamefont {E.}~\bibnamefont {Fradkin}},\
  }\bibfield  {title} {\bibinfo {title} {Topological pair-density-wave
  superconducting states},\ }\href
  {https://doi.org/10.1103/PhysRevLett.113.256405} {\bibfield  {journal}
  {\bibinfo  {journal} {Phys. Rev. Lett.}\ }\textbf {\bibinfo {volume} {113}},\
  \bibinfo {pages} {256405} (\bibinfo {year} {2014})}\BibitemShut {NoStop}%
\bibitem [{\citenamefont {Wang}\ \emph {et~al.}(2015)\citenamefont {Wang},
  \citenamefont {Agterberg},\ and\ \citenamefont
  {Chubukov}}]{phenomenological4}%
  \BibitemOpen
  \bibfield  {author} {\bibinfo {author} {\bibfnamefont {Y.}~\bibnamefont
  {Wang}}, \bibinfo {author} {\bibfnamefont {D.~F.}\ \bibnamefont
  {Agterberg}},\ and\ \bibinfo {author} {\bibfnamefont {A.}~\bibnamefont
  {Chubukov}},\ }\bibfield  {title} {\bibinfo {title} {Coexistence of
  charge-density-wave and pair-density-wave orders in underdoped cuprates},\
  }\href {https://doi.org/10.1103/PhysRevLett.114.197001} {\bibfield  {journal}
  {\bibinfo  {journal} {Phys. Rev. Lett.}\ }\textbf {\bibinfo {volume} {114}},\
  \bibinfo {pages} {197001} (\bibinfo {year} {2015})}\BibitemShut {NoStop}%
\bibitem [{\citenamefont {Wang}\ \emph {et~al.}(2018)\citenamefont {Wang},
  \citenamefont {Edkins}, \citenamefont {Hamidian}, \citenamefont {Davis},
  \citenamefont {Fradkin},\ and\ \citenamefont {Kivelson}}]{phenomenological5}%
  \BibitemOpen
  \bibfield  {author} {\bibinfo {author} {\bibfnamefont {Y.}~\bibnamefont
  {Wang}}, \bibinfo {author} {\bibfnamefont {S.~D.}\ \bibnamefont {Edkins}},
  \bibinfo {author} {\bibfnamefont {M.~H.}\ \bibnamefont {Hamidian}}, \bibinfo
  {author} {\bibfnamefont {J.~C.~S.}\ \bibnamefont {Davis}}, \bibinfo {author}
  {\bibfnamefont {E.}~\bibnamefont {Fradkin}},\ and\ \bibinfo {author}
  {\bibfnamefont {S.~A.}\ \bibnamefont {Kivelson}},\ }\bibfield  {title}
  {\bibinfo {title} {Pair density waves in superconducting vortex halos},\
  }\href {https://doi.org/10.1103/PhysRevB.97.174510} {\bibfield  {journal}
  {\bibinfo  {journal} {Phys. Rev. B}\ }\textbf {\bibinfo {volume} {97}},\
  \bibinfo {pages} {174510} (\bibinfo {year} {2018})}\BibitemShut {NoStop}%
\bibitem [{\citenamefont {Jin}\ \emph {et~al.}(2022)\citenamefont {Jin},
  \citenamefont {Jiang}, \citenamefont {Yao},\ and\ \citenamefont
  {Zhou}}]{interplay_yizhou_2022}%
  \BibitemOpen
  \bibfield  {author} {\bibinfo {author} {\bibfnamefont {J.-T.}\ \bibnamefont
  {Jin}}, \bibinfo {author} {\bibfnamefont {K.}~\bibnamefont {Jiang}}, \bibinfo
  {author} {\bibfnamefont {H.}~\bibnamefont {Yao}},\ and\ \bibinfo {author}
  {\bibfnamefont {Y.}~\bibnamefont {Zhou}},\ }\bibfield  {title} {\bibinfo
  {title} {Interplay between pair density wave and a nested fermi surface},\
  }\href {https://doi.org/10.1103/PhysRevLett.129.167001} {\bibfield  {journal}
  {\bibinfo  {journal} {Phys. Rev. Lett.}\ }\textbf {\bibinfo {volume} {129}},\
  \bibinfo {pages} {167001} (\bibinfo {year} {2022})}\BibitemShut {NoStop}%
\bibitem [{\citenamefont {Zhou}\ and\ \citenamefont
  {Wang}(2022)}]{Ziqiang-2022}%
  \BibitemOpen
  \bibfield  {author} {\bibinfo {author} {\bibfnamefont {S.}~\bibnamefont
  {Zhou}}\ and\ \bibinfo {author} {\bibfnamefont {Z.}~\bibnamefont {Wang}},\
  }\bibfield  {title} {\bibinfo {title} {Chern fermi pocket, topological pair
  density wave, and charge-4e and charge-6e superconductivity in kagomé
  superconductors},\ }\href {https://doi.org/10.1038/s41467-022-34832-2}
  {\bibfield  {journal} {\bibinfo  {journal} {Nat. Commun.}\ }\textbf {\bibinfo
  {volume} {13}},\ \bibinfo {pages} {7288} (\bibinfo {year}
  {2022})}\BibitemShut {NoStop}%
\bibitem [{\citenamefont {Zhang}(1998)}]{microscopic_1998}%
  \BibitemOpen
  \bibfield  {author} {\bibinfo {author} {\bibfnamefont {S.-C.}\ \bibnamefont
  {Zhang}},\ }\bibfield  {title} {\bibinfo {title} {Recent developments in the
  {SO}(5) theory of high {Tc} superconductivity},\ }\href
  {https://doi.org/10.1016/S0022-3697(98)00096-1} {\bibfield  {journal}
  {\bibinfo  {journal} {J. Phys. Chem. Solids}\ }\textbf {\bibinfo {volume}
  {59}},\ \bibinfo {pages} {1774} (\bibinfo {year} {1998})}\BibitemShut
  {NoStop}%
\bibitem [{\citenamefont {Himeda}\ \emph {et~al.}(2002)\citenamefont {Himeda},
  \citenamefont {Kato},\ and\ \citenamefont {Ogata}}]{microscopic_2002}%
  \BibitemOpen
  \bibfield  {author} {\bibinfo {author} {\bibfnamefont {A.}~\bibnamefont
  {Himeda}}, \bibinfo {author} {\bibfnamefont {T.}~\bibnamefont {Kato}},\ and\
  \bibinfo {author} {\bibfnamefont {M.}~\bibnamefont {Ogata}},\ }\bibfield
  {title} {\bibinfo {title} {Stripe {States} with {Spatially} {Oscillating} d
  -{Wave} {Superconductivity} in the {Two}-{Dimensional}
  $t\text{\ensuremath{-}}t'\text{\ensuremath{-}}{J}$ {Model}},\ }\href
  {https://doi.org/10.1103/PhysRevLett.88.117001} {\bibfield  {journal}
  {\bibinfo  {journal} {Phys. Rev. Lett.}\ }\textbf {\bibinfo {volume} {88}},\
  \bibinfo {pages} {117001} (\bibinfo {year} {2002})}\BibitemShut {NoStop}%
\bibitem [{\citenamefont {Raczkowski}\ \emph {et~al.}(2007)\citenamefont
  {Raczkowski}, \citenamefont {Capello}, \citenamefont {Poilblanc},
  \citenamefont {Fr\'esard},\ and\ \citenamefont {Ole\ifmmode~\acute{s}\else
  \'{s}\fi{}}}]{microscopic_2007}%
  \BibitemOpen
  \bibfield  {author} {\bibinfo {author} {\bibfnamefont {M.}~\bibnamefont
  {Raczkowski}}, \bibinfo {author} {\bibfnamefont {M.}~\bibnamefont {Capello}},
  \bibinfo {author} {\bibfnamefont {D.}~\bibnamefont {Poilblanc}}, \bibinfo
  {author} {\bibfnamefont {R.}~\bibnamefont {Fr\'esard}},\ and\ \bibinfo
  {author} {\bibfnamefont {A.~M.}\ \bibnamefont {Ole\ifmmode~\acute{s}\else
  \'{s}\fi{}}},\ }\bibfield  {title} {\bibinfo {title} {Unidirectional $d$-wave
  superconducting domains in the two-dimensional $t\text{\ensuremath{-}}{J}$
  model},\ }\href {https://doi.org/10.1103/PhysRevB.76.140505} {\bibfield
  {journal} {\bibinfo  {journal} {Phys. Rev. B}\ }\textbf {\bibinfo {volume}
  {76}},\ \bibinfo {pages} {140505} (\bibinfo {year} {2007})}\BibitemShut
  {NoStop}%
\bibitem [{\citenamefont {Capello}\ \emph {et~al.}(2008)\citenamefont
  {Capello}, \citenamefont {Raczkowski},\ and\ \citenamefont
  {Poilblanc}}]{microscopic_2008}%
  \BibitemOpen
  \bibfield  {author} {\bibinfo {author} {\bibfnamefont {M.}~\bibnamefont
  {Capello}}, \bibinfo {author} {\bibfnamefont {M.}~\bibnamefont
  {Raczkowski}},\ and\ \bibinfo {author} {\bibfnamefont {D.}~\bibnamefont
  {Poilblanc}},\ }\bibfield  {title} {\bibinfo {title} {Stability of rvb hole
  stripes in high-temperature superconductors},\ }\href
  {https://doi.org/10.1103/PhysRevB.77.224502} {\bibfield  {journal} {\bibinfo
  {journal} {Phys. Rev. B}\ }\textbf {\bibinfo {volume} {77}},\ \bibinfo
  {pages} {224502} (\bibinfo {year} {2008})}\BibitemShut {NoStop}%
\bibitem [{\citenamefont {Yang}\ \emph {et~al.}(2009)\citenamefont {Yang},
  \citenamefont {Chen}, \citenamefont {Rice}, \citenamefont {Sigrist},\ and\
  \citenamefont {Zhang}}]{microscopic_2009}%
  \BibitemOpen
  \bibfield  {author} {\bibinfo {author} {\bibfnamefont {K.-Y.}\ \bibnamefont
  {Yang}}, \bibinfo {author} {\bibfnamefont {W.~Q.}\ \bibnamefont {Chen}},
  \bibinfo {author} {\bibfnamefont {T.~M.}\ \bibnamefont {Rice}}, \bibinfo
  {author} {\bibfnamefont {M.}~\bibnamefont {Sigrist}},\ and\ \bibinfo {author}
  {\bibfnamefont {F.-C.}\ \bibnamefont {Zhang}},\ }\bibfield  {title} {\bibinfo
  {title} {Nature of stripes in the generalized \textit{t}-\textit{{J}} model
  applied to the cuprate superconductors},\ }\href
  {https://doi.org/10.1088/1367-2630/11/5/055053} {\bibfield  {journal}
  {\bibinfo  {journal} {New J. Phys.}\ }\textbf {\bibinfo {volume} {11}},\
  \bibinfo {pages} {055053} (\bibinfo {year} {2009})}\BibitemShut {NoStop}%
\bibitem [{\citenamefont {Loder}\ \emph {et~al.}(2010)\citenamefont {Loder},
  \citenamefont {Kampf},\ and\ \citenamefont {Kopp}}]{microscopic_2010_1}%
  \BibitemOpen
  \bibfield  {author} {\bibinfo {author} {\bibfnamefont {F.}~\bibnamefont
  {Loder}}, \bibinfo {author} {\bibfnamefont {A.~P.}\ \bibnamefont {Kampf}},\
  and\ \bibinfo {author} {\bibfnamefont {T.}~\bibnamefont {Kopp}},\ }\bibfield
  {title} {\bibinfo {title} {Superconducting state with a finite-momentum
  pairing mechanism in zero external magnetic field},\ }\href
  {https://doi.org/10.1103/PhysRevB.81.020511} {\bibfield  {journal} {\bibinfo
  {journal} {Phys. Rev. B}\ }\textbf {\bibinfo {volume} {81}},\ \bibinfo
  {pages} {020511} (\bibinfo {year} {2010})}\BibitemShut {NoStop}%
\bibitem [{\citenamefont {Berg}\ \emph {et~al.}(2010)\citenamefont {Berg},
  \citenamefont {Fradkin},\ and\ \citenamefont
  {Kivelson}}]{microscopic_2010_2}%
  \BibitemOpen
  \bibfield  {author} {\bibinfo {author} {\bibfnamefont {E.}~\bibnamefont
  {Berg}}, \bibinfo {author} {\bibfnamefont {E.}~\bibnamefont {Fradkin}},\ and\
  \bibinfo {author} {\bibfnamefont {S.~A.}\ \bibnamefont {Kivelson}},\
  }\bibfield  {title} {\bibinfo {title} {Pair-density-wave correlations in the
  kondo-heisenberg model},\ }\href
  {https://doi.org/10.1103/PhysRevLett.105.146403} {\bibfield  {journal}
  {\bibinfo  {journal} {Phys. Rev. Lett.}\ }\textbf {\bibinfo {volume} {105}},\
  \bibinfo {pages} {146403} (\bibinfo {year} {2010})}\BibitemShut {NoStop}%
\bibitem [{\citenamefont {Loder}\ \emph {et~al.}(2011)\citenamefont {Loder},
  \citenamefont {Graser}, \citenamefont {Kampf},\ and\ \citenamefont
  {Kopp}}]{microscopic_2011}%
  \BibitemOpen
  \bibfield  {author} {\bibinfo {author} {\bibfnamefont {F.}~\bibnamefont
  {Loder}}, \bibinfo {author} {\bibfnamefont {S.}~\bibnamefont {Graser}},
  \bibinfo {author} {\bibfnamefont {A.~P.}\ \bibnamefont {Kampf}},\ and\
  \bibinfo {author} {\bibfnamefont {T.}~\bibnamefont {Kopp}},\ }\bibfield
  {title} {\bibinfo {title} {Mean-field pairing theory for the charge-stripe
  phase of high-temperature cuprate superconductors},\ }\href
  {https://doi.org/10.1103/PhysRevLett.107.187001} {\bibfield  {journal}
  {\bibinfo  {journal} {Phys. Rev. Lett.}\ }\textbf {\bibinfo {volume} {107}},\
  \bibinfo {pages} {187001} (\bibinfo {year} {2011})}\BibitemShut {NoStop}%
\bibitem [{\citenamefont {Jaefari}\ and\ \citenamefont
  {Fradkin}(2012)}]{microscopic_2012}%
  \BibitemOpen
  \bibfield  {author} {\bibinfo {author} {\bibfnamefont {A.}~\bibnamefont
  {Jaefari}}\ and\ \bibinfo {author} {\bibfnamefont {E.}~\bibnamefont
  {Fradkin}},\ }\bibfield  {title} {\bibinfo {title} {Pair-density-wave
  superconducting order in two-leg ladders},\ }\href
  {https://doi.org/10.1103/PhysRevB.85.035104} {\bibfield  {journal} {\bibinfo
  {journal} {Phys. Rev. B}\ }\textbf {\bibinfo {volume} {85}},\ \bibinfo
  {pages} {035104} (\bibinfo {year} {2012})}\BibitemShut {NoStop}%
\bibitem [{\citenamefont {Soto-Garrido}\ and\ \citenamefont
  {Fradkin}(2014)}]{microscopic_2014}%
  \BibitemOpen
  \bibfield  {author} {\bibinfo {author} {\bibfnamefont {R.}~\bibnamefont
  {Soto-Garrido}}\ and\ \bibinfo {author} {\bibfnamefont {E.}~\bibnamefont
  {Fradkin}},\ }\bibfield  {title} {\bibinfo {title} {Pair-density-wave
  superconducting states and electronic liquid-crystal phases},\ }\href
  {https://doi.org/10.1103/PhysRevB.89.165126} {\bibfield  {journal} {\bibinfo
  {journal} {Phys. Rev. B}\ }\textbf {\bibinfo {volume} {89}},\ \bibinfo
  {pages} {165126} (\bibinfo {year} {2014})}\BibitemShut {NoStop}%
\bibitem [{\citenamefont {Soto-Garrido}\ \emph {et~al.}(2015)\citenamefont
  {Soto-Garrido}, \citenamefont {Cho},\ and\ \citenamefont
  {Fradkin}}]{microscopic_2015}%
  \BibitemOpen
  \bibfield  {author} {\bibinfo {author} {\bibfnamefont {R.}~\bibnamefont
  {Soto-Garrido}}, \bibinfo {author} {\bibfnamefont {G.~Y.}\ \bibnamefont
  {Cho}},\ and\ \bibinfo {author} {\bibfnamefont {E.}~\bibnamefont {Fradkin}},\
  }\bibfield  {title} {\bibinfo {title} {Quasi-one-dimensional pair density
  wave superconducting state},\ }\href
  {https://doi.org/10.1103/PhysRevB.91.195102} {\bibfield  {journal} {\bibinfo
  {journal} {Phys. Rev. B}\ }\textbf {\bibinfo {volume} {91}},\ \bibinfo
  {pages} {195102} (\bibinfo {year} {2015})}\BibitemShut {NoStop}%
\bibitem [{\citenamefont {W\aa{}rdh}\ and\ \citenamefont
  {Granath}(2017)}]{microscopic_2017}%
  \BibitemOpen
  \bibfield  {author} {\bibinfo {author} {\bibfnamefont {J.}~\bibnamefont
  {W\aa{}rdh}}\ and\ \bibinfo {author} {\bibfnamefont {M.}~\bibnamefont
  {Granath}},\ }\bibfield  {title} {\bibinfo {title} {Effective model for a
  supercurrent in a pair-density wave},\ }\href
  {https://doi.org/10.1103/PhysRevB.96.224503} {\bibfield  {journal} {\bibinfo
  {journal} {Phys. Rev. B}\ }\textbf {\bibinfo {volume} {96}},\ \bibinfo
  {pages} {224503} (\bibinfo {year} {2017})}\BibitemShut {NoStop}%
\bibitem [{\citenamefont {W\aa{}rdh}\ \emph {et~al.}(2018)\citenamefont
  {W\aa{}rdh}, \citenamefont {Andersen},\ and\ \citenamefont
  {Granath}}]{microscopic_2018_1}%
  \BibitemOpen
  \bibfield  {author} {\bibinfo {author} {\bibfnamefont {J.}~\bibnamefont
  {W\aa{}rdh}}, \bibinfo {author} {\bibfnamefont {B.~M.}\ \bibnamefont
  {Andersen}},\ and\ \bibinfo {author} {\bibfnamefont {M.}~\bibnamefont
  {Granath}},\ }\bibfield  {title} {\bibinfo {title} {Suppression of superfluid
  stiffness near a lifshitz-point instability to finite-momentum
  superconductivity},\ }\href {https://doi.org/10.1103/PhysRevB.98.224501}
  {\bibfield  {journal} {\bibinfo  {journal} {Phys. Rev. B}\ }\textbf {\bibinfo
  {volume} {98}},\ \bibinfo {pages} {224501} (\bibinfo {year}
  {2018})}\BibitemShut {NoStop}%
\bibitem [{\citenamefont {Sarkar}\ \emph {et~al.}(2018)\citenamefont {Sarkar},
  \citenamefont {Greene},\ and\ \citenamefont
  {Das~Sarma}}]{microscopic_2018_2}%
  \BibitemOpen
  \bibfield  {author} {\bibinfo {author} {\bibfnamefont {T.}~\bibnamefont
  {Sarkar}}, \bibinfo {author} {\bibfnamefont {R.~L.}\ \bibnamefont {Greene}},\
  and\ \bibinfo {author} {\bibfnamefont {S.}~\bibnamefont {Das~Sarma}},\
  }\bibfield  {title} {\bibinfo {title} {Anomalous normal-state resistivity in
  superconducting
  {$\mathrm{L}{\mathrm{a}}_{2\ensuremath{-}x}\mathrm{C}{\mathrm{e}}_{x}\mathrm{Cu}{\mathrm{O}}_{4}$}:
  Fermi liquid or strange metal?},\ }\href
  {https://doi.org/10.1103/PhysRevB.98.224503} {\bibfield  {journal} {\bibinfo
  {journal} {Phys. Rev. B}\ }\textbf {\bibinfo {volume} {98}},\ \bibinfo
  {pages} {224503} (\bibinfo {year} {2018})}\BibitemShut {NoStop}%
\bibitem [{\citenamefont {Xu}\ \emph {et~al.}(2019)\citenamefont {Xu},
  \citenamefont {Law},\ and\ \citenamefont {Lee}}]{microscopic_2019}%
  \BibitemOpen
  \bibfield  {author} {\bibinfo {author} {\bibfnamefont {X.~Y.}\ \bibnamefont
  {Xu}}, \bibinfo {author} {\bibfnamefont {K.~T.}\ \bibnamefont {Law}},\ and\
  \bibinfo {author} {\bibfnamefont {P.~A.}\ \bibnamefont {Lee}},\ }\bibfield
  {title} {\bibinfo {title} {Pair density wave in the doped {$t-J$} model with
  ring exchange on a triangular lattice},\ }\href
  {https://doi.org/10.1103/PhysRevLett.122.167001} {\bibfield  {journal}
  {\bibinfo  {journal} {Phys. Rev. Lett.}\ }\textbf {\bibinfo {volume} {122}},\
  \bibinfo {pages} {167001} (\bibinfo {year} {2019})}\BibitemShut {NoStop}%
\bibitem [{\citenamefont {Huang}\ \emph {et~al.}(2022)\citenamefont {Huang},
  \citenamefont {Han}, \citenamefont {Kivelson},\ and\ \citenamefont
  {Yao}}]{microscopic_2022_1}%
  \BibitemOpen
  \bibfield  {author} {\bibinfo {author} {\bibfnamefont {K.~S.}\ \bibnamefont
  {Huang}}, \bibinfo {author} {\bibfnamefont {Z.}~\bibnamefont {Han}}, \bibinfo
  {author} {\bibfnamefont {S.~A.}\ \bibnamefont {Kivelson}},\ and\ \bibinfo
  {author} {\bibfnamefont {H.}~\bibnamefont {Yao}},\ }\bibfield  {title}
  {\bibinfo {title} {Pair-density-wave in the strong coupling limit of the
  {Holstein}-{Hubbard} model},\ }\href
  {https://doi.org/10.1038/s41535-022-00426-w} {\bibfield  {journal} {\bibinfo
  {journal} {npj Quantum Mater.}\ }\textbf {\bibinfo {volume} {7}},\ \bibinfo
  {pages} {17} (\bibinfo {year} {2022})}\BibitemShut {NoStop}%
\bibitem [{\citenamefont {Han}\ and\ \citenamefont
  {Kivelson}(2022)}]{microscopic_2022_2}%
  \BibitemOpen
  \bibfield  {author} {\bibinfo {author} {\bibfnamefont {Z.}~\bibnamefont
  {Han}}\ and\ \bibinfo {author} {\bibfnamefont {S.~A.}\ \bibnamefont
  {Kivelson}},\ }\bibfield  {title} {\bibinfo {title} {Pair density wave and
  reentrant superconducting tendencies originating from valley polarization},\
  }\href {https://doi.org/10.1103/PhysRevB.105.L100509} {\bibfield  {journal}
  {\bibinfo  {journal} {Phys. Rev. B}\ }\textbf {\bibinfo {volume} {105}},\
  \bibinfo {pages} {L100509} (\bibinfo {year} {2022})}\BibitemShut {NoStop}%
\bibitem [{\citenamefont {Wu}\ \emph {et~al.}(2023{\natexlab{a}})\citenamefont
  {Wu}, \citenamefont {Nosov}, \citenamefont {Patel},\ and\ \citenamefont
  {Raghu}}]{microscopic_2023_1_raghu_prl}%
  \BibitemOpen
  \bibfield  {author} {\bibinfo {author} {\bibfnamefont {Y.-M.}\ \bibnamefont
  {Wu}}, \bibinfo {author} {\bibfnamefont {P.~A.}\ \bibnamefont {Nosov}},
  \bibinfo {author} {\bibfnamefont {A.~A.}\ \bibnamefont {Patel}},\ and\
  \bibinfo {author} {\bibfnamefont {S.}~\bibnamefont {Raghu}},\ }\bibfield
  {title} {\bibinfo {title} {Pair density wave order from electron repulsion},\
  }\href {https://doi.org/10.1103/PhysRevLett.130.026001} {\bibfield  {journal}
  {\bibinfo  {journal} {Phys. Rev. Lett.}\ }\textbf {\bibinfo {volume} {130}},\
  \bibinfo {pages} {026001} (\bibinfo {year} {2023}{\natexlab{a}})}\BibitemShut
  {NoStop}%
\bibitem [{\citenamefont {Wu}\ \emph {et~al.}(2023{\natexlab{b}})\citenamefont
  {Wu}, \citenamefont {Wu},\ and\ \citenamefont {Wu}}]{microscopic_2023_2}%
  \BibitemOpen
  \bibfield  {author} {\bibinfo {author} {\bibfnamefont {Z.}~\bibnamefont
  {Wu}}, \bibinfo {author} {\bibfnamefont {Y.-M.}\ \bibnamefont {Wu}},\ and\
  \bibinfo {author} {\bibfnamefont {F.}~\bibnamefont {Wu}},\ }\bibfield
  {title} {\bibinfo {title} {Pair density wave and loop current promoted by van
  hove singularities in moir\'e systems},\ }\href
  {https://doi.org/10.1103/PhysRevB.107.045122} {\bibfield  {journal} {\bibinfo
   {journal} {Phys. Rev. B}\ }\textbf {\bibinfo {volume} {107}},\ \bibinfo
  {pages} {045122} (\bibinfo {year} {2023}{\natexlab{b}})}\BibitemShut
  {NoStop}%
\bibitem [{\citenamefont {Wu}\ \emph {et~al.}(2023{\natexlab{c}})\citenamefont
  {Wu}, \citenamefont {Wu},\ and\ \citenamefont {Yao}}]{microscopic_2023_3}%
  \BibitemOpen
  \bibfield  {author} {\bibinfo {author} {\bibfnamefont {Y.-M.}\ \bibnamefont
  {Wu}}, \bibinfo {author} {\bibfnamefont {Z.}~\bibnamefont {Wu}},\ and\
  \bibinfo {author} {\bibfnamefont {H.}~\bibnamefont {Yao}},\ }\bibfield
  {title} {\bibinfo {title} {Pair-density-wave and chiral superconductivity in
  twisted bilayer transition metal dichalcogenides},\ }\href
  {https://doi.org/10.1103/PhysRevLett.130.126001} {\bibfield  {journal}
  {\bibinfo  {journal} {Phys. Rev. Lett.}\ }\textbf {\bibinfo {volume} {130}},\
  \bibinfo {pages} {126001} (\bibinfo {year} {2023}{\natexlab{c}})}\BibitemShut
  {NoStop}%
\bibitem [{\citenamefont {Setty}\ \emph
  {et~al.}(2023{\natexlab{a}})\citenamefont {Setty}, \citenamefont
  {Fanfarillo},\ and\ \citenamefont {Hirschfeld}}]{microscopic_2023_4}%
  \BibitemOpen
  \bibfield  {author} {\bibinfo {author} {\bibfnamefont {C.}~\bibnamefont
  {Setty}}, \bibinfo {author} {\bibfnamefont {L.}~\bibnamefont {Fanfarillo}},\
  and\ \bibinfo {author} {\bibfnamefont {P.~J.}\ \bibnamefont {Hirschfeld}},\
  }\bibfield  {title} {\bibinfo {title} {Mechanism for fluctuating pair density
  wave},\ }\href {https://doi.org/10.1038/s41467-023-38956-x} {\bibfield
  {journal} {\bibinfo  {journal} {Nat. Commun.}\ }\textbf {\bibinfo {volume}
  {14}},\ \bibinfo {pages} {3181} (\bibinfo {year}
  {2023}{\natexlab{a}})}\BibitemShut {NoStop}%
\bibitem [{\citenamefont {Jiang}(2023)}]{microscopic_2023_5}%
  \BibitemOpen
  \bibfield  {author} {\bibinfo {author} {\bibfnamefont {H.-C.}\ \bibnamefont
  {Jiang}},\ }\bibfield  {title} {\bibinfo {title} {Pair density wave in the
  doped three-band hubbard model on two-leg square cylinders},\ }\href
  {https://doi.org/10.1103/PhysRevB.107.214504} {\bibfield  {journal} {\bibinfo
   {journal} {Phys. Rev. B}\ }\textbf {\bibinfo {volume} {107}},\ \bibinfo
  {pages} {214504} (\bibinfo {year} {2023})}\BibitemShut {NoStop}%
\bibitem [{\citenamefont {Shaffer}\ \emph {et~al.}(2023)\citenamefont
  {Shaffer}, \citenamefont {Burnell},\ and\ \citenamefont
  {Fernandes}}]{microscopic_2023_6}%
  \BibitemOpen
  \bibfield  {author} {\bibinfo {author} {\bibfnamefont {D.}~\bibnamefont
  {Shaffer}}, \bibinfo {author} {\bibfnamefont {F.~J.}\ \bibnamefont
  {Burnell}},\ and\ \bibinfo {author} {\bibfnamefont {R.~M.}\ \bibnamefont
  {Fernandes}},\ }\bibfield  {title} {\bibinfo {title} {Weak-coupling theory of
  pair density wave instabilities in transition metal dichalcogenides},\ }\href
  {https://doi.org/10.1103/PhysRevB.107.224516} {\bibfield  {journal} {\bibinfo
   {journal} {Phys. Rev. B}\ }\textbf {\bibinfo {volume} {107}},\ \bibinfo
  {pages} {224516} (\bibinfo {year} {2023})}\BibitemShut {NoStop}%
\bibitem [{\citenamefont {Shaffer}\ and\ \citenamefont
  {Santos}(2023)}]{microscopic_2023_7}%
  \BibitemOpen
  \bibfield  {author} {\bibinfo {author} {\bibfnamefont {D.}~\bibnamefont
  {Shaffer}}\ and\ \bibinfo {author} {\bibfnamefont {L.~H.}\ \bibnamefont
  {Santos}},\ }\bibfield  {title} {\bibinfo {title} {Triplet pair density wave
  superconductivity on the $\ensuremath{\pi}$-flux square lattice},\ }\href
  {https://doi.org/10.1103/PhysRevB.108.035135} {\bibfield  {journal} {\bibinfo
   {journal} {Phys. Rev. B}\ }\textbf {\bibinfo {volume} {108}},\ \bibinfo
  {pages} {035135} (\bibinfo {year} {2023})}\BibitemShut {NoStop}%
\bibitem [{\citenamefont {Jiang}\ and\ \citenamefont
  {Yao}(2024)}]{microscopic_2023_8_yao}%
  \BibitemOpen
  \bibfield  {author} {\bibinfo {author} {\bibfnamefont {Y.-F.}\ \bibnamefont
  {Jiang}}\ and\ \bibinfo {author} {\bibfnamefont {H.}~\bibnamefont {Yao}},\
  }\bibfield  {title} {\bibinfo {title} {Pair-density-wave superconductivity: A
  microscopic model on the 2d honeycomb lattice},\ }\href
  {https://doi.org/10.1103/PhysRevLett.133.176501} {\bibfield  {journal}
  {\bibinfo  {journal} {Phys. Rev. Lett.}\ }\textbf {\bibinfo {volume} {133}},\
  \bibinfo {pages} {176501} (\bibinfo {year} {2024})}\BibitemShut {NoStop}%
\bibitem [{\citenamefont {Wu}\ \emph {et~al.}(2023{\natexlab{d}})\citenamefont
  {Wu}, \citenamefont {Thomale},\ and\ \citenamefont
  {Raghu}}]{microscopic_2023_9_raghu_prb}%
  \BibitemOpen
  \bibfield  {author} {\bibinfo {author} {\bibfnamefont {Y.-M.}\ \bibnamefont
  {Wu}}, \bibinfo {author} {\bibfnamefont {R.}~\bibnamefont {Thomale}},\ and\
  \bibinfo {author} {\bibfnamefont {S.}~\bibnamefont {Raghu}},\ }\bibfield
  {title} {\bibinfo {title} {Sublattice interference promotes pair density wave
  order in kagome metals},\ }\href
  {https://doi.org/10.1103/PhysRevB.108.L081117} {\bibfield  {journal}
  {\bibinfo  {journal} {Phys. Rev. B}\ }\textbf {\bibinfo {volume} {108}},\
  \bibinfo {pages} {L081117} (\bibinfo {year}
  {2023}{\natexlab{d}})}\BibitemShut {NoStop}%
\bibitem [{\citenamefont {Zhang}\ \emph {et~al.}(2023)\citenamefont {Zhang},
  \citenamefont {Sun},\ and\ \citenamefont {Weng}}]{microscopic_2023_10}%
  \BibitemOpen
  \bibfield  {author} {\bibinfo {author} {\bibfnamefont {H.-K.}\ \bibnamefont
  {Zhang}}, \bibinfo {author} {\bibfnamefont {R.-Y.}\ \bibnamefont {Sun}},\
  and\ \bibinfo {author} {\bibfnamefont {Z.-Y.}\ \bibnamefont {Weng}},\
  }\bibfield  {title} {\bibinfo {title} {Pair density wave characterized by a
  hidden string order parameter},\ }\href
  {https://doi.org/10.1103/PhysRevB.108.115136} {\bibfield  {journal} {\bibinfo
   {journal} {Phys. Rev. B}\ }\textbf {\bibinfo {volume} {108}},\ \bibinfo
  {pages} {115136} (\bibinfo {year} {2023})}\BibitemShut {NoStop}%
\bibitem [{\citenamefont {Chen}\ and\ \citenamefont
  {Sheng}(2023)}]{microscopic_2023_11}%
  \BibitemOpen
  \bibfield  {author} {\bibinfo {author} {\bibfnamefont {F.}~\bibnamefont
  {Chen}}\ and\ \bibinfo {author} {\bibfnamefont {D.~N.}\ \bibnamefont
  {Sheng}},\ }\bibfield  {title} {\bibinfo {title} {Singlet, triplet, and pair
  density wave superconductivity in the doped triangular-lattice moir\'e
  system},\ }\href {https://doi.org/10.1103/PhysRevB.108.L201110} {\bibfield
  {journal} {\bibinfo  {journal} {Phys. Rev. B}\ }\textbf {\bibinfo {volume}
  {108}},\ \bibinfo {pages} {L201110} (\bibinfo {year} {2023})}\BibitemShut
  {NoStop}%
\bibitem [{\citenamefont {Setty}\ \emph
  {et~al.}(2023{\natexlab{b}})\citenamefont {Setty}, \citenamefont {Zhao},
  \citenamefont {Fanfarillo}, \citenamefont {Huang}, \citenamefont
  {Hirschfeld}, \citenamefont {Phillips},\ and\ \citenamefont
  {Yang}}]{microscopic_2023_12}%
  \BibitemOpen
  \bibfield  {author} {\bibinfo {author} {\bibfnamefont {C.}~\bibnamefont
  {Setty}}, \bibinfo {author} {\bibfnamefont {J.}~\bibnamefont {Zhao}},
  \bibinfo {author} {\bibfnamefont {L.}~\bibnamefont {Fanfarillo}}, \bibinfo
  {author} {\bibfnamefont {E.~W.}\ \bibnamefont {Huang}}, \bibinfo {author}
  {\bibfnamefont {P.~J.}\ \bibnamefont {Hirschfeld}}, \bibinfo {author}
  {\bibfnamefont {P.~W.}\ \bibnamefont {Phillips}},\ and\ \bibinfo {author}
  {\bibfnamefont {K.}~\bibnamefont {Yang}},\ }\bibfield  {title} {\bibinfo
  {title} {Exact solution for finite center-of-mass momentum cooper pairing},\
  }\href {https://doi.org/10.1103/PhysRevB.108.174506} {\bibfield  {journal}
  {\bibinfo  {journal} {Phys. Rev. B}\ }\textbf {\bibinfo {volume} {108}},\
  \bibinfo {pages} {174506} (\bibinfo {year} {2023}{\natexlab{b}})}\BibitemShut
  {NoStop}%
\bibitem [{\citenamefont {Jiang}\ and\ \citenamefont
  {Devereaux}(2023)}]{microscopic_2023_13}%
  \BibitemOpen
  \bibfield  {author} {\bibinfo {author} {\bibfnamefont {H.-C.}\ \bibnamefont
  {Jiang}}\ and\ \bibinfo {author} {\bibfnamefont {T.~P.}\ \bibnamefont
  {Devereaux}},\ }\bibfield  {title} {\bibinfo {title} {Pair density wave and
  superconductivity in a kinetically frustrated doped {Emery} model on a square
  lattice},\ }\href {https://doi.org/10.3389/femat.2023.1323404} {\bibfield
  {journal} {\bibinfo  {journal} {Front. Electron. Mater.}\ }\textbf {\bibinfo
  {volume} {3}},\ \bibinfo {pages} {1323404} (\bibinfo {year}
  {2023})}\BibitemShut {NoStop}%
\bibitem [{\citenamefont {Liu}\ and\ \citenamefont
  {Han}(2024)}]{microscopic_2024_1}%
  \BibitemOpen
  \bibfield  {author} {\bibinfo {author} {\bibfnamefont {F.}~\bibnamefont
  {Liu}}\ and\ \bibinfo {author} {\bibfnamefont {Z.}~\bibnamefont {Han}},\
  }\bibfield  {title} {\bibinfo {title} {Pair density wave and $s \pm id$
  superconductivity in a strongly coupled lightly doped kondo insulator},\
  }\href {https://doi.org/10.1103/PhysRevB.109.L121101} {\bibfield  {journal}
  {\bibinfo  {journal} {Phys. Rev. B}\ }\textbf {\bibinfo {volume} {109}},\
  \bibinfo {pages} {L121101} (\bibinfo {year} {2024})}\BibitemShut {NoStop}%
\bibitem [{\citenamefont {Bose}\ \emph {et~al.}(2024)\citenamefont {Bose},
  \citenamefont {Vadnais},\ and\ \citenamefont
  {Paramekanti}}]{microscopic_2024_2}%
  \BibitemOpen
  \bibfield  {author} {\bibinfo {author} {\bibfnamefont {A.}~\bibnamefont
  {Bose}}, \bibinfo {author} {\bibfnamefont {S.}~\bibnamefont {Vadnais}},\ and\
  \bibinfo {author} {\bibfnamefont {A.}~\bibnamefont {Paramekanti}},\
  }\bibfield  {title} {\bibinfo {title} {Altermagnetism and superconductivity
  in a multiorbital {$t\ensuremath{-}J$} model},\ }\href
  {https://doi.org/10.1103/PhysRevB.110.205120} {\bibfield  {journal} {\bibinfo
   {journal} {Phys. Rev. B}\ }\textbf {\bibinfo {volume} {110}},\ \bibinfo
  {pages} {205120} (\bibinfo {year} {2024})}\BibitemShut {NoStop}%
\bibitem [{\citenamefont {Wang}\ \emph {et~al.}(2024)\citenamefont {Wang},
  \citenamefont {Sun}, \citenamefont {Wang}, \citenamefont {Han}, \citenamefont
  {Kivelson},\ and\ \citenamefont {Yao}}]{microscopic_2024_3}%
  \BibitemOpen
  \bibfield  {author} {\bibinfo {author} {\bibfnamefont {J.}~\bibnamefont
  {Wang}}, \bibinfo {author} {\bibfnamefont {W.}~\bibnamefont {Sun}}, \bibinfo
  {author} {\bibfnamefont {H.-X.}\ \bibnamefont {Wang}}, \bibinfo {author}
  {\bibfnamefont {Z.}~\bibnamefont {Han}}, \bibinfo {author} {\bibfnamefont
  {S.~A.}\ \bibnamefont {Kivelson}},\ and\ \bibinfo {author} {\bibfnamefont
  {H.}~\bibnamefont {Yao}},\ }\href@noop {} {\bibinfo {title} {Pair density
  waves in the strong-coupling two-dimensional holstein-hubbard model: a
  variational monte carlo study}} (\bibinfo {year} {2024}),\ \Eprint
  {https://arxiv.org/abs/2404.11950} {arXiv:2404.11950 [cond-mat.str-el]}
  \BibitemShut {NoStop}%
\bibitem [{\citenamefont {Ortiz}\ \emph {et~al.}(2019)\citenamefont {Ortiz},
  \citenamefont {Gomes}, \citenamefont {Morey}, \citenamefont {Winiarski},
  \citenamefont {Bordelon}, \citenamefont {Mangum}, \citenamefont {Oswald},
  \citenamefont {Rodriguez-Rivera}, \citenamefont {Neilson}, \citenamefont
  {Wilson}, \citenamefont {Ertekin}, \citenamefont {McQueen},\ and\
  \citenamefont {Toberer}}]{Ortiz2019}%
  \BibitemOpen
  \bibfield  {author} {\bibinfo {author} {\bibfnamefont {B.~R.}\ \bibnamefont
  {Ortiz}}, \bibinfo {author} {\bibfnamefont {L.~C.}\ \bibnamefont {Gomes}},
  \bibinfo {author} {\bibfnamefont {J.~R.}\ \bibnamefont {Morey}}, \bibinfo
  {author} {\bibfnamefont {M.}~\bibnamefont {Winiarski}}, \bibinfo {author}
  {\bibfnamefont {M.}~\bibnamefont {Bordelon}}, \bibinfo {author}
  {\bibfnamefont {J.~S.}\ \bibnamefont {Mangum}}, \bibinfo {author}
  {\bibfnamefont {I.~W.~H.}\ \bibnamefont {Oswald}}, \bibinfo {author}
  {\bibfnamefont {J.~A.}\ \bibnamefont {Rodriguez-Rivera}}, \bibinfo {author}
  {\bibfnamefont {J.~R.}\ \bibnamefont {Neilson}}, \bibinfo {author}
  {\bibfnamefont {S.~D.}\ \bibnamefont {Wilson}}, \bibinfo {author}
  {\bibfnamefont {E.}~\bibnamefont {Ertekin}}, \bibinfo {author} {\bibfnamefont
  {T.~M.}\ \bibnamefont {McQueen}},\ and\ \bibinfo {author} {\bibfnamefont
  {E.~S.}\ \bibnamefont {Toberer}},\ }\bibfield  {title} {\bibinfo {title} {New
  kagome prototype materials: discovery of {KV$_3$Sb$_5$, RbV$_3$Sb$_5$, and
  CsV$_3$Sb$_5$}},\ }\href {https://doi.org/10.1103/PhysRevMaterials.3.094407}
  {\bibfield  {journal} {\bibinfo  {journal} {Phys. Rev. Mater.}\ }\textbf
  {\bibinfo {volume} {3}},\ \bibinfo {pages} {094407} (\bibinfo {year}
  {2019})}\BibitemShut {NoStop}%
\bibitem [{\citenamefont {Ortiz}\ \emph {et~al.}(2020)\citenamefont {Ortiz},
  \citenamefont {Teicher}, \citenamefont {Hu}, \citenamefont {Zuo},
  \citenamefont {Sarte}, \citenamefont {Schueller}, \citenamefont {Abeykoon},
  \citenamefont {Krogstad}, \citenamefont {Rosenkranz}, \citenamefont {Osborn},
  \citenamefont {Seshadri}, \citenamefont {Balents}, \citenamefont {He},\ and\
  \citenamefont {Wilson}}]{CsV3Sb5-1}%
  \BibitemOpen
  \bibfield  {author} {\bibinfo {author} {\bibfnamefont {B.~R.}\ \bibnamefont
  {Ortiz}}, \bibinfo {author} {\bibfnamefont {S.~M.~L.}\ \bibnamefont
  {Teicher}}, \bibinfo {author} {\bibfnamefont {Y.}~\bibnamefont {Hu}},
  \bibinfo {author} {\bibfnamefont {J.~L.}\ \bibnamefont {Zuo}}, \bibinfo
  {author} {\bibfnamefont {P.~M.}\ \bibnamefont {Sarte}}, \bibinfo {author}
  {\bibfnamefont {E.~C.}\ \bibnamefont {Schueller}}, \bibinfo {author}
  {\bibfnamefont {A.~M.~M.}\ \bibnamefont {Abeykoon}}, \bibinfo {author}
  {\bibfnamefont {M.~J.}\ \bibnamefont {Krogstad}}, \bibinfo {author}
  {\bibfnamefont {S.}~\bibnamefont {Rosenkranz}}, \bibinfo {author}
  {\bibfnamefont {R.}~\bibnamefont {Osborn}}, \bibinfo {author} {\bibfnamefont
  {R.}~\bibnamefont {Seshadri}}, \bibinfo {author} {\bibfnamefont
  {L.}~\bibnamefont {Balents}}, \bibinfo {author} {\bibfnamefont
  {J.}~\bibnamefont {He}},\ and\ \bibinfo {author} {\bibfnamefont {S.~D.}\
  \bibnamefont {Wilson}},\ }\bibfield  {title} {\bibinfo {title}
  {{$\mathrm{Cs}{\mathrm{V}}_{3}{\mathrm{Sb}}_{5}$: A ${\mathbb{Z}}_{2}$
  Topological Kagome Metal with a Superconducting Ground State}},\ }\href
  {https://doi.org/10.1103/PhysRevLett.125.247002} {\bibfield  {journal}
  {\bibinfo  {journal} {Phys. Rev. Lett.}\ }\textbf {\bibinfo {volume} {125}},\
  \bibinfo {pages} {247002} (\bibinfo {year} {2020})}\BibitemShut {NoStop}%
\bibitem [{\citenamefont {Shumiya}\ \emph {et~al.}(2021)\citenamefont
  {Shumiya}, \citenamefont {Hossain}, \citenamefont {Yin}, \citenamefont
  {Jiang}, \citenamefont {Ortiz}, \citenamefont {Liu}, \citenamefont {Shi},
  \citenamefont {Yin}, \citenamefont {Lei}, \citenamefont {Zhang},
  \citenamefont {Chang}, \citenamefont {Zhang}, \citenamefont {Cochran},
  \citenamefont {Multer}, \citenamefont {Litskevich}, \citenamefont {Cheng},
  \citenamefont {Yang}, \citenamefont {Guguchia}, \citenamefont {Wilson},\ and\
  \citenamefont {Hasan}}]{RbV3Sb5-1}%
  \BibitemOpen
  \bibfield  {author} {\bibinfo {author} {\bibfnamefont {N.}~\bibnamefont
  {Shumiya}}, \bibinfo {author} {\bibfnamefont {M.~S.}\ \bibnamefont
  {Hossain}}, \bibinfo {author} {\bibfnamefont {J.-X.}\ \bibnamefont {Yin}},
  \bibinfo {author} {\bibfnamefont {Y.-X.}\ \bibnamefont {Jiang}}, \bibinfo
  {author} {\bibfnamefont {B.~R.}\ \bibnamefont {Ortiz}}, \bibinfo {author}
  {\bibfnamefont {H.}~\bibnamefont {Liu}}, \bibinfo {author} {\bibfnamefont
  {Y.}~\bibnamefont {Shi}}, \bibinfo {author} {\bibfnamefont {Q.}~\bibnamefont
  {Yin}}, \bibinfo {author} {\bibfnamefont {H.}~\bibnamefont {Lei}}, \bibinfo
  {author} {\bibfnamefont {S.~S.}\ \bibnamefont {Zhang}}, \bibinfo {author}
  {\bibfnamefont {G.}~\bibnamefont {Chang}}, \bibinfo {author} {\bibfnamefont
  {Q.}~\bibnamefont {Zhang}}, \bibinfo {author} {\bibfnamefont {T.~A.}\
  \bibnamefont {Cochran}}, \bibinfo {author} {\bibfnamefont {D.}~\bibnamefont
  {Multer}}, \bibinfo {author} {\bibfnamefont {M.}~\bibnamefont {Litskevich}},
  \bibinfo {author} {\bibfnamefont {Z.-J.}\ \bibnamefont {Cheng}}, \bibinfo
  {author} {\bibfnamefont {X.~P.}\ \bibnamefont {Yang}}, \bibinfo {author}
  {\bibfnamefont {Z.}~\bibnamefont {Guguchia}}, \bibinfo {author}
  {\bibfnamefont {S.~D.}\ \bibnamefont {Wilson}},\ and\ \bibinfo {author}
  {\bibfnamefont {M.~Z.}\ \bibnamefont {Hasan}},\ }\bibfield  {title} {\bibinfo
  {title} {Intrinsic nature of chiral charge order in the kagome superconductor
  {$\mathrm{Rb}{\mathrm{V}}_{3}{\mathrm{Sb}}_{5}$}},\ }\href
  {https://doi.org/10.1103/PhysRevB.104.035131} {\bibfield  {journal} {\bibinfo
   {journal} {Phys. Rev. B}\ }\textbf {\bibinfo {volume} {104}},\ \bibinfo
  {pages} {035131} (\bibinfo {year} {2021})}\BibitemShut {NoStop}%
\bibitem [{\citenamefont {Ortiz}\ \emph {et~al.}(2021)\citenamefont {Ortiz},
  \citenamefont {Sarte}, \citenamefont {Kenney}, \citenamefont {Graf},
  \citenamefont {Teicher}, \citenamefont {Seshadri},\ and\ \citenamefont
  {Wilson}}]{KV3Sb5-1}%
  \BibitemOpen
  \bibfield  {author} {\bibinfo {author} {\bibfnamefont {B.~R.}\ \bibnamefont
  {Ortiz}}, \bibinfo {author} {\bibfnamefont {P.~M.}\ \bibnamefont {Sarte}},
  \bibinfo {author} {\bibfnamefont {E.~M.}\ \bibnamefont {Kenney}}, \bibinfo
  {author} {\bibfnamefont {M.~J.}\ \bibnamefont {Graf}}, \bibinfo {author}
  {\bibfnamefont {S.~M.~L.}\ \bibnamefont {Teicher}}, \bibinfo {author}
  {\bibfnamefont {R.}~\bibnamefont {Seshadri}},\ and\ \bibinfo {author}
  {\bibfnamefont {S.~D.}\ \bibnamefont {Wilson}},\ }\bibfield  {title}
  {\bibinfo {title} {{Superconductivity in the ${\mathbb{Z}}_{2}$ kagome metal
  ${\mathrm{KV}}_{3}{\mathrm{Sb}}_{5}$}},\ }\href
  {https://doi.org/10.1103/PhysRevMaterials.5.034801} {\bibfield  {journal}
  {\bibinfo  {journal} {Phys. Rev. Mater.}\ }\textbf {\bibinfo {volume} {5}},\
  \bibinfo {pages} {034801} (\bibinfo {year} {2021})}\BibitemShut {NoStop}%
\bibitem [{\citenamefont {Yin}\ \emph {et~al.}(2022)\citenamefont {Yin},
  \citenamefont {Lian},\ and\ \citenamefont {Hasan}}]{review_Yin2022}%
  \BibitemOpen
  \bibfield  {author} {\bibinfo {author} {\bibfnamefont {J.-X.}\ \bibnamefont
  {Yin}}, \bibinfo {author} {\bibfnamefont {B.}~\bibnamefont {Lian}},\ and\
  \bibinfo {author} {\bibfnamefont {M.~Z.}\ \bibnamefont {Hasan}},\ }\bibfield
  {title} {\bibinfo {title} {Topological kagome magnets and superconductors},\
  }\href {https://doi.org/10.1038/s41586-022-05516-0} {\bibfield  {journal}
  {\bibinfo  {journal} {Nature}\ }\textbf {\bibinfo {volume} {612}},\ \bibinfo
  {pages} {647} (\bibinfo {year} {2022})}\BibitemShut {NoStop}%
\bibitem [{\citenamefont {Neupert}\ \emph {et~al.}(2022)\citenamefont
  {Neupert}, \citenamefont {Denner}, \citenamefont {Yin}, \citenamefont
  {Thomale},\ and\ \citenamefont {Hasan}}]{review_Neupert2022}%
  \BibitemOpen
  \bibfield  {author} {\bibinfo {author} {\bibfnamefont {T.}~\bibnamefont
  {Neupert}}, \bibinfo {author} {\bibfnamefont {M.~M.}\ \bibnamefont {Denner}},
  \bibinfo {author} {\bibfnamefont {J.-X.}\ \bibnamefont {Yin}}, \bibinfo
  {author} {\bibfnamefont {R.}~\bibnamefont {Thomale}},\ and\ \bibinfo {author}
  {\bibfnamefont {M.~Z.}\ \bibnamefont {Hasan}},\ }\bibfield  {title} {\bibinfo
  {title} {Charge order and superconductivity in kagome materials},\ }\href
  {https://doi.org/10.1038/s41567-021-01404-y} {\bibfield  {journal} {\bibinfo
  {journal} {Nat. Phys.}\ }\textbf {\bibinfo {volume} {18}},\ \bibinfo {pages}
  {137} (\bibinfo {year} {2022})}\BibitemShut {NoStop}%
\bibitem [{\citenamefont {Jiang}\ \emph {et~al.}(2023)\citenamefont {Jiang},
  \citenamefont {Wu}, \citenamefont {Yin}, \citenamefont {Wang}, \citenamefont
  {Hasan}, \citenamefont {Wilson}, \citenamefont {Chen},\ and\ \citenamefont
  {Hu}}]{review_Jiang2023}%
  \BibitemOpen
  \bibfield  {author} {\bibinfo {author} {\bibfnamefont {K.}~\bibnamefont
  {Jiang}}, \bibinfo {author} {\bibfnamefont {T.}~\bibnamefont {Wu}}, \bibinfo
  {author} {\bibfnamefont {J.-X.}\ \bibnamefont {Yin}}, \bibinfo {author}
  {\bibfnamefont {Z.}~\bibnamefont {Wang}}, \bibinfo {author} {\bibfnamefont
  {M.~Z.}\ \bibnamefont {Hasan}}, \bibinfo {author} {\bibfnamefont {S.~D.}\
  \bibnamefont {Wilson}}, \bibinfo {author} {\bibfnamefont {X.}~\bibnamefont
  {Chen}},\ and\ \bibinfo {author} {\bibfnamefont {J.}~\bibnamefont {Hu}},\
  }\bibfield  {title} {\bibinfo {title} {Kagome superconductors {AV$_3$Sb$_5$}
  ({A} = {K}, {Rb}, {Cs})},\ }\href {https://doi.org/10.1093/nsr/nwac199}
  {\bibfield  {journal} {\bibinfo  {journal} {Natl. Sci. Rev.}\ }\textbf
  {\bibinfo {volume} {10}},\ \bibinfo {pages} {nwac199} (\bibinfo {year}
  {2023})}\BibitemShut {NoStop}%
\bibitem [{\citenamefont {Wilson}\ and\ \citenamefont
  {Ortiz}(2024)}]{review_Wilson2024}%
  \BibitemOpen
  \bibfield  {author} {\bibinfo {author} {\bibfnamefont {S.~D.}\ \bibnamefont
  {Wilson}}\ and\ \bibinfo {author} {\bibfnamefont {B.~R.}\ \bibnamefont
  {Ortiz}},\ }\bibfield  {title} {\bibinfo {title} {{AV$_3$Sb$_5$} kagome
  superconductors},\ }\href {https://doi.org/10.1038/s41578-024-00677-y}
  {\bibfield  {journal} {\bibinfo  {journal} {Nat. Rev. Mater.}\ }\textbf
  {\bibinfo {volume} {9}},\ \bibinfo {pages} {420} (\bibinfo {year}
  {2024})}\BibitemShut {NoStop}%
\bibitem [{\citenamefont {Yu}\ and\ \citenamefont
  {Li}(2012)}]{matrix-element-ef1}%
  \BibitemOpen
  \bibfield  {author} {\bibinfo {author} {\bibfnamefont {S.-L.}\ \bibnamefont
  {Yu}}\ and\ \bibinfo {author} {\bibfnamefont {J.-X.}\ \bibnamefont {Li}},\
  }\bibfield  {title} {\bibinfo {title} {Chiral superconducting phase and
  chiral spin-density-wave phase in a hubbard model on the kagome lattice},\
  }\href {https://doi.org/10.1103/PhysRevB.85.144402} {\bibfield  {journal}
  {\bibinfo  {journal} {Phys. Rev. B}\ }\textbf {\bibinfo {volume} {85}},\
  \bibinfo {pages} {144402} (\bibinfo {year} {2012})}\BibitemShut {NoStop}%
\bibitem [{\citenamefont {Kiesel}\ and\ \citenamefont
  {Thomale}(2012)}]{matrix-element-ef2}%
  \BibitemOpen
  \bibfield  {author} {\bibinfo {author} {\bibfnamefont {M.~L.}\ \bibnamefont
  {Kiesel}}\ and\ \bibinfo {author} {\bibfnamefont {R.}~\bibnamefont
  {Thomale}},\ }\bibfield  {title} {\bibinfo {title} {Sublattice interference
  in the kagome hubbard model},\ }\href
  {https://doi.org/10.1103/PhysRevB.86.121105} {\bibfield  {journal} {\bibinfo
  {journal} {Phys. Rev. B}\ }\textbf {\bibinfo {volume} {86}},\ \bibinfo
  {pages} {121105} (\bibinfo {year} {2012})}\BibitemShut {NoStop}%
\bibitem [{\citenamefont {Wang}\ \emph {et~al.}(2013)\citenamefont {Wang},
  \citenamefont {Li}, \citenamefont {Xiang},\ and\ \citenamefont
  {Wang}}]{matrix-element-ef3}%
  \BibitemOpen
  \bibfield  {author} {\bibinfo {author} {\bibfnamefont {W.-S.}\ \bibnamefont
  {Wang}}, \bibinfo {author} {\bibfnamefont {Z.-Z.}\ \bibnamefont {Li}},
  \bibinfo {author} {\bibfnamefont {Y.-Y.}\ \bibnamefont {Xiang}},\ and\
  \bibinfo {author} {\bibfnamefont {Q.-H.}\ \bibnamefont {Wang}},\ }\bibfield
  {title} {\bibinfo {title} {Competing electronic orders on kagome lattices at
  van hove filling},\ }\href {https://doi.org/10.1103/PhysRevB.87.115135}
  {\bibfield  {journal} {\bibinfo  {journal} {Phys. Rev. B}\ }\textbf {\bibinfo
  {volume} {87}},\ \bibinfo {pages} {115135} (\bibinfo {year}
  {2013})}\BibitemShut {NoStop}%
\bibitem [{\citenamefont {Kiesel}\ \emph {et~al.}(2013)\citenamefont {Kiesel},
  \citenamefont {Platt},\ and\ \citenamefont {Thomale}}]{matrix-element-ef4}%
  \BibitemOpen
  \bibfield  {author} {\bibinfo {author} {\bibfnamefont {M.~L.}\ \bibnamefont
  {Kiesel}}, \bibinfo {author} {\bibfnamefont {C.}~\bibnamefont {Platt}},\ and\
  \bibinfo {author} {\bibfnamefont {R.}~\bibnamefont {Thomale}},\ }\bibfield
  {title} {\bibinfo {title} {Unconventional fermi surface instabilities in the
  kagome hubbard model},\ }\href
  {https://doi.org/10.1103/PhysRevLett.110.126405} {\bibfield  {journal}
  {\bibinfo  {journal} {Phys. Rev. Lett.}\ }\textbf {\bibinfo {volume} {110}},\
  \bibinfo {pages} {126405} (\bibinfo {year} {2013})}\BibitemShut {NoStop}%
\bibitem [{\citenamefont {Dong}\ \emph {et~al.}(2023)\citenamefont {Dong},
  \citenamefont {Wang},\ and\ \citenamefont {Zhou}}]{Ziqiang-2023}%
  \BibitemOpen
  \bibfield  {author} {\bibinfo {author} {\bibfnamefont {J.-W.}\ \bibnamefont
  {Dong}}, \bibinfo {author} {\bibfnamefont {Z.}~\bibnamefont {Wang}},\ and\
  \bibinfo {author} {\bibfnamefont {S.}~\bibnamefont {Zhou}},\ }\bibfield
  {title} {\bibinfo {title} {Loop-current charge density wave driven by
  long-range coulomb repulsion on the kagom\'e lattice},\ }\href
  {https://doi.org/10.1103/PhysRevB.107.045127} {\bibfield  {journal} {\bibinfo
   {journal} {Phys. Rev. B}\ }\textbf {\bibinfo {volume} {107}},\ \bibinfo
  {pages} {045127} (\bibinfo {year} {2023})}\BibitemShut {NoStop}%
\bibitem [{\citenamefont {Yang}\ \emph {et~al.}(2024)\citenamefont {Yang},
  \citenamefont {Yao}, \citenamefont {Wang},\ and\ \citenamefont
  {Wang}}]{QGY-MY}%
  \BibitemOpen
  \bibfield  {author} {\bibinfo {author} {\bibfnamefont {Q.-G.}\ \bibnamefont
  {Yang}}, \bibinfo {author} {\bibfnamefont {M.}~\bibnamefont {Yao}}, \bibinfo
  {author} {\bibfnamefont {D.}~\bibnamefont {Wang}},\ and\ \bibinfo {author}
  {\bibfnamefont {Q.-H.}\ \bibnamefont {Wang}},\ }\bibfield  {title} {\bibinfo
  {title} {Charge bond order and $s$-wave superconductivity in the kagome
  lattice with electron-phonon coupling and electron-electron interaction},\
  }\href {https://doi.org/10.1103/PhysRevB.109.075130} {\bibfield  {journal}
  {\bibinfo  {journal} {Phys. Rev. B}\ }\textbf {\bibinfo {volume} {109}},\
  \bibinfo {pages} {075130} (\bibinfo {year} {2024})}\BibitemShut {NoStop}%
\bibitem [{\citenamefont {Liu}\ \emph {et~al.}(2024)\citenamefont {Liu},
  \citenamefont {Liu}, \citenamefont {Wang}, \citenamefont {Wang},\ and\
  \citenamefont {Wang}}]{YQL}%
  \BibitemOpen
  \bibfield  {author} {\bibinfo {author} {\bibfnamefont {Y.-Q.}\ \bibnamefont
  {Liu}}, \bibinfo {author} {\bibfnamefont {Y.-B.}\ \bibnamefont {Liu}},
  \bibinfo {author} {\bibfnamefont {W.-S.}\ \bibnamefont {Wang}}, \bibinfo
  {author} {\bibfnamefont {D.}~\bibnamefont {Wang}},\ and\ \bibinfo {author}
  {\bibfnamefont {Q.-H.}\ \bibnamefont {Wang}},\ }\bibfield  {title} {\bibinfo
  {title} {Electronic orders on the kagome lattice at the lower van hove
  filling},\ }\href {https://doi.org/10.1103/PhysRevB.109.075127} {\bibfield
  {journal} {\bibinfo  {journal} {Phys. Rev. B}\ }\textbf {\bibinfo {volume}
  {109}},\ \bibinfo {pages} {075127} (\bibinfo {year} {2024})}\BibitemShut
  {NoStop}%
\bibitem [{Note1()}]{Note1}%
  \BibitemOpen
  \bibinfo {note} {See Supplemental Material at [URL] for theoretical technique
  details, further numerical results, and a microscopic derivations of the GL
  theory. The Supplemental Material also contains Refs. \cite
  {scanningJosephson,Jiang-2021-unconventional, Xu-2021-Multiband,
  zhao-2021-cascade, luo_unique_2023}.}\BibitemShut {Stop}%
\bibitem [{\citenamefont {Mine}\ \emph {et~al.}(2024)\citenamefont {Mine},
  \citenamefont {Zhong}, \citenamefont {Liu}, \citenamefont {Suzuki},
  \citenamefont {Najafzadeh}, \citenamefont {Uchiyama}, \citenamefont {Yin},
  \citenamefont {Wu}, \citenamefont {Shi}, \citenamefont {Wang}, \citenamefont
  {Yao},\ and\ \citenamefont {Okazaki}}]{gapless-ARPES}%
  \BibitemOpen
  \bibfield  {author} {\bibinfo {author} {\bibfnamefont {A.}~\bibnamefont
  {Mine}}, \bibinfo {author} {\bibfnamefont {Y.}~\bibnamefont {Zhong}},
  \bibinfo {author} {\bibfnamefont {J.}~\bibnamefont {Liu}}, \bibinfo {author}
  {\bibfnamefont {T.}~\bibnamefont {Suzuki}}, \bibinfo {author} {\bibfnamefont
  {S.}~\bibnamefont {Najafzadeh}}, \bibinfo {author} {\bibfnamefont
  {T.}~\bibnamefont {Uchiyama}}, \bibinfo {author} {\bibfnamefont {J.-X.}\
  \bibnamefont {Yin}}, \bibinfo {author} {\bibfnamefont {X.}~\bibnamefont
  {Wu}}, \bibinfo {author} {\bibfnamefont {X.}~\bibnamefont {Shi}}, \bibinfo
  {author} {\bibfnamefont {Z.}~\bibnamefont {Wang}}, \bibinfo {author}
  {\bibfnamefont {Y.}~\bibnamefont {Yao}},\ and\ \bibinfo {author}
  {\bibfnamefont {K.}~\bibnamefont {Okazaki}},\ }\href@noop {} {\bibinfo
  {title} {Direct observation of anisotropic cooper pairing in kagome
  superconductor {CsV$_3$Sb$_5$}}} (\bibinfo {year} {2024}),\ \Eprint
  {https://arxiv.org/abs/2404.18472} {arXiv:2404.18472 [cond-mat.supr-con]}
  \BibitemShut {NoStop}%
\bibitem [{\citenamefont {Sivan}\ and\ \citenamefont
  {Imry}(1986)}]{thermalHall}%
  \BibitemOpen
  \bibfield  {author} {\bibinfo {author} {\bibfnamefont {U.}~\bibnamefont
  {Sivan}}\ and\ \bibinfo {author} {\bibfnamefont {Y.}~\bibnamefont {Imry}},\
  }\bibfield  {title} {\bibinfo {title} {Multichannel landauer formula for
  thermoelectric transport with application to thermopower near the mobility
  edge},\ }\href {https://doi.org/10.1103/PhysRevB.33.551} {\bibfield
  {journal} {\bibinfo  {journal} {Phys. Rev. B}\ }\textbf {\bibinfo {volume}
  {33}},\ \bibinfo {pages} {551} (\bibinfo {year} {1986})}\BibitemShut
  {NoStop}%
\bibitem [{\citenamefont {Feng}\ \emph {et~al.}(2021)\citenamefont {Feng},
  \citenamefont {Jiang}, \citenamefont {Wang},\ and\ \citenamefont
  {Hu}}]{cdw-1}%
  \BibitemOpen
  \bibfield  {author} {\bibinfo {author} {\bibfnamefont {X.}~\bibnamefont
  {Feng}}, \bibinfo {author} {\bibfnamefont {K.}~\bibnamefont {Jiang}},
  \bibinfo {author} {\bibfnamefont {Z.}~\bibnamefont {Wang}},\ and\ \bibinfo
  {author} {\bibfnamefont {J.}~\bibnamefont {Hu}},\ }\bibfield  {title}
  {\bibinfo {title} {Chiral flux phase in the kagome superconductor
  {AV$_3$Sb$_5$}},\ }\href
  {https://doi.org/https://doi.org/10.1016/j.scib.2021.04.043} {\bibfield
  {journal} {\bibinfo  {journal} {Science Bulletin}\ }\textbf {\bibinfo
  {volume} {66}},\ \bibinfo {pages} {1384} (\bibinfo {year}
  {2021})}\BibitemShut {NoStop}%
\bibitem [{\citenamefont {Denner}\ \emph {et~al.}(2021)\citenamefont {Denner},
  \citenamefont {Thomale},\ and\ \citenamefont {Neupert}}]{cdw-2}%
  \BibitemOpen
  \bibfield  {author} {\bibinfo {author} {\bibfnamefont {M.~M.}\ \bibnamefont
  {Denner}}, \bibinfo {author} {\bibfnamefont {R.}~\bibnamefont {Thomale}},\
  and\ \bibinfo {author} {\bibfnamefont {T.}~\bibnamefont {Neupert}},\
  }\bibfield  {title} {\bibinfo {title} {Analysis of charge order in the kagome
  metal {$A{\mathrm{V}}_{3}{\mathrm{Sb}}_{5}$
  ($A=\mathrm{K},\mathrm{Rb},\mathrm{Cs}$)}},\ }\href
  {https://doi.org/10.1103/PhysRevLett.127.217601} {\bibfield  {journal}
  {\bibinfo  {journal} {Phys. Rev. Lett.}\ }\textbf {\bibinfo {volume} {127}},\
  \bibinfo {pages} {217601} (\bibinfo {year} {2021})}\BibitemShut {NoStop}%
\bibitem [{\citenamefont {Lin}\ and\ \citenamefont
  {Nandkishore}(2021)}]{cdw-3}%
  \BibitemOpen
  \bibfield  {author} {\bibinfo {author} {\bibfnamefont {Y.-P.}\ \bibnamefont
  {Lin}}\ and\ \bibinfo {author} {\bibfnamefont {R.~M.}\ \bibnamefont
  {Nandkishore}},\ }\bibfield  {title} {\bibinfo {title} {{Complex charge
  density waves at Van Hove singularity on hexagonal lattices: Haldane-model
  phase diagram and potential realization in the kagome metals
  $A{V}_{3}{\mathrm{Sb}}_{5}$ ($A$=K, Rb, Cs)}},\ }\href
  {https://doi.org/10.1103/PhysRevB.104.045122} {\bibfield  {journal} {\bibinfo
   {journal} {Phys. Rev. B}\ }\textbf {\bibinfo {volume} {104}},\ \bibinfo
  {pages} {045122} (\bibinfo {year} {2021})}\BibitemShut {NoStop}%
\bibitem [{\citenamefont {Park}\ \emph {et~al.}(2021)\citenamefont {Park},
  \citenamefont {Ye},\ and\ \citenamefont {Balents}}]{cdw-4}%
  \BibitemOpen
  \bibfield  {author} {\bibinfo {author} {\bibfnamefont {T.}~\bibnamefont
  {Park}}, \bibinfo {author} {\bibfnamefont {M.}~\bibnamefont {Ye}},\ and\
  \bibinfo {author} {\bibfnamefont {L.}~\bibnamefont {Balents}},\ }\bibfield
  {title} {\bibinfo {title} {Electronic instabilities of kagome metals: Saddle
  points and landau theory},\ }\href
  {https://doi.org/10.1103/PhysRevB.104.035142} {\bibfield  {journal} {\bibinfo
   {journal} {Phys. Rev. B}\ }\textbf {\bibinfo {volume} {104}},\ \bibinfo
  {pages} {035142} (\bibinfo {year} {2021})}\BibitemShut {NoStop}%
\bibitem [{\citenamefont {Christensen}\ \emph {et~al.}(2021)\citenamefont
  {Christensen}, \citenamefont {Birol}, \citenamefont {Andersen},\ and\
  \citenamefont {Fernandes}}]{cdw-5}%
  \BibitemOpen
  \bibfield  {author} {\bibinfo {author} {\bibfnamefont {M.~H.}\ \bibnamefont
  {Christensen}}, \bibinfo {author} {\bibfnamefont {T.}~\bibnamefont {Birol}},
  \bibinfo {author} {\bibfnamefont {B.~M.}\ \bibnamefont {Andersen}},\ and\
  \bibinfo {author} {\bibfnamefont {R.~M.}\ \bibnamefont {Fernandes}},\
  }\bibfield  {title} {\bibinfo {title} {Theory of the charge density wave in
  $a{\mathrm{v}}_{3}{\mathrm{sb}}_{5}$ kagome metals},\ }\href
  {https://doi.org/10.1103/PhysRevB.104.214513} {\bibfield  {journal} {\bibinfo
   {journal} {Phys. Rev. B}\ }\textbf {\bibinfo {volume} {104}},\ \bibinfo
  {pages} {214513} (\bibinfo {year} {2021})}\BibitemShut {NoStop}%
\bibitem [{\citenamefont {Christensen}\ \emph {et~al.}(2022)\citenamefont
  {Christensen}, \citenamefont {Birol}, \citenamefont {Andersen},\ and\
  \citenamefont {Fernandes}}]{cdw-6}%
  \BibitemOpen
  \bibfield  {author} {\bibinfo {author} {\bibfnamefont {M.~H.}\ \bibnamefont
  {Christensen}}, \bibinfo {author} {\bibfnamefont {T.}~\bibnamefont {Birol}},
  \bibinfo {author} {\bibfnamefont {B.~M.}\ \bibnamefont {Andersen}},\ and\
  \bibinfo {author} {\bibfnamefont {R.~M.}\ \bibnamefont {Fernandes}},\
  }\bibfield  {title} {\bibinfo {title} {Loop currents in
  {$A{\mathrm{V}}_{3}{\mathrm{Sb}}_{5}$} kagome metals: Multipolar and toroidal
  magnetic orders},\ }\href {https://doi.org/10.1103/PhysRevB.106.144504}
  {\bibfield  {journal} {\bibinfo  {journal} {Phys. Rev. B}\ }\textbf {\bibinfo
  {volume} {106}},\ \bibinfo {pages} {144504} (\bibinfo {year}
  {2022})}\BibitemShut {NoStop}%
\bibitem [{\citenamefont {Kang}\ \emph {et~al.}(2022)\citenamefont {Kang},
  \citenamefont {Fang}, \citenamefont {Kim}, \citenamefont {Ortiz},
  \citenamefont {Ryu}, \citenamefont {Kim}, \citenamefont {Yoo}, \citenamefont
  {Sangiovanni}, \citenamefont {Di~Sante}, \citenamefont {Park}, \citenamefont
  {Jozwiak}, \citenamefont {Bostwick}, \citenamefont {Rotenberg}, \citenamefont
  {Kaxiras}, \citenamefont {Wilson}, \citenamefont {Park},\ and\ \citenamefont
  {Comin}}]{ARPES-2022-1}%
  \BibitemOpen
  \bibfield  {author} {\bibinfo {author} {\bibfnamefont {M.}~\bibnamefont
  {Kang}}, \bibinfo {author} {\bibfnamefont {S.}~\bibnamefont {Fang}}, \bibinfo
  {author} {\bibfnamefont {J.-K.}\ \bibnamefont {Kim}}, \bibinfo {author}
  {\bibfnamefont {B.~R.}\ \bibnamefont {Ortiz}}, \bibinfo {author}
  {\bibfnamefont {S.~H.}\ \bibnamefont {Ryu}}, \bibinfo {author} {\bibfnamefont
  {J.}~\bibnamefont {Kim}}, \bibinfo {author} {\bibfnamefont {J.}~\bibnamefont
  {Yoo}}, \bibinfo {author} {\bibfnamefont {G.}~\bibnamefont {Sangiovanni}},
  \bibinfo {author} {\bibfnamefont {D.}~\bibnamefont {Di~Sante}}, \bibinfo
  {author} {\bibfnamefont {B.-G.}\ \bibnamefont {Park}}, \bibinfo {author}
  {\bibfnamefont {C.}~\bibnamefont {Jozwiak}}, \bibinfo {author} {\bibfnamefont
  {A.}~\bibnamefont {Bostwick}}, \bibinfo {author} {\bibfnamefont
  {E.}~\bibnamefont {Rotenberg}}, \bibinfo {author} {\bibfnamefont
  {E.}~\bibnamefont {Kaxiras}}, \bibinfo {author} {\bibfnamefont {S.~D.}\
  \bibnamefont {Wilson}}, \bibinfo {author} {\bibfnamefont {J.-H.}\
  \bibnamefont {Park}},\ and\ \bibinfo {author} {\bibfnamefont
  {R.}~\bibnamefont {Comin}},\ }\bibfield  {title} {\bibinfo {title} {Twofold
  van {Hove} singularity and origin of charge order in topological kagome
  superconductor {CsV$_3$Sb$_5$}},\ }\href
  {https://doi.org/10.1038/s41567-021-01451-5} {\bibfield  {journal} {\bibinfo
  {journal} {Nat. Phys.}\ }\textbf {\bibinfo {volume} {18}},\ \bibinfo {pages}
  {301} (\bibinfo {year} {2022})}\BibitemShut {NoStop}%
\bibitem [{\citenamefont {Hu}\ \emph {et~al.}(2022)\citenamefont {Hu},
  \citenamefont {Wu}, \citenamefont {Ortiz}, \citenamefont {Ju}, \citenamefont
  {Han}, \citenamefont {Ma}, \citenamefont {Plumb}, \citenamefont {Radovic},
  \citenamefont {Thomale}, \citenamefont {Wilson}, \citenamefont {Schnyder},\
  and\ \citenamefont {Shi}}]{ARPES-2022-2}%
  \BibitemOpen
  \bibfield  {author} {\bibinfo {author} {\bibfnamefont {Y.}~\bibnamefont
  {Hu}}, \bibinfo {author} {\bibfnamefont {X.}~\bibnamefont {Wu}}, \bibinfo
  {author} {\bibfnamefont {B.~R.}\ \bibnamefont {Ortiz}}, \bibinfo {author}
  {\bibfnamefont {S.}~\bibnamefont {Ju}}, \bibinfo {author} {\bibfnamefont
  {X.}~\bibnamefont {Han}}, \bibinfo {author} {\bibfnamefont {J.}~\bibnamefont
  {Ma}}, \bibinfo {author} {\bibfnamefont {N.~C.}\ \bibnamefont {Plumb}},
  \bibinfo {author} {\bibfnamefont {M.}~\bibnamefont {Radovic}}, \bibinfo
  {author} {\bibfnamefont {R.}~\bibnamefont {Thomale}}, \bibinfo {author}
  {\bibfnamefont {S.~D.}\ \bibnamefont {Wilson}}, \bibinfo {author}
  {\bibfnamefont {A.~P.}\ \bibnamefont {Schnyder}},\ and\ \bibinfo {author}
  {\bibfnamefont {M.}~\bibnamefont {Shi}},\ }\bibfield  {title} {\bibinfo
  {title} {Rich nature of {Van} {Hove} singularities in {Kagome} superconductor
  {CsV$_3$Sb$_5$}},\ }\href {https://doi.org/10.1038/s41467-022-29828-x}
  {\bibfield  {journal} {\bibinfo  {journal} {Nat. Commun.}\ }\textbf {\bibinfo
  {volume} {13}},\ \bibinfo {pages} {2220} (\bibinfo {year}
  {2022})}\BibitemShut {NoStop}%
\bibitem [{\citenamefont {Li}\ \emph {et~al.}(2023)\citenamefont {Li},
  \citenamefont {Oh}, \citenamefont {Kang}, \citenamefont {Zhao}, \citenamefont
  {Ortiz}, \citenamefont {Oey}, \citenamefont {Fang}, \citenamefont {Ren},
  \citenamefont {Jozwiak}, \citenamefont {Bostwick}, \citenamefont {Rotenberg},
  \citenamefont {Checkelsky}, \citenamefont {Wang}, \citenamefont {Wilson},
  \citenamefont {Comin},\ and\ \citenamefont {Zeljkovic}}]{li-prx-2023}%
  \BibitemOpen
  \bibfield  {author} {\bibinfo {author} {\bibfnamefont {H.}~\bibnamefont
  {Li}}, \bibinfo {author} {\bibfnamefont {D.}~\bibnamefont {Oh}}, \bibinfo
  {author} {\bibfnamefont {M.}~\bibnamefont {Kang}}, \bibinfo {author}
  {\bibfnamefont {H.}~\bibnamefont {Zhao}}, \bibinfo {author} {\bibfnamefont
  {B.~R.}\ \bibnamefont {Ortiz}}, \bibinfo {author} {\bibfnamefont
  {Y.}~\bibnamefont {Oey}}, \bibinfo {author} {\bibfnamefont {S.}~\bibnamefont
  {Fang}}, \bibinfo {author} {\bibfnamefont {Z.}~\bibnamefont {Ren}}, \bibinfo
  {author} {\bibfnamefont {C.}~\bibnamefont {Jozwiak}}, \bibinfo {author}
  {\bibfnamefont {A.}~\bibnamefont {Bostwick}}, \bibinfo {author}
  {\bibfnamefont {E.}~\bibnamefont {Rotenberg}}, \bibinfo {author}
  {\bibfnamefont {J.~G.}\ \bibnamefont {Checkelsky}}, \bibinfo {author}
  {\bibfnamefont {Z.}~\bibnamefont {Wang}}, \bibinfo {author} {\bibfnamefont
  {S.~D.}\ \bibnamefont {Wilson}}, \bibinfo {author} {\bibfnamefont
  {R.}~\bibnamefont {Comin}},\ and\ \bibinfo {author} {\bibfnamefont
  {I.}~\bibnamefont {Zeljkovic}},\ }\bibfield  {title} {\bibinfo {title} {Small
  fermi pockets intertwined with charge stripes and pair density wave order in
  a kagome superconductor},\ }\href
  {https://doi.org/10.1103/PhysRevX.13.031030} {\bibfield  {journal} {\bibinfo
  {journal} {Phys. Rev. X}\ }\textbf {\bibinfo {volume} {13}},\ \bibinfo
  {pages} {031030} (\bibinfo {year} {2023})}\BibitemShut {NoStop}%
\bibitem [{\citenamefont {Hu}\ \emph {et~al.}(2023)\citenamefont {Hu},
  \citenamefont {Wu}, \citenamefont {Schnyder},\ and\ \citenamefont
  {Shi}}]{review-1}%
  \BibitemOpen
  \bibfield  {author} {\bibinfo {author} {\bibfnamefont {Y.}~\bibnamefont
  {Hu}}, \bibinfo {author} {\bibfnamefont {X.}~\bibnamefont {Wu}}, \bibinfo
  {author} {\bibfnamefont {A.~P.}\ \bibnamefont {Schnyder}},\ and\ \bibinfo
  {author} {\bibfnamefont {M.}~\bibnamefont {Shi}},\ }\bibfield  {title}
  {\bibinfo {title} {Electronic landscape of kagome superconductors
  {{AV$_3$Sb$_5$} ({A} = {K}, {Rb}, {Cs})} from angle-resolved photoemission
  spectroscopy},\ }\href {https://doi.org/10.1038/s41535-023-00599-y}
  {\bibfield  {journal} {\bibinfo  {journal} {npj Quantum Mater.}\ }\textbf
  {\bibinfo {volume} {8}},\ \bibinfo {pages} {67} (\bibinfo {year}
  {2023})}\BibitemShut {NoStop}%
\bibitem [{\citenamefont {Deng}\ \emph {et~al.}(2024)\citenamefont {Deng},
  \citenamefont {Qin}, \citenamefont {Liu}, \citenamefont {Yang}, \citenamefont
  {Fu}, \citenamefont {Zhang}, \citenamefont {Wu}, \citenamefont {Wang},
  \citenamefont {Shi}, \citenamefont {Liu}, \citenamefont {Liu}, \citenamefont
  {Yan}, \citenamefont {Song}, \citenamefont {Xu}, \citenamefont {Zhao},
  \citenamefont {Yi}, \citenamefont {Xu}, \citenamefont {Hohmann},
  \citenamefont {Holbek}, \citenamefont {Dürrnagel}, \citenamefont {Zhou},
  \citenamefont {Chang}, \citenamefont {Yao}, \citenamefont {Wang},
  \citenamefont {Guguchia}, \citenamefont {Neupert}, \citenamefont {Thomale},
  \citenamefont {Fischer},\ and\ \citenamefont {Yin}}]{Deng_N_2024}%
  \BibitemOpen
  \bibfield  {author} {\bibinfo {author} {\bibfnamefont {H.}~\bibnamefont
  {Deng}}, \bibinfo {author} {\bibfnamefont {H.}~\bibnamefont {Qin}}, \bibinfo
  {author} {\bibfnamefont {G.}~\bibnamefont {Liu}}, \bibinfo {author}
  {\bibfnamefont {T.}~\bibnamefont {Yang}}, \bibinfo {author} {\bibfnamefont
  {R.}~\bibnamefont {Fu}}, \bibinfo {author} {\bibfnamefont {Z.}~\bibnamefont
  {Zhang}}, \bibinfo {author} {\bibfnamefont {X.}~\bibnamefont {Wu}}, \bibinfo
  {author} {\bibfnamefont {Z.}~\bibnamefont {Wang}}, \bibinfo {author}
  {\bibfnamefont {Y.}~\bibnamefont {Shi}}, \bibinfo {author} {\bibfnamefont
  {J.}~\bibnamefont {Liu}}, \bibinfo {author} {\bibfnamefont {H.}~\bibnamefont
  {Liu}}, \bibinfo {author} {\bibfnamefont {X.-Y.}\ \bibnamefont {Yan}},
  \bibinfo {author} {\bibfnamefont {W.}~\bibnamefont {Song}}, \bibinfo {author}
  {\bibfnamefont {X.}~\bibnamefont {Xu}}, \bibinfo {author} {\bibfnamefont
  {Y.}~\bibnamefont {Zhao}}, \bibinfo {author} {\bibfnamefont {M.}~\bibnamefont
  {Yi}}, \bibinfo {author} {\bibfnamefont {G.}~\bibnamefont {Xu}}, \bibinfo
  {author} {\bibfnamefont {H.}~\bibnamefont {Hohmann}}, \bibinfo {author}
  {\bibfnamefont {S.~C.}\ \bibnamefont {Holbek}}, \bibinfo {author}
  {\bibfnamefont {M.}~\bibnamefont {Dürrnagel}}, \bibinfo {author}
  {\bibfnamefont {S.}~\bibnamefont {Zhou}}, \bibinfo {author} {\bibfnamefont
  {G.}~\bibnamefont {Chang}}, \bibinfo {author} {\bibfnamefont
  {Y.}~\bibnamefont {Yao}}, \bibinfo {author} {\bibfnamefont {Q.}~\bibnamefont
  {Wang}}, \bibinfo {author} {\bibfnamefont {Z.}~\bibnamefont {Guguchia}},
  \bibinfo {author} {\bibfnamefont {T.}~\bibnamefont {Neupert}}, \bibinfo
  {author} {\bibfnamefont {R.}~\bibnamefont {Thomale}}, \bibinfo {author}
  {\bibfnamefont {M.~H.}\ \bibnamefont {Fischer}},\ and\ \bibinfo {author}
  {\bibfnamefont {J.-X.}\ \bibnamefont {Yin}},\ }\bibfield  {title} {\bibinfo
  {title} {Chiral kagome superconductivity modulations with residual {Fermi}
  arcs},\ }\href {https://doi.org/10.1038/s41586-024-07798-y} {\bibfield
  {journal} {\bibinfo  {journal} {Nature}\ }\textbf {\bibinfo {volume} {632}},\
  \bibinfo {pages} {775} (\bibinfo {year} {2024})}\BibitemShut {NoStop}%
\bibitem [{\citenamefont {Sigrist}\ \emph {et~al.}(1989)\citenamefont
  {Sigrist}, \citenamefont {Rice},\ and\ \citenamefont {Ueda}}]{Sigrist1989}%
  \BibitemOpen
  \bibfield  {author} {\bibinfo {author} {\bibfnamefont {M.}~\bibnamefont
  {Sigrist}}, \bibinfo {author} {\bibfnamefont {T.~M.}\ \bibnamefont {Rice}},\
  and\ \bibinfo {author} {\bibfnamefont {K.}~\bibnamefont {Ueda}},\ }\bibfield
  {title} {\bibinfo {title} {Low-field magnetic response of complex
  superconductors},\ }\href {https://doi.org/10.1103/PhysRevLett.63.1727}
  {\bibfield  {journal} {\bibinfo  {journal} {Phys. Rev. Lett.}\ }\textbf
  {\bibinfo {volume} {63}},\ \bibinfo {pages} {1727} (\bibinfo {year}
  {1989})}\BibitemShut {NoStop}%
\bibitem [{\citenamefont {Babaev}(2002)}]{Babaev2002}%
  \BibitemOpen
  \bibfield  {author} {\bibinfo {author} {\bibfnamefont {E.}~\bibnamefont
  {Babaev}},\ }\bibfield  {title} {\bibinfo {title} {Vortices with {{Fractional
  Flux}} in {{Two-Gap Superconductors}} and in {{Extended Faddeev Model}}},\
  }\href {https://doi.org/10.1103/PhysRevLett.89.067001} {\bibfield  {journal}
  {\bibinfo  {journal} {Phys. Rev. Lett.}\ }\textbf {\bibinfo {volume} {89}},\
  \bibinfo {pages} {067001} (\bibinfo {year} {2002})}\BibitemShut {NoStop}%
\bibitem [{\citenamefont {Ge}\ \emph {et~al.}(2024)\citenamefont {Ge},
  \citenamefont {Wang}, \citenamefont {Xing}, \citenamefont {Yin},
  \citenamefont {Wang}, \citenamefont {Shen}, \citenamefont {Lei},
  \citenamefont {Wang},\ and\ \citenamefont {Wang}}]{4e6e_PRX_2024}%
  \BibitemOpen
  \bibfield  {author} {\bibinfo {author} {\bibfnamefont {J.}~\bibnamefont
  {Ge}}, \bibinfo {author} {\bibfnamefont {P.}~\bibnamefont {Wang}}, \bibinfo
  {author} {\bibfnamefont {Y.}~\bibnamefont {Xing}}, \bibinfo {author}
  {\bibfnamefont {Q.}~\bibnamefont {Yin}}, \bibinfo {author} {\bibfnamefont
  {A.}~\bibnamefont {Wang}}, \bibinfo {author} {\bibfnamefont {J.}~\bibnamefont
  {Shen}}, \bibinfo {author} {\bibfnamefont {H.}~\bibnamefont {Lei}}, \bibinfo
  {author} {\bibfnamefont {Z.}~\bibnamefont {Wang}},\ and\ \bibinfo {author}
  {\bibfnamefont {J.}~\bibnamefont {Wang}},\ }\bibfield  {title} {\bibinfo
  {title} {Charge-$4e$ and charge-$6e$ flux quantization and higher charge
  superconductivity in kagome superconductor ring devices},\ }\href
  {https://doi.org/10.1103/PhysRevX.14.021025} {\bibfield  {journal} {\bibinfo
  {journal} {Phys. Rev. X}\ }\textbf {\bibinfo {volume} {14}},\ \bibinfo
  {pages} {021025} (\bibinfo {year} {2024})}\BibitemShut {NoStop}%
\bibitem [{\citenamefont {\ifmmode~\check{S}\else \v{S}\fi{}makov}\ \emph
  {et~al.}(2001)\citenamefont {\ifmmode~\check{S}\else \v{S}\fi{}makov},
  \citenamefont {Martin},\ and\ \citenamefont {Balatsky}}]{scanningJosephson}%
  \BibitemOpen
  \bibfield  {author} {\bibinfo {author} {\bibfnamefont {J.}~\bibnamefont
  {\ifmmode~\check{S}\else \v{S}\fi{}makov}}, \bibinfo {author} {\bibfnamefont
  {I.}~\bibnamefont {Martin}},\ and\ \bibinfo {author} {\bibfnamefont {A.~V.}\
  \bibnamefont {Balatsky}},\ }\bibfield  {title} {\bibinfo {title} {Josephson
  scanning tunneling microscopy},\ }\href
  {https://doi.org/10.1103/PhysRevB.64.212506} {\bibfield  {journal} {\bibinfo
  {journal} {Phys. Rev. B}\ }\textbf {\bibinfo {volume} {64}},\ \bibinfo
  {pages} {212506} (\bibinfo {year} {2001})}\BibitemShut {NoStop}%
\bibitem [{\citenamefont {Jiang}\ \emph {et~al.}(2021)\citenamefont {Jiang},
  \citenamefont {Yin}, \citenamefont {Denner}, \citenamefont {Shumiya},
  \citenamefont {Ortiz}, \citenamefont {Xu}, \citenamefont {Guguchia},
  \citenamefont {He}, \citenamefont {Hossain}, \citenamefont {Liu} \emph
  {et~al.}}]{Jiang-2021-unconventional}%
  \BibitemOpen
  \bibfield  {author} {\bibinfo {author} {\bibfnamefont {Y.-X.}\ \bibnamefont
  {Jiang}}, \bibinfo {author} {\bibfnamefont {J.-X.}\ \bibnamefont {Yin}},
  \bibinfo {author} {\bibfnamefont {M.~M.}\ \bibnamefont {Denner}}, \bibinfo
  {author} {\bibfnamefont {N.}~\bibnamefont {Shumiya}}, \bibinfo {author}
  {\bibfnamefont {B.~R.}\ \bibnamefont {Ortiz}}, \bibinfo {author}
  {\bibfnamefont {G.}~\bibnamefont {Xu}}, \bibinfo {author} {\bibfnamefont
  {Z.}~\bibnamefont {Guguchia}}, \bibinfo {author} {\bibfnamefont
  {J.}~\bibnamefont {He}}, \bibinfo {author} {\bibfnamefont {M.~S.}\
  \bibnamefont {Hossain}}, \bibinfo {author} {\bibfnamefont {X.}~\bibnamefont
  {Liu}}, \emph {et~al.},\ }\bibfield  {title} {\bibinfo {title}
  {Unconventional chiral charge order in kagome superconductor
  {KV$_3$Sb$_5$}},\ }\href
  {https://doi.org/https://doi.org/10.1038/s41563-021-01034-y} {\bibfield
  {journal} {\bibinfo  {journal} {Nat. Mater.}\ }\textbf {\bibinfo {volume}
  {20}},\ \bibinfo {pages} {1353} (\bibinfo {year} {2021})}\BibitemShut
  {NoStop}%
\bibitem [{\citenamefont {Xu}\ \emph {et~al.}(2021)\citenamefont {Xu},
  \citenamefont {Yan}, \citenamefont {Yin}, \citenamefont {Xia}, \citenamefont
  {Fang}, \citenamefont {Chen}, \citenamefont {Li}, \citenamefont {Yang},
  \citenamefont {Guo},\ and\ \citenamefont {Feng}}]{Xu-2021-Multiband}%
  \BibitemOpen
  \bibfield  {author} {\bibinfo {author} {\bibfnamefont {H.-S.}\ \bibnamefont
  {Xu}}, \bibinfo {author} {\bibfnamefont {Y.-J.}\ \bibnamefont {Yan}},
  \bibinfo {author} {\bibfnamefont {R.}~\bibnamefont {Yin}}, \bibinfo {author}
  {\bibfnamefont {W.}~\bibnamefont {Xia}}, \bibinfo {author} {\bibfnamefont
  {S.}~\bibnamefont {Fang}}, \bibinfo {author} {\bibfnamefont {Z.}~\bibnamefont
  {Chen}}, \bibinfo {author} {\bibfnamefont {Y.}~\bibnamefont {Li}}, \bibinfo
  {author} {\bibfnamefont {W.}~\bibnamefont {Yang}}, \bibinfo {author}
  {\bibfnamefont {Y.}~\bibnamefont {Guo}},\ and\ \bibinfo {author}
  {\bibfnamefont {D.-L.}\ \bibnamefont {Feng}},\ }\bibfield  {title} {\bibinfo
  {title} {Multiband superconductivity with sign-preserving order parameter in
  kagome superconductor {${\mathrm{CsV}}_{3}{\mathrm{Sb}}_{5}$}},\ }\href
  {https://doi.org/10.1103/PhysRevLett.127.187004} {\bibfield  {journal}
  {\bibinfo  {journal} {Phys. Rev. Lett.}\ }\textbf {\bibinfo {volume} {127}},\
  \bibinfo {pages} {187004} (\bibinfo {year} {2021})}\BibitemShut {NoStop}%
\bibitem [{\citenamefont {Zhao}\ \emph {et~al.}(2021)\citenamefont {Zhao},
  \citenamefont {Li}, \citenamefont {Ortiz}, \citenamefont {Teicher},
  \citenamefont {Park}, \citenamefont {Ye}, \citenamefont {Wang}, \citenamefont
  {Balents}, \citenamefont {Wilson},\ and\ \citenamefont
  {Zeljkovic}}]{zhao-2021-cascade}%
  \BibitemOpen
  \bibfield  {author} {\bibinfo {author} {\bibfnamefont {H.}~\bibnamefont
  {Zhao}}, \bibinfo {author} {\bibfnamefont {H.}~\bibnamefont {Li}}, \bibinfo
  {author} {\bibfnamefont {B.~R.}\ \bibnamefont {Ortiz}}, \bibinfo {author}
  {\bibfnamefont {S.~M.}\ \bibnamefont {Teicher}}, \bibinfo {author}
  {\bibfnamefont {T.}~\bibnamefont {Park}}, \bibinfo {author} {\bibfnamefont
  {M.}~\bibnamefont {Ye}}, \bibinfo {author} {\bibfnamefont {Z.}~\bibnamefont
  {Wang}}, \bibinfo {author} {\bibfnamefont {L.}~\bibnamefont {Balents}},
  \bibinfo {author} {\bibfnamefont {S.~D.}\ \bibnamefont {Wilson}},\ and\
  \bibinfo {author} {\bibfnamefont {I.}~\bibnamefont {Zeljkovic}},\ }\bibfield
  {title} {\bibinfo {title} {Cascade of correlated electron states in the
  kagome superconductor {CsV$_3$Sb$_5$}},\ }\href
  {https://doi.org/https://doi.org/10.1038/s41586-021-03946-w} {\bibfield
  {journal} {\bibinfo  {journal} {Nature}\ }\textbf {\bibinfo {volume} {599}},\
  \bibinfo {pages} {216} (\bibinfo {year} {2021})}\BibitemShut {NoStop}%
\bibitem [{\citenamefont {Luo}\ \emph {et~al.}(2023)\citenamefont {Luo},
  \citenamefont {Han}, \citenamefont {Liu}, \citenamefont {Chen}, \citenamefont
  {Huang}, \citenamefont {Huai}, \citenamefont {Li}, \citenamefont {Wang},
  \citenamefont {Shen}, \citenamefont {Ding}, \citenamefont {Li}, \citenamefont
  {Peng}, \citenamefont {Wei}, \citenamefont {Miao}, \citenamefont {Sun},
  \citenamefont {Ou}, \citenamefont {Xiang}, \citenamefont {Hashimoto},
  \citenamefont {Lu}, \citenamefont {Yao}, \citenamefont {Yang}, \citenamefont
  {Chen}, \citenamefont {Gao}, \citenamefont {Qiao}, \citenamefont {Wang},\
  and\ \citenamefont {He}}]{luo_unique_2023}%
  \BibitemOpen
  \bibfield  {author} {\bibinfo {author} {\bibfnamefont {Y.}~\bibnamefont
  {Luo}}, \bibinfo {author} {\bibfnamefont {Y.}~\bibnamefont {Han}}, \bibinfo
  {author} {\bibfnamefont {J.}~\bibnamefont {Liu}}, \bibinfo {author}
  {\bibfnamefont {H.}~\bibnamefont {Chen}}, \bibinfo {author} {\bibfnamefont
  {Z.}~\bibnamefont {Huang}}, \bibinfo {author} {\bibfnamefont
  {L.}~\bibnamefont {Huai}}, \bibinfo {author} {\bibfnamefont {H.}~\bibnamefont
  {Li}}, \bibinfo {author} {\bibfnamefont {B.}~\bibnamefont {Wang}}, \bibinfo
  {author} {\bibfnamefont {J.}~\bibnamefont {Shen}}, \bibinfo {author}
  {\bibfnamefont {S.}~\bibnamefont {Ding}}, \bibinfo {author} {\bibfnamefont
  {Z.}~\bibnamefont {Li}}, \bibinfo {author} {\bibfnamefont {S.}~\bibnamefont
  {Peng}}, \bibinfo {author} {\bibfnamefont {Z.}~\bibnamefont {Wei}}, \bibinfo
  {author} {\bibfnamefont {Y.}~\bibnamefont {Miao}}, \bibinfo {author}
  {\bibfnamefont {X.}~\bibnamefont {Sun}}, \bibinfo {author} {\bibfnamefont
  {Z.}~\bibnamefont {Ou}}, \bibinfo {author} {\bibfnamefont {Z.}~\bibnamefont
  {Xiang}}, \bibinfo {author} {\bibfnamefont {M.}~\bibnamefont {Hashimoto}},
  \bibinfo {author} {\bibfnamefont {D.}~\bibnamefont {Lu}}, \bibinfo {author}
  {\bibfnamefont {Y.}~\bibnamefont {Yao}}, \bibinfo {author} {\bibfnamefont
  {H.}~\bibnamefont {Yang}}, \bibinfo {author} {\bibfnamefont {X.}~\bibnamefont
  {Chen}}, \bibinfo {author} {\bibfnamefont {H.-J.}\ \bibnamefont {Gao}},
  \bibinfo {author} {\bibfnamefont {Z.}~\bibnamefont {Qiao}}, \bibinfo {author}
  {\bibfnamefont {Z.}~\bibnamefont {Wang}},\ and\ \bibinfo {author}
  {\bibfnamefont {J.}~\bibnamefont {He}},\ }\bibfield  {title} {\bibinfo
  {title} {A unique van {Hove} singularity in kagome superconductor
  {$\mathrm{CsV}_{3-x}\mathrm{Ta}_x\mathrm{Sb}_5$ } with enhanced
  superconductivity},\ }\href {https://doi.org/10.1038/s41467-023-39500-7}
  {\bibfield  {journal} {\bibinfo  {journal} {Nat. Commun.}\ }\textbf {\bibinfo
  {volume} {14}},\ \bibinfo {pages} {3819} (\bibinfo {year}
  {2023})}\BibitemShut {NoStop}%
\end{thebibliography}%
\end{document}